\documentclass{article}
\usepackage{kr}

\usepackage[page,header]{appendix}
\usepackage{titletoc}

\usepackage{times}
\usepackage{soul}
\usepackage[lowtilde]{url} 
\usepackage[hidelinks]{hyperref}
\usepackage[utf8]{inputenc}
\usepackage[small]{caption}
\usepackage{graphicx}
\usepackage{amsmath, latexsym,amsthm}
\usepackage{amsfonts, amssymb, stmaryrd}
\usepackage{booktabs}
\usepackage{algorithm}
\usepackage{algorithmic}
\usepackage{thm-restate} 
\usepackage{xspace}
\usepackage{multirow}

\usepackage{graphicx}
\usepackage{caption}
\usepackage{subcaption}
\usepackage{pgfplotstable}
\usepackage{color}
\usepackage{colortbl}

\definecolor{Lgray}{rgb}{0.86, 0.84, 0.82}

\usepackage{tikz}
\tikzset{>=latex}

\newtheorem{theorem}{Theorem}
\newtheorem{lemma}{Lemma}
\newtheorem{proposition}{Proposition}

\theoremstyle{definition} 
\newtheorem{definition}{Definition}
\newtheorem{example}{Example}
\newtheorem{remark}{Remark}

\allowdisplaybreaks

\newcommand{\algosat}{\mn{Simple}\xspace}
\newcommand{\algomaxsat}{\mn{All\text{-}MaxSAT}\xspace}
\newcommand{\algomuses}{\mn{All\text{-}MUSes}\xspace}
\newcommand{\algoassump}{\mn{Assumptions}\xspace}
\newcommand{\algocauses}{\mn{Cause\text{-}by\text{-}cause}\xspace}
\newcommand{\algoIARcauses}{\mn{IAR\text{-}causes}\xspace}
\newcommand{\algoIARfacts}{\mn{IAR\text{-}facts}\xspace}

\newcommand{\cqaprienc}{$\mn{neg}_1$\xspace}
\newcommand{\cavsatenc}{$\mn{neg}_2$\xspace}

\def\dllitecore{DL-Lite$_{\mn{core}}$\xspace}

\newcommand{\tup}[1]{\langle #1\rangle}
\newcommand{\ans}{\vec{a}}

\newcommand{\sem}{\text{Sem}\xspace}
\newcommand{\semmodels}[1]{\models_{\sem}^{#1}}
\newcommand{\bravemodels}[1]{\models_{\text{brave}}^{#1}}
\newcommand{\armodels}[1]{\models_{\text{AR}}^{#1}}
\newcommand{\iarmodels}[1]{\models_{\text{IAR}}^{#1}}

\newcommand{\conflicts}[1]{\mi{Conf}(#1)}

\newcommand{\causes}[1]{\mi{Causes}(#1)}

\newcommand{\reps}[1]{\mi{SRep}(#1)}

\newcommand{\preps}[1]{\mi{PRep}(#1)}
\newcommand{\creps}[1]{\mi{CRep}(#1)}
\newcommand{\xreps}[1]{\mi{XRep}(#1)}

\def\braveans{\mi{PotAns}}

\def\confof{\bot}

\def\facts{\mathsf{facts}}

\def\ptime{\textsc{PTime}\xspace}
\def\np{\textsc{NP}\xspace}
\def\conp{co\textsc{NP}\xspace}

\def\true{\ensuremath{\mathsf{true}}}
\def\false{\ensuremath{\mathsf{false}}}

\newcommand{\mn}[1]{\ensuremath{\mathsf{#1}}}
\newcommand{\mi}[1]{\ensuremath{\mathit{#1}}}

\newcommand{\Bmc}{\ensuremath{\mathcal{B}}}
\newcommand{\Cmc}{\ensuremath{\mathcal{C}}}
\newcommand{\Dmc}{\ensuremath{\mathcal{D}}}

\newcommand{\Lmc}{\ensuremath{\mathcal{L}}}
\newcommand{\Kmc}{\ensuremath{\mathcal{K}}}

\newcommand{\Rmc}{\ensuremath{\mathcal{R}}}

\newcommand{\Tmc}{\ensuremath{\mathcal{T}}}

\newcommand{\eg}{e.g.,~}

\newcommand{\ie}{i.e.,~}
\newcommand{\wrt}{w.r.t.~}
\newcommand{\cf}{cf.~}

\def\reachr{\mathsf{R}}

\title{Querying Inconsistent Prioritized Data with {\sc \textbf{orbits}}: \\ Algorithms, Implementation, and Experiments\footnote{This is an extended version of a paper appearing at the 19th International Conference on Principles of Knowledge Representation and Reasoning (KR 2022). }}

\author{%
Meghyn Bienvenu$^1$\and
Camille Bourgaux$^2$
\affiliations
$^1$ CNRS \& University of Bordeaux, France\\
$^2$ DI ENS, ENS, CNRS, PSL University \& Inria, Paris, France\\  
\emails
meghyn.bienvenu@labri.fr,
camille.bourgaux@ens.fr
}

\begin{document}

\maketitle

\begin{abstract}
We investigate practical algorithms for inconsistency-tolerant query answering over prioritized knowledge bases, which consist of a logical theory, a set of facts, and a priority relation between conflicting facts. We consider three well-known semantics (AR, IAR and brave) based upon two notions of optimal repairs (Pareto and completion). Deciding whether a query answer holds under these semantics is (co)NP-complete in data complexity for a large class of logical theories, and SAT-based procedures have been devised for repair-based semantics when there is no priority relation, or the relation has a special structure. The present paper introduces the first SAT encodings for Pareto- and completion-optimal repairs w.r.t.\ general priority relations and proposes several ways of employing existing and new encodings to compute answers under (optimal) repair-based semantics, by exploiting different reasoning modes of SAT solvers. The comprehensive experimental evaluation of our implementation compares both (i) the impact of adopting semantics based on different kinds of repairs, and (ii) the relative performances of alternative procedures for the same semantics. 
\end{abstract}

\section{Introduction}
The question of 
how to handle data that is inconsistent w.r.t.\ expressed 
constraints, be they given by database dependencies or ontologies, has great practical relevance. 
Data cleaning addresses this problem by 
modifying datasets so that they satisfy the constraints, often using 
heuristics to decide how 
to resolve contradictions, which may result in wrong facts being kept, or true facts removed (as discussed e.g.\ in \cite{DBLP:journals/sigmod/Fan15}).  
An alternative, more principled, 
approach is to 
adopt inconsistency-tolerant semantics in order to extract 
meaningful information from the contradictory data. 

In the database setting, such an approach goes by the name of \emph{consistent query answering} (CQA)
and has been extensively studied since the seminal work of \citeauthor{ArenasBC99} \shortcite{ArenasBC99}, see e.g.\ the recent survey by \citeauthor{DBLP:journals/sigmod/Wijsen19} \shortcite{DBLP:journals/sigmod/Wijsen19}. 
A central notion is that of a \emph{(subset) repair}, defined as a maximal subset of the dataset 
that satisfies the 
constraints. Intuitively, 
repairs represent all different ways of minimally modifying the data 
to satisfy the constraints.
As we do not know which repair corresponds to the true part of the data, 
the CQA semantics stipulates that a tuple is a query answer if 
it is an answer w.r.t.\ every repair 
(in line with how skeptical inference is defined in many KR settings). 

Inconsistency-tolerant semantics have also drawn considerable interest in the setting of ontology-mediated query answering (OMQA) \cite{DBLP:journals/jods/PoggiLCGLR08,DBLP:conf/rweb/BienvenuO15,DBLP:conf/ijcai/XiaoCKLPRZ18},
where the ontology not only specifies constraints on the data but also captures other forms of domain knowledge, which can be exploited at query time. 
In addition to the \emph{AR semantics} (the OMQA analog of the CQA semantics), several other inconsistency-tolerant semantics 
have been proposed (see \cite{DBLP:conf/rweb/BienvenuB16,DBLP:journals/ki/Bienvenu20} for surveys and references), among which:
the \emph{brave semantics} \cite{Bienvenu_TractableApproximation_long}, which only requires a tuple to be an answer w.r.t.\ some repair,
provides a natural notion of possible answer, and the \emph{IAR semantics} \cite{LemboLRRS10}, which answers queries over the intersection of the repairs, 
identifies the most reliable answers. 

The basic notion of repair can be refined by exploiting preference information. 
A prominent approach, introduced by
\citeauthor{DBLP:journals/amai/StaworkoCM12} (\citeyear{DBLP:journals/amai/StaworkoCM12}) in the database setting 
and recently explored in the OMQA setting \cite{DBLP:conf/kr/BienvenuB20}, 
assumes that preferences are given by 
a binary \emph{priority relation} between conflicting facts. 
Three notions of `best' repairs w.r.t.\ a priority relation were proposed, namely,
Pareto-optimal, globally-optimal, and completion-optimal repairs, and 
 can be used in place of subset repairs in any repair-based semantics. 


The complexity of answering queries under (optimal) repair-based semantics has 
been extensively studied in the database and OMQA settings, refer to \cite{DBLP:journals/sigmod/Wijsen19,DBLP:conf/rweb/BienvenuB16} for an overview and references. We can briefly summarize these (many!) complexity results as follows: 
query answering under the AR (or CQA) semantics is 
\conp-hard in data complexity even in the simplest of settings (e.g. key constraints, class disjointness), 
and adopting optimal repairs in place of subset repairs leads to (co)\np-hardness for the brave and IAR 
semantics as well. Membership in (co)\np holds for AR, brave, and IAR semantics w.r.t.\ subset, Pareto-optimal, and completion-optimal repairs
in the most commonly considered settings i.e. for database constraints given by primary keys or more generally, functional dependencies (FDs),
and for ontologies formulated in data-tractable description logics such as those of the DL-Lite family \cite{calvaneseetal:dllite}. 

The preceding (co)\np\ complexity results naturally suggest 
the interest of employing SAT solvers. Two recent systems, \textsf{CQAPri} and \textsf{CAvSAT}, have begun to explore such an approach. 
\textsf{CQAPri} 
(\citeauthor{DBLP:conf/aaai/BienvenuBG14} \citeyear{DBLP:conf/aaai/BienvenuBG14}; \citeyear{DBLP:journals/jair/BienvenuBG19}) 
uses tractable approximations together with calls to SAT solvers to 
answer queries over inconsistent DL-Lite knowledge bases, 
under the AR, brave, and IAR semantics,
w.r.t.\ subset repairs as well as optimal repairs for the restricted class of score-structured priority relations. 
\textsf{CAvSAT} \cite{DBLP:conf/sat/DixitK19}  
targets relational databases equipped with denial constraints (which include FDs as a special case)
and computes query answers under the AR semantics w.r.t.\  
subset repairs. 
While geared to different forms of constraints, the two systems solve essentially the same problem,
yet they employ SAT solvers in different ways. \textsf{CQAPri} makes a single SAT call for each candidate query answer,
whereas \textsf{CAvSAT} treats all candidate answers at the same time via calls to a weighted MaxSAT solver. 

This paper presents a comprehensive study of the use of SAT-based approaches for 
inconsistency-tolerant query answering, which abstracts from the particular setting and 
provides a solid foundation for the future development of such systems. Our contributions can be summarized as follows. 
In Section 3, we provide propositional encodings of the AR, brave, and IAR semantics, including the first encodings 
for Pareto- and completion-optimal repairs. Our encodings are generic and 
are built in a modular manner from a core set of basic formulas. 
Based upon these encodings, we develop in Section 4 several SAT-based algorithms, which 
utilize different functionalities of modern SAT solvers: weighted MaxSAT, MUS enumeration, iterative SAT calls with assumptions.  
 In Section 5, we present our implemented system, {\sc orbits}, which computes query answers under the chosen semantics using the selected encoding and algorithm. 
Section 6 presents the result of our extensive experimental evaluation, using existing OMQA and database benchmarks, 
aimed at comparing the different semantics, 
and understanding the relative performances of different encodings and/or algorithms for the same semantics. We conclude the paper in Section 7 and briefly mention some directions for future work. 
Proofs, pseudo-code for algorithms, and details on the experimental evaluation are provided in the appendix. 

\section{Querying Inconsistent Knowledge Bases}\label{sec:prelim}
We introduce notation and terminology for talking about knowledge bases and 
then recall different inconsistency-tolerant semantics. 
We shall assume throughout the paper that readers are familiar with 
propositional and first-order logic. Key notions are illustrated in Example \ref{example}.

\subsubsection*{Knowledge bases}
By \emph{knowledge base (KB)}, we mean a pair $\Kmc=(\Dmc,\Tmc)$ consisting of a \emph{dataset} $\Dmc$
and a \emph{logical theory}~$\Tmc$. The dataset $\mathcal{D}$ is a finite set of ground \emph{facts}, i.e. atoms $P(c_1, \ldots, c_n)$ 
where $P$ is an $n$-ary predicate and each $c_i$ is a constant. The theory~$\Tmc$ is a finite set of first-order logic (FOL) sentences.
An \emph{$\Lmc$ KB} is a KB whose theory is formulated in the $\Lmc$ fragment of FOL.  
Typically, 
$\Tmc$ will be either 
an \emph{ontology} (with $\Lmc$ a description logic or decidable class of existential rules)
or a set of \emph{database constraints}, for instance, letting 
$\Lmc$ be the class of \emph{denial constraints}, which take the form $\forall \vec{x} \neg (\alpha_1 \wedge \ldots \wedge \alpha_n)$, where each $\alpha_i$ is a relational or inequality atom whose variables are among $\vec{x}$. Denial constraints generalize the more well-known functional dependencies
(FDs) and (primary) key constraints. 
 


A KB $\Kmc=(\Dmc,\Tmc)$ is \emph{consistent}, and its dataset $\Dmc$ is called \emph{$\Tmc$-consistent}, if $\Dmc \cup \Tmc$ has at least one model. Otherwise, $\Kmc$ is \emph{inconsistent}, denoted $\Kmc \models \bot$. 
To identify the reasons for a KB $\Kmc=(\Dmc,\Tmc)$ being inconsistent, we use the notion of 
a \emph{conflict} of $\Kmc$, defined as an inclusion-minimal subset 
$\Dmc' \subseteq \Dmc$ such that $(\Dmc',\Tmc) \models \bot$. 
We use $\conflicts{\Kmc}$, or $\conflicts{\Dmc,\Tmc}$, to refer to the set of all conflicts of $\Kmc=(\Dmc,\Tmc)$. Note that 
$\conflicts{\Dmc,\Tmc} \subseteq \conflicts{\Dmc',\Tmc}$ whenever $\Dmc \subseteq \Dmc'$. 
In particular, this means that 
adding more facts cannot render an inconsistent KB consistent.  


We will be interested in answering queries over KBs. 
In this paper, when we speak of queries, we mean \emph{conjunctive queries}, 
which take the form of conjunctions of relational atoms $P(t_1, \ldots, t_n)$ (with each $t_i$ a constant or variable),
where some variables may be existentially quantified. 
Given a query $q(\vec{x})$, with free variables~$\vec{x}=(x_1, \ldots, x_k)$,
and a tuple of constants $\vec{a}=(a_1, \ldots, a_k) $,
we denote by $q(\vec{a})$ the first-order sentence obtained by replacing each variable in~$\vec{x}$
by the corresponding constant in~$\vec{a}$. A \emph{(certain) answer} to $q(\vec{x})$ 
over $\Kmc=(\Dmc,\Tmc)$ is a tuple of constants~$\vec{a}$ from $\Dmc$ 
such that $q(\vec{a})$ holds in every model of $\Kmc$. We use $\Kmc \models q(\vec{a})$ to indicate
that $\vec{a}$ is a certain answer to $q(\vec{x})$ over $\Kmc$.  

Certain answers are preserved under the addition of facts:
if $(\Dmc,\Tmc) \models q(\vec{a})$ and $\Dmc \subseteq \Dmc'$, then $(\Dmc',\Tmc) \models q(\vec{a})$. 
It thus makes sense to consider the minimal subsets of the data responsible for an answer.  
We call a $\Tmc$-consistent subset $\Cmc \subseteq \Dmc$ a \emph{cause} for $q(\vec{a})$ w.r.t.\ $\Kmc=(\Dmc,\Tmc)$
if $(\Cmc, \Tmc) \models q(\vec{a})$ and $(\Cmc',\Tmc) \not \models q(\vec{a})$  for every $\Cmc' \subsetneq \Cmc$; 
the set of causes for $q(\vec{a})$ w.r.t.\ $\Kmc$ is denoted by $\causes{q(\ans),\Kmc} $. 



Observe that if $\Kmc$ is inconsistent, then $\Kmc \models q(\ans)$ for \emph{every} candidate answer $\ans$, which is uninformative and motivates   
the need for alternative inconsistency-tolerant semantics. 

\subsubsection*{Repairs} 
In order to extract meaningful information from an inconsistent KB, 
it is useful to consider the parts of the data that are consistent with the theory. 
This can be 
formalized using the notion of repair: 

\begin{definition} A \emph{(subset) repair} of a KB $\Kmc=(\Dmc,\Tmc)$ is an inclusion-maximal subset $\Rmc \subseteq \Dmc$
such that $(\Rmc,\Tmc) \not \models \bot$. We use $\reps{\Kmc}$ to denote the set of repairs of KB $\Kmc$.
\end{definition}


Subset repairs 
treats all facts equally. 
However,  preferences between conflicting facts should be taken into account
when they are available. 
Following  \cite{DBLP:journals/amai/StaworkoCM12,DBLP:conf/kr/BienvenuB20}, 
we assume 
such 
preferences are given as a priority relation: 

\begin{definition}
A \emph{priority relation} $\succ$ for a KB $\Kmc=(\Dmc,\Tmc)$ is an acyclic binary relation over the facts of $\Dmc$ such that if $\alpha\succ\beta$, then there exists $\Cmc\in\conflicts{\Kmc}$ such that $\{\alpha,\beta\}\subseteq \Cmc$. 
We say that~$\succ$ is \emph{total} if for every pair $\alpha\neq\beta$ such that $\{\alpha,\beta\}\subseteq \Cmc$ for some $\Cmc\in\conflicts{\Kmc}$, either $\alpha\succ\beta$ or $\beta\succ\alpha$. 
A \emph{completion} of $\succ$ is a total priority relation $\succ'\ \supseteq \ \,\succ$.
\end{definition}

\begin{definition}
A \emph{prioritized KB} $\Kmc_\succ$ consists of a KB $\Kmc=(\Dmc,\Tmc)$ and a priority relation $\succ$ for $\Kmc$.  
\end{definition}

We recall two\footnote{Staworko et al. (\citeyear{DBLP:journals/amai/StaworkoCM12}) defined a third notion, \emph{globally-optimal repairs}, which we do not consider due to their higher complexity. } natural ways of refining the notion of repair to exploit priority relations \cite{DBLP:journals/amai/StaworkoCM12,DBLP:conf/kr/BienvenuB20}: 

\begin{definition} Consider a prioritized KB $\Kmc_\succ$ with $\Kmc=(\Dmc,\Tmc)$, and let $\Rmc \in \reps{\Kmc}$.
\begin{itemize}
\item A \emph{Pareto improvement} of $\Rmc$ 
 is a $\Tmc$-consistent $\Bmc\subseteq\Dmc$ such that there is 
 $ \beta\in\Bmc\setminus\Rmc$ with $\beta\succ\alpha$ for every $\alpha\in\Rmc\setminus\Bmc$.
\end{itemize}
The repair $\Rmc$ is a:
\begin{itemize}
\item \emph{Pareto-optimal repair} of $\Kmc_\succ$ if there is no Pareto improvement of $\Rmc$. 
\item \emph{completion-optimal repair} of $\Kmc_\succ$
if $\Rmc$ is a Pareto-optimal repair of $\Kmc_{\succ'}$, for some completion $\succ'$ of $\succ$.
\end{itemize}
We denote by 
$\preps{\Kmc_\succ}$ and $\creps{\Kmc_\succ}$ the sets of 
Pareto- and completion-optimal repairs.
\end{definition}


It is known that $\creps{\Kmc_\succ} 
\subseteq \preps{\Kmc_\succ}\subseteq \reps{\Kmc}$, 
with each of the inclusions potentially strict. Interestingly, however,
if  $\succ$ 
 is induced by assigning scores to facts, then 
Pareto- and completion-optimal repairs coincide: 

\begin{definition}
A priority relation $\succ$ for $\Kmc=(\Dmc,\Tmc)$ is \emph{score-structured} if there exists 
a scoring function $s:\Dmc \rightarrow \mathbb{N}$ such that 
for every pair of facts $\alpha$ and $\beta$ that appear together in a conflict of $\Kmc$, we have
$\alpha\succ\beta$ iff $s(\alpha)>s(\beta)$. 
\end{definition}


\begin{theorem}\cite{DBLP:conf/pods/LivshitsK17,DBLP:phd/hal/Bourgaux16}
Let $\Kmc_\succ$ be a prioritized KB  such that $\succ$ is score-structured.
Then $\creps{\Kmc_\succ}= \preps{\Kmc_\succ}$. 
\end{theorem}

\citeauthor{DBLP:phd/hal/Bourgaux16} (\citeyear{DBLP:phd/hal/Bourgaux16}) further 
shows that for score-structured priorities, 
Pareto- and completion-optimal repairs 
also coincide with the 
$\subseteq_P$-repairs from \cite{DBLP:conf/aaai/BienvenuBG14} based upon lexicographic set inclusion. 


\subsubsection*{Repair-based semantics} 
We next recall three prominent inconsistency-tolerant semantics (brave, AR, and IAR),
which we parameterize by the considered type of repair: 
\begin{definition}
Fix X $\in \{S,P,C\}$ and 
consider a prioritized KB $\Kmc_\succ$ with $\Kmc=(\Dmc,\Tmc)$, query $q(\vec{x})$, and tuple of constants $\ans$ from $\Dmc$ with $|\vec{x}|=|\ans|$. 
Then $\ans$ is an answer to $q$ over $\Kmc_\succ$ 
\begin{itemize}
\item 
 under \emph{X-brave semantics}, denoted $\Kmc_\succ \bravemodels{X} q(\ans)$, if $(\Rmc,\Tmc) \models q(\ans)$ for some $\Rmc \in \xreps{\Kmc_\succ}$
\item under \emph{X-AR semantics}, denoted $\Kmc_\succ \armodels{X} q(\ans)$, if $(\Rmc,\Tmc) \models q(\ans)$ for every $\Rmc \in \xreps{\Kmc_\succ}$
\item under \emph{X-IAR semantics}, denoted $\Kmc_\succ \iarmodels{X} q(\ans)$, if $(\Bmc,\Tmc) \models q(\ans)$ where $\Bmc=\bigcap_{\Rmc \in \xreps{\Kmc_\succ}} \Rmc$
\end{itemize}
\end{definition}
The AR semantics is arguably the most natural way of defining \emph{plausible} query answers, and 
it is the semantics used for consistent query answering in databases \cite{DBLP:series/synthesis/2011Bertossi}. 
The brave semantics captures the notion of \emph{possible} answers, 
while the IAR semantics identifies the answers that can be obtained using only the \emph{most reliable} facts. 
Observe that 
$\Kmc_\succ \iarmodels{X} q \Rightarrow \Kmc_\succ \armodels{X} q  \Rightarrow \Kmc_\succ \bravemodels{X} q.$

\begin{example}\label{example}
Let $\Kmc=(\Dmc,\Tmc)$ where $\Dmc$ contains four 
facts $\alpha=R(a,b)$, $\beta=R(a,c)$, $\gamma=S(d,c)$, and $\delta=S(d,b)$, and $\Tmc$ contains FDs $\forall x,y,z \neg(R(x,y)\wedge R(x,z) \wedge y\neq z)$, $\forall x,y,z \neg(S(x,y)\wedge S(x,z) \wedge y\neq z)$ and the denial constraint $\forall x,y,z \neg(R(y,x)\wedge S(z,x))$. 

The conflicts of $\Kmc$ are $\{\alpha,\beta\}$, $\{\gamma,\delta\}$, $\{\alpha, \delta\}$ and $\{\beta,\gamma\}$, hence 
$\reps{\Kmc}=\{\{\alpha,\gamma\}, \{\beta,\delta\}\}$. 
If we define $\succ$ by $\alpha\succ\beta$ and $\gamma\succ\delta$, we obtain  
$\preps{\Kmc_\succ}=\reps{\Kmc}$ but $\creps{\Kmc_\succ}=\{\{\alpha,\gamma\}\}$.  Indeed, any completion $\succ'$ of $\succ$ is such that $\alpha\succ'\delta$ or $\gamma\succ'\beta$ by acyclicity.

If $q(x)=\exists y R(x,y)$, $\causes{q(a),\Kmc}=\{\{\alpha\},\{\beta\}\}$. Hence, $\Kmc_\succ \iarmodels{C} q(a)$, $\Kmc_\succ \armodels{P} q(a)$ but $\Kmc_\succ \not\iarmodels{P} q(a)$. 
\end{example}



We briefly recall what is known about the complexity of query answering under (optimal) repair-based semantics. 
Note that 
\emph{when we speak of complexity, we mean data complexity}, 
measured solely in terms of the size of the dataset.

Theorems \ref{ub-thm} and \ref{lb-thm} summarize upper and lower bounds from the database and ontology settings. All of them are known 
\cite{DBLP:journals/amai/StaworkoCM12,DBLP:conf/kr/BienvenuB20,DBLP:conf/ijcai/Rosati11}, except the lower bounds for X-brave and X-IAR semantics in the case of FDs, 
proven in the appendix. 
%
We say that a logic $\Lmc$ 
\emph{enjoys \ptime\ consistency checking (resp.\ query entailment)} if 
the problem of deciding whether $\Kmc \models \bot$ (resp.\ $\Kmc \models q(\ans)$)
for an input $\Lmc$ KB
is in \ptime. 


\begin{theorem} \label{ub-thm} 
Let $\Lmc$ be any FOL fragment 
that enjoys \ptime\ consistency checking and query entailment, and let X $\in \{S,P,C\}$.  
Then query entailment for $\Lmc$ KBs  is 
\begin{itemize}
\item in $\np$ 
under X-brave semantics, and 
\item in $\conp$ 
under X-AR and X-IAR semantics. 
\end{itemize}
\end{theorem}



\begin{theorem}\label{lb-thm}
Query entailment for $\Lmc$ KBs is 
\begin{itemize}
\item $\np$-hard under X-brave semantics (X $\in \{P,C\}$) 
\item $\conp$-hard under X-AR semantics  (X  $\in \{S,P,C\}$)
\item $\conp$-hard under X-IAR semantics (X $\in \{P,C\}$)
\end{itemize}
for any $\Lmc$ that extends \dllitecore, $\mathcal{EL}_\bot$, or FDs. 
%
\end{theorem}

\begin{remark}
Query entailment under S-brave and S-IAR is tractable both for DL-Lite ontologies and denial constraints \cite{Bienvenu_TractableApproximation_long}. 
\end{remark}

\section{SAT Encodings}\label{sec:encodings}
The (co)NP complexity results from the previous section suggest 
a SAT-based approach  
to query entailment under (optimal) repair-based semantics. While our work is not the first
to explore SAT encodings for inconsistency-tolerant semantics, 
our contribution is a uniform approach that covers a wide range of semantics and settings, and 
provides the first encodings for 
Pareto- and completion-optimal repairs. 

\subsection{Overview}\label{enc-overview}
Our propositional encodings are built from: 
\begin{itemize}
\item the set $\conflicts{\Kmc}$ of conflicts of the considered KB $\Kmc$, 
\item a set $\braveans$ of \emph{potential answers}, 
and for each $\ans \in \braveans$, the (non-empty\footnote{If $\causes{q(\ans),\Kmc}\! = \!\emptyset$, $q(\ans)$ holds for none of our semantics.}) set $\causes{q(\ans),\Kmc}$, 
\item the priority relation $\succ$ of $\Kmc$,
\end{itemize}
and 
are of polynomial size w.r.t.\ to these inputs. Note that 
for 
DL-Lite ontologies 
and denial constraints, the sets of conflicts, candidate answers, and their causes,  
can be computed in \ptime\ via database query evaluation,
so our encodings will yield procedures of the expected co(NP) complexity. 

To simplify the treatment, \emph{we shall assume that every conflict contains at most two facts}, a property that is
 satisfied by the most common DL-Lite dialects and for FDs. 
(The extension to 
non-binary conflicts is discussed later in the section.)
By restricting our attention to binary conflicts, 
we can \emph{use a convenient notation $\alpha \confof \beta$ in place of $\{\alpha,\beta\}\in\conflicts{\Kmc}$},
and define a graph representation of conflicts and priorities.
The \emph{directed conflict graph} $\mathcal{G}_{\Kmc_\succ}$ has facts from $\conflicts{\Kmc}$ as nodes and an edge from $\alpha$ to $\beta$
iff $\alpha \confof \beta$ 
and $\alpha{\not\succ}\beta$. 

Our encodings will use variables of the form $x_\alpha$ 
to indicate whether fact $\alpha$ appears in a (partial) repair. Including one such variable for each fact in $\Dmc$
would yield prohibitively large encodings, which is why we use 
%
%
 $\mathcal{G}_{\Kmc_\succ}$ to focus on relevant facts: 
given any $F \subseteq \Dmc$, we let $\reachr(F)$ be the set of all \emph{facts 
that are reachable in} $\mathcal{G}_{\Kmc_\succ}$ from some $\alpha \in F$.

Before proceeding to the details, let us 
give a high-level overview of our encodings. 
All of the encodings try to construct a set of facts that is consistent with the theory (i.e. does not contain any conflicts)
and that 
can be extended to an 
 optimal repair. 
For the X-AR semantics, we are trying to build an optimal repair that does \emph{not} entail the considered query answer(s), which can be done by including facts that contradict each of the answer(s) causes.
For the X-brave semantics, we must ensure that
 the 
 repair entails the considered query answer(s), 
achieved by requiring the presence of some cause in the chosen subset.  Finally, for the X-IAR semantics, we ensure the existence of optimal repairs that omit a given cause or fact appearing 
in a cause 
by including facts that 
contradict the considered cause or fact. 

In the next section, we will present SAT-based algorithms that consider each potential answer in turn,
as well as algorithms that treat all potential answers conjointly. For that reason, in what follows,
we will present encodings both for a single answer and for several answers at a time. 

\subsection{Basic Building Blocks}
Our various encodings rely upon some common ingredients, which are presented next. 
%
%
\subsubsection*{Absence or presence of causes} 
To contradict a specific cause $\Cmc$, we use 
$\varphi_{\neg \Cmc}$, with two alternative definitions inspired respectively by the \textsf{CQAPri} and \textsf{CAvSAT} encodings:
\begin{align*}
\varphi_{\neg \Cmc} = & \bigvee_{\alpha\in \Cmc} \, \bigvee_{\alpha \confof \beta, \alpha\not\succ\beta} x_\beta &&(\text{\cqaprienc})\\
\varphi_{\neg \Cmc} = & \bigvee_{\alpha\in \Cmc} \neg x_\alpha \wedge\, \bigwedge_{\alpha\in\Cmc} (x_\alpha\vee \!\!\!\! \bigvee_{\alpha \confof \beta, \alpha\not\succ\beta} \!\! x_\beta)&&(\text{\cavsatenc})
\end{align*}
For encodings with multiple answers, we use a variant $\varphi'_{\neg \Cmc}(y)$ obtained by adding $\neg y$ as a disjunct of the (first) clause
of $\varphi_{\neg \Cmc}$, so it is 
 only `active' 
when the given variable $y$ is true.
%
To block all causes of the BCQ $q(\ans)$, we can use 
$$\varphi_{\neg q(\ans)} = \bigwedge_{\Cmc\in \causes{q(\ans),\Kmc}} \varphi_{\neg \Cmc}
$$
%
%
\noindent or, in the case of encodings for several answers, a variant $\varphi'_{\neg q(\ans)}$ obtained by replacing $\varphi_{\neg \Cmc}$ by $\varphi'_{\neg \Cmc}(x_{\ans})$. 

If instead we want to force that cause $\Cmc$ holds, we use: 
\begin{align*}
\varphi_{\Cmc} = & \bigwedge_{\alpha\in \Cmc} \, x_\alpha
\end{align*}
and to ensure \emph{some} cause for BCQ $q(\ans)$ is present, we use: 
\begin{align*} 
\varphi_{q(\ans)} = & (\!\!\!\bigvee_{\Cmc\in \causes{q(\ans),\Kmc}} \!\!\! x_\Cmc)\,\wedge\!\!\! \bigwedge_{\Cmc\in \causes{q(\ans),\Kmc}}\, \bigwedge_{\alpha\in \Cmc} \! \neg x_\Cmc\vee x_\alpha
\end{align*}
or, for multi-answer encodings, a variant $\varphi'_{q(\ans)}$ obtained from $\varphi_{q(\ans)}$ by adding $\neg x_{\ans}$ as a disjunct of the first clause.   

\subsubsection{Consistency}
We use $\varphi_\mn{cons}(F)$ to ensure that 
the valuation of $\{x_\alpha \mid \alpha \in F\}$ corresponds to a \Tmc-consistent set of facts:
\begin{align*}
\varphi_\mn{cons}(F) = & \bigwedge_{\alpha,\beta\in F, \alpha \confof \beta} \, 
\Big(\neg x_\alpha \vee \neg x_\beta\Big).
\end{align*}

\begin{figure*}
\begin{align*}
\varphi_\mn{pref}=&\bigwedge_{\alpha\in \reachr(F)} \Big (x_\alpha \vee \bigvee_{\beta\in \reachr(F),\alpha \confof \beta} x_{\beta\rightarrow\alpha} \Big ) \,\, \wedge \,\,\bigwedge_{\alpha,\beta\in \reachr(F), \alpha \confof \beta} \Big ( (\neg x_{\beta\rightarrow\alpha} \vee x_\beta) \wedge (\neg x_{\beta\rightarrow\alpha} \vee x_{\beta\succ'\alpha}) \Big )
\\
\varphi_\mn{compl}=& \bigwedge_{\alpha,\beta\in \reachr(F),\alpha \confof \beta} \Big (x_{\alpha\succ'\beta} \vee x_{\beta\succ'\alpha} \Big )\,\, \wedge \,\,\bigwedge_{\alpha,\beta\in \reachr(F),\alpha \confof \beta} \Big ( \neg x_{\alpha\succ'\beta} \vee \neg x_{\beta\succ'\alpha} \Big ) \,\,
\wedge \,\, \bigwedge_{\alpha,\beta\in \reachr(F),\alpha\succ\beta} x_{\alpha\succ'\beta} 
\\
\varphi_\mn{acyc}=& \bigwedge_{\alpha,\beta\in \reachr(F), \alpha \confof \beta} \Big (\neg x_{\alpha\succ'\beta} \vee t_{\alpha,\beta} \Big ) \,\, \wedge  \bigwedge_{\alpha,\beta\in \reachr(F), \alpha \confof \beta} \Big (\neg x_{\alpha\succ'\beta} \vee \neg t_{\beta,\alpha}\Big ) \,\,
\wedge \bigwedge_{\alpha,\beta,\gamma\in \reachr(F), \beta \confof\gamma
} \Big (\neg t_{\alpha,\beta} \vee \neg x_{\beta\succ'\gamma} \vee t_{\alpha,\gamma}\Big )
\end{align*}
\caption{Subformulas of $\varphi_\mn{C\text{-}max}(F)$, which is used to ensure it is possible to extend the selected subset of $F$ to a completion-optimal repair. }
\label{completion-fig}
\end{figure*}

\subsubsection{Extension to optimal repair}
The most intricate part of the encoding is ensuring that the selected 
set of facts can be extended to a repair of the desired type. 
To this end, we introduce formulas of the form $\varphi_\mn{X\text{-}max}(F)$, where $F$ provides the facts that may appear, 
and X is the type of repair. 
\begin{itemize}
\item Subset repairs: 
we can simply set $\varphi_\mn{S\text{-}max}(F) = \top$, as every consistent set of facts extends to some S-repair. 
\item Pareto-optimal repairs: we prove that the following encoding of maximality for
$\subseteq_P$-repairs (w.r.t.\ score-structured $\succ$) from Bienvenu et al. (\citeyear{DBLP:conf/aaai/BienvenuBG14}) 
in fact also works for Pareto-optimal repairs and arbitrary $\succ$: 
$$ \varphi_\mn{P_{1}\text{-}max}(F) = \bigwedge_{\alpha\in \reachr(F)} (x_\alpha \vee \bigvee_{\alpha \confof \beta, \alpha\not\succ\beta} x_\beta)$$ 
Essentially, it states that a relevant fact can only be omitted if we include a non-dominated contradicting fact. 
We also propose an alternative encoding with fewer variables:
$$ \varphi_\mn{P_{2}\text{-}max}(F) =\bigwedge_{\alpha \in \reachr^-(F)} 
\bigwedge_{\beta\succ\alpha} (\neg x_\alpha\vee \bigvee_{\
\beta \confof \gamma, \beta\not\succ\gamma} x_\gamma )$$ 
where $\reachr^-(F)= \bigcup_{i=1}^\infty \reachr_i$ with $\reachr_0=F$ and 
$$\reachr_{i+1}=\reachr_i \cup \{\gamma \mid \exists \alpha \in \reachr_i, \beta\succ\alpha, \beta \confof \gamma, \beta\not\succ\gamma\}. $$ Intuitively, 
to include $\alpha \in F$, we must contradict every more preferred fact that conflicts with $\alpha$, then do the same for selected contradicting facts. 
%
\item Completion-optimal repairs (w.r.t. arbitrary $\succ$): we use
\begin{align*}
\varphi_\mn{C\text{-}max}(F) =&\varphi_\mn{pref}\wedge\varphi_\mn{compl}\wedge\varphi_\mn{acyc}
\end{align*}
where the subformulas $\varphi_\mn{pref}$, $\varphi_\mn{compl}$, and $\varphi_\mn{acyc}$ are defined in Figure \ref{completion-fig}. 
The formula $\varphi_\mn{pref}$ states that if $x_\alpha$ is omitted, then we must include a contradicting fact $\beta$ that is preferred to $\alpha$
according to $\succ'$. We use $\varphi_\mn{compl}$ to ensure that $\succ'$ compares all contradicting facts and extends $\succ$, while 
the acyclicity of $\succ'$ is ensured by $\varphi_\mn{acyc}$, which uses variables $t_{\alpha,\beta}$ to compute the transitive closure  of $\succ'$.  
\end{itemize}

\subsubsection*{Non-binary conflicts} We briefly discuss how to modify the preceding formulas 
to handle non-binary conflicts. The most essential difference is that instead of choosing a single contradicting fact, 
we may need to choose a conjunction of facts, and use additional variables to refer to them. 
We must also redefine $\mathcal{G}_{\Kmc_\succ}$ as a directed hypergraph, 
and use hypergraph reachability to define $\reachr(F)$. Details of the required modifications are provided in the appendix.



\subsection{Propositional Encodings}
We now present our 
encodings, built from the 
preceding components following the intuitions given in Section \ref{enc-overview}. 

Note that the following results hold no matter which variant of $\varphi_{\neg \Cmc}$
we use and with $\varphi_\mn{P\text{-}max}$ 
instantiated as either $\varphi_\mn{P_{1}\text{-}max}$ or $\varphi_\mn{P_{2}\text{-}max}$. 
The notation $\facts(\varphi)$ will be used for the set of facts $\alpha$ such that 
the 
variable $x_\alpha$ occurs in $\varphi$. 

\subsubsection*{X-AR semantics} 
Formula $\Phi_{X\text{-}AR}(q(\ans))$ in Figure \ref{sem-encodings-fig}  is used to 
test whether a particular tuple $\ans$ holds. Roughly speaking, it 
selects a way to contradict every clause, checking that the resulting set of facts 
is contained in a $X$-optimal repair. The second formula 
$\Psi_{X\text{-}AR}(\mi{PotAns})$ simultaneously handles 
all tuples in $\mi{PotAns}$. Clauses of the form $x_{\ans}$ can be added to activate $\varphi'_{\neg q(\ans)}$. 


The correctness of our encodings is given in the next result, which applies to all 
X $\in \{S, P, C\}$ and $\ans \in \mi{PotAns}$: 

\begin{theorem} The following are equivalent : 
(i) $\Kmc_\succ \armodels{X} q(\ans)$, 
(ii) $\Phi_{X\text{-}AR}(q(\ans))$ is unsatisfiable, and
(iii) $x_{\ans}$ is false in every satisfying assignment of $\Psi_{X\text{-}AR}(\mi{PotAns})$. 
\end{theorem}

\subsubsection*{X-brave semantics}
Figure \ref{sem-encodings-fig} presents our encodings for the X-brave semantics. 
Formula $\Phi_{X\text{-}brave}(q(\ans))$ checks a particular tuple $\ans$ and is essentially the same 
as $\Phi_{X\text{-}AR}(q(\ans))$, but with $\varphi_{q(\ans)}$ in place of $\varphi_{\neg q(\ans)}$. 
A variant $\Phi_{X\text{-}brave}(\Cmc)$ can be used to check whether a particular cause $\Cmc$ holds in some X-optimal repair. 
For a multi-answer encodings, 
we use the formula $\Psi_{X\text{-}brave}(\mi{PotAns})$, again adding clauses of the form $x_{\ans}$ to activate $\varphi'_{q(\ans)}$. 
%
We obtain an analogous correctness result: 

\begin{theorem} The following are equivalent: 
(i) $\Kmc_\succ \bravemodels{X} q(\ans)$, 
(ii) $\Phi_{X\text{-}brave}(q(\ans))$ is satisfiable,
(iii) $\Phi_{X\text{-}brave}(\Cmc)$ is satisfiable for some $\Cmc\in \causes{q(\ans),\Kmc}$,
and 
(iv) $x_{\ans}$ is true in some satisfying assignment of  $\Psi_{X\text{-}brave}(\mi{PotAns})$.
\end{theorem}

\subsubsection*{X-IAR semantics} 
Formula $\Phi_{X\text{-}IAR}(\Cmc)$ from Figure \ref{sem-encodings-fig} can be used to test whether 
there exists a X-optimal repair that excludes cause $\Cmc$. To check a potential answer $\ans$,
we use $\Phi_{X\text{-}IAR}(q(\ans))$, which is essentially the conjunction of $\Phi_{X\text{-}IAR}(\Cmc)$ 
for every cause $\Cmc$ of $q(\ans)$, but where the \emph{conjuncts for different causes use distinct variables} 
($x^\Cmc_\alpha$ in place of $x_\alpha$). A multi-answer encoding $\Psi_{X\text{-}IAR}(\mi{PotAns})$ can be obtained 
by taking the conjunction of $\Phi_{X\text{-}IAR}(q(\ans))$ for all $\ans \in \mi{PotAns}$, but with $\varphi^\Cmc_{\neg \Cmc}$ replaced by $\varphi'^\Cmc_{\neg \Cmc}(x_{\ans})$ 
(see the appendix for details). We obtain the following: 

\begin{theorem} The following are equivalent: 
(i) $\Kmc_\succ \iarmodels{X} q(\ans)$, 
(ii) $\Phi_{X\text{-}IAR}(q(\ans))$ is unsatisfiable, 
(iii) $\Phi_{X\text{-}IAR}(\Cmc)$ is unsatisfiable for some $\Cmc\in \causes{q(\ans),\Kmc}$, 
and
(iv) $x_{\ans}$ is false in every satisfying assignment of $\Psi_{X\text{-}IAR}(\mi{PotAns})$.
\end{theorem}

We shall also require encodings for individual facts, using 
$$\Phi_{X\text{-}IAR}(\alpha)=\Phi_{X\text{-}IAR}(\{\alpha\})$$
to test if $\alpha$ holds in all X-optimal repairs. We also consider a multi-fact version $\Psi_{X\text{-}IAR}(\mi{Rel})$, parameterized by a set of facts $\mi{Rel}$, and to which we add clause $y_{\alpha}$ to activate $\varphi'_{\neg \{\alpha\}}(y_\alpha)$.

\begin{theorem}
For every 
$\alpha\in\Dmc$, the following are equivalent: (i) $\Kmc_\succ \iarmodels{X} \alpha$, (ii) $\Phi_{X\text{-}IAR}(\alpha)$ is unsatisfiable, and 
(iii) if $\alpha\in\mi{Rel}$ then $y_{\alpha}$ is false in every satisfying assignment of $\Psi_{X\text{-}IAR}(\mi{Rel})$.
\end{theorem}

\begin{figure*}[ht]
\begin{align*}
\Phi_{X\text{-}AR}(q(\ans))&=\varphi_{\neg q(\ans)} \wedge \varphi_\mn{X\text{-}max}(F_1)\wedge \varphi_\mn{cons}(F_2) &&  F_1=\facts(\varphi_{\neg q(\ans)}), F_2= F_1 \cup  \facts(\varphi_\mn{X\text{-}max}(F_1))\\
\Psi_{X\text{-}AR}(\mi{PotAns})&=\!\!\! \bigwedge_{\ans\in\mi{PotAns}} \!\!\! \varphi'_{\neg q(\ans)}\wedge 
\varphi_\mn{X\text{-}max}(F'_1) \wedge \varphi_\mn{cons}(F'_2) && F'_1 = \facts(\!\!\!\!\!\!\bigwedge_{\ans\in\mi{PotAns}}\!\!\!\!\!\!\!\! \varphi'_{\neg q(\ans)}), 
F'_2\!= \!F'_1 \cup \facts(\varphi_\mn{X\text{-}max}(F'_1))\\
%
\Phi_{X\text{-}brave}(q(\ans))&=\varphi_{q(\ans)} \wedge \varphi_\mn{X\text{-}max}(G_1)\wedge \varphi_\mn{cons}(G_2) 
&& G_1=\facts(\varphi_{q(\ans)}), G_2= G_1 \cup  \facts(\varphi_\mn{X\text{-}max}(G_1))\\
\Phi_{X\text{-}brave}(\Cmc)&=\varphi_{\Cmc} \wedge \varphi_\mn{X\text{-}max}(G^*_1)\wedge \varphi_\mn{cons}(G^*_2)    
&& G^*_1 = \facts(\varphi_{\Cmc}), G^*_2 = G^*_1 \cup \facts(\varphi_\mn{X\text{-}max}(G^*_1))\\
\Psi_{X\text{-}brave}(\mi{PotAns})&= \!\!\! \bigwedge_{\ans\in\mi{PotAns}} \!\!\! \varphi'_{q(\ans)} \wedge 
\varphi_\mn{X\text{-}max}(G'_1) \wedge \varphi_\mn{cons}(G'_2) 
&& G'_1 = \facts(\!\!\!\!\!\!\bigwedge_{\ans\in\mi{PotAns}}\!\!\!\!\! \!\! \varphi'_{q(\ans)}), 
G'_2\!= \!G'_1 \cup \facts(\varphi_\mn{X\text{-}max}(G'_1))\\
%
\Phi_{X\text{-}IAR}(\Cmc)&=\varphi_{\neg \Cmc} \wedge \varphi_\mn{X\text{-}max}(H_1)\wedge \varphi_\mn{cons}(H_2) &&
H_1=\facts(\varphi_{\neg \Cmc}), H_2= H_1 \cup \facts(\varphi_\mn{X\text{-}max}(H_1))\\
\Phi_{X\text{-}IAR}(q(\ans))&= \!\!\!\!\!\!
\bigwedge_{\Cmc\in \causes{q(\ans),\Kmc}} \!\!\!\!\!\Big (\varphi^\Cmc_{\neg \Cmc} \wedge \varphi^\Cmc_\mn{X\text{-}max}(H_1^\Cmc)\wedge \varphi^\Cmc_\mn{cons}(H_2^\Cmc) \Big)
&& H_1^\Cmc=\facts(\varphi^\Cmc_{\neg \Cmc}), H_2^\Cmc=H_1^\Cmc \cup \facts(\varphi^\Cmc_\mn{X\text{-}max}(H_1^\Cmc))
\end{align*}
\caption{Single- and multi-answer encodings for X-AR (top) and X-brave (middle) semantics, single-answer encodings for X-IAR (bottom). }
\label{sem-encodings-fig}
\end{figure*}

\section{Algorithms}\label{sec:algos}

Inspired by the different use of SAT solvers made by \textsf{CQAPri} and \textsf{CAvSAT}, 
we propose several algorithms based on the encodings of Section \ref{sec:encodings}. 
The pseudo code of all algorithms is available in the appendix.

Before describing the algorithms, note that we can show that the encodings of Section \ref{sec:encodings} are still valid if $\causes{q(\ans),\Kmc}$ is actually \emph{a superset of the causes} such that every superfluous $\Bmc$ in it either (1) includes a real cause of $q(\ans)$ or (2) contains two distinct facts that form a conflict. 
Since checking consistency and minimality to obtain the real causes can be costly in practice, our algorithms accept such `sets of causes'. They actually even accept the presence of self-inconsistent facts in the `causes' (which would make the encodings for X-AR and X-IAR not applicable) and handle them in a preprocessing step. 

Our high-level algorithm takes as input a semantics $\sem$~$\in\{$brave, AR, IAR$\}$, a repair notion X~$\in \{S,P,C\}$, a directed conflict graph $\mathcal{G}_{\Kmc_\succ}$, and a set of potential answers $\braveans$ with their causes, and outputs the set of tuples from $PotAns$ that are answers 
under the X-$\sem$ semantics. 

An initial preprocessing step serves to (1) check whether $\mathcal{G}_{\Kmc_\succ}$ contains some self-inconsistent facts, remove them from $\mathcal{G}_{\Kmc_\succ}$, discard all causes that contain such facts, then all answers that do not have any cause left, and (2) find some answers that trivially hold under X-$\sem$.  To do so, it removes from the causes all facts that do not have any outgoing edge in the directed conflict graph, and thus trivially belong to all optimal repairs.  
The \emph{trivial answers} that have some cause that contains only such facts hold under X-$\sem$ semantics and are filtered during this step. 

It then remains to filter the remaining potential answers. 
The four first algorithms we propose to do so are generic in the sense that they can be used for all semantics. 
\begin{itemize}
\item \algosat is similar to the algorithm used by \textsf{CQAPri}. For each answer to filter $\ans$, it checks whether $\Phi_{X\text{-}\sem}(q(\ans))$ is satisfiable, which decides whether $\Kmc_\succ \semmodels{X} q(\ans)$.

\item \algomaxsat is similar to the \textsf{CAvSAT} algorithm. It constructs a weighted MaxSAT instance $\Psi_{X\text{-}\sem}(\mi{PotAns})\wedge \bigwedge_{\ans\in \mi{PotAns}} x_{\ans}$, where the $x_{\ans}$ are soft clauses, and all other clauses are hard. 
It then relies on the solver to maximize the number of soft clauses satisfied, which filters the corresponding answers. 
After each iteration, 
the $\neg x_{\ans}$ corresponding to the satisfied soft clauses are added to the set of assumed literals for the next iteration.

\item \algomuses is based on the observation that if $x_{\ans}$ is false in every satisfying assignment of $\Psi_{X\text{-}\sem}(\mi{PotAns})$, then $\{x_{\ans}\}$ is a minimal unsatisfiable subset (MUS) of $\bigwedge_{\ans\in \mi{PotAns}} x_{\ans}$ \wrt $\Psi_{X\text{-}\sem}(\mi{PotAns})$. 
\algomuses relies on the solver to compute all MUSes, and decides whether $\Kmc_\succ \semmodels{X} q(\ans)$ by looking at those of size one.

\item \algoassump iteratively evaluates $\Psi_{X\text{-}\sem}(\mi{PotAns})$, treating the variables $x_{\ans}$, with $\ans\in\mi{PotAns}$ as assumptions. If $\Psi_{X\text{-}\sem}(\mi{PotAns})[x_{\ans}]$ is satisfiable, there exists a satisfying assignment of $\Psi_{X\text{-}\sem}(\mi{PotAns})$ in which $x_{\ans}$ is true, which decides whether $\Kmc_\succ \semmodels{X} q(\ans)$.
\end{itemize} 
While we may need to consider all causes to decide whether an answer holds under X-AR semantics, in the X-brave or X-IAR case it is sufficient to find a single cause that belongs to some or all optimal repairs. Moreover, the encoding $\Phi_{X\text{-}IAR}(q(\ans))$ is a conjunction of independent sub-problems built on distinct variables for each cause, which does not seem very fit for a SAT solver. We hence propose algorithms specific to these cases.
\begin{itemize}
\item \algocauses can be used for X-brave and X-IAR. 
For each answer to filter $\ans$, it checks whether there is a cause $\Cmc$ of $q(\ans)$ such that $\Phi_{X\text{-}\sem}(\Cmc)$ is (un)satisfiable: if $\sem=$ brave and $\Phi_{X\text{-}\sem}(\Cmc)$ is satisfiable, or $\sem=$ IAR and $\Phi_{X\text{-}\sem}(\Cmc)$ is unsatisfiable, then $\Kmc_\succ \semmodels{X} q(\ans)$; if no cause witnesses $\Kmc_\succ \semmodels{X} q(\ans)$, then $\Kmc_\succ \not\semmodels{X} q(\ans)$.

\item \algoIARcauses is specific to X-IAR. It considers the answers in turn while maintaining two sets of facts: the X-IAR facts that belong to the intersection of the optimal repairs and the non-X-IAR facts. For each cause $\Cmc$ of $q(\ans)$ that does not contain any known non X-IAR fact, it removes the known X-IAR facts from $\Cmc$. If $\Cmc$ becomes empty, then $\Kmc_\succ \iarmodels{X} q(\ans)$. Otherwise, for each remaining $\alpha\in\Cmc$, it checks whether $\Kmc_\succ \iarmodels{X} \alpha$ using $\Phi_{X\text{-}IAR}(\alpha)$ and adds $\alpha$ to the corresponding set of facts. If every $\alpha\in\Cmc$ is such that $\Kmc_\succ \iarmodels{X} \alpha$, then $\Kmc_\succ \iarmodels{X} q(\ans)$.

\item \algoIARfacts is also specific to X-IAR and considers the answers in turn while maintaining the two sets of X-IAR and non X-IAR facts. The difference is that for each answer, it uses $\Psi_{X\text{-}IAR}(\mi{Rel})\wedge \bigwedge_{\alpha\in \mi{Rel}} y_{\alpha}$ and a weighted MaxSAT solver to decide which facts hold under X-IAR among the set $\mi{Rel}$ of facts that belong to some cause and have not already been assigned to one of the two sets. Then it checks whether there is a cause that only contains X-IAR facts.
\end{itemize}

%
\section{Implementation \& Experimental Setting}

We implemented the algorithms 
presented in Section \ref{sec:algos} in java 11. 
Our system {\sc orbits} (Optimal Repair-Based Inconsistency-Tolerant Semantics) takes as input two JSON files containing
the directed conflict graph $\mathcal{G}_{\Kmc_\succ}$, and 
the potential answers $\braveans$ of the query associated with their causes. 
The user specifies a semantics (AR, IAR, or brave), 
a value X (among S, P$_1$, P$_2$, or C) to use in $\varphi_\mn{X\text{-}max}(F)$, the desired encoding 
for $\varphi_{\neg \Cmc}$ (\cqaprienc or \cavsatenc), and the algorithm to use to compute the answers w.r.t.\ 
the chosen semantics. 
The set of answers is output as a JSON file. 

{\sc orbits} relies on the Sat4j java library (version 2.3.4) to solve the SAT, weighted MaxSAT, and MUS 
 enumeration problems \cite{DBLP:journals/jsat/BerreP10}. 
In principle, a standalone 
solver could be used, but we found that the time needed to print out the encoding to pass it to an external solver tends to be prohibitive compared to using Sat4j. 

The source code of {\sc orbits} is available at \url{https://github.com/bourgaux/orbits}, the inputs files we used in the experiments at \url{https://zenodo.org/record/5946827}, and details on the experimental setting in the appendix. 

\subsubsection*{Experimental Environment}
All experiments were run with 16GB of RAM in a cluster node running CentOS 7.9 with linux kernel 3.10.0, with processor 2x Cascade Lake Intel Xeon 5218 16 cores, 2.4GHz. 
Reported times are averaged over 5 runs, with 
a 30 minutes time-out. 
Since we aim at comparing our different algorithms and encodings, in what follows, we focus on the time needed to filter the candidate answers, excluding the time needed to load the inputs from the JSON files or serialize the output. 
This input loading time is generally below 1 second and never exceeds a few seconds. 
However, 
in real-world applications, we would not use {\sc orbits} as a standalone tool, but rather make it a library to be integrated in a full query answering system.

\subsubsection*{Test KBs}
We evaluate {\sc orbits} on three (sets of) KBs. 
The first is the CQAPri benchmark \cite{DBLP:phd/hal/Bourgaux16}, 
 a synthetic benchmark crafted to evaluate inconsistency-tolerant query answering over DL-Lite KBs, 
 adapted from the LUBM$^\exists_{20}$ benchmark \cite{DBLP:conf/semweb/LutzSTW13}. 
The two others, called Food Inspection and Physicians, are real-world datasets 
built from public open data, which have already been used to evaluate data cleaning and consistent query answering systems 
\cite{DBLP:journals/pvldb/RekatsinasCIR17,DBLP:conf/sat/DixitK19}. They consist of relational databases built from the original csv files, on which typical 
integrity constraints have been added. 
We briefly summarize their main characteristics below. 

We use the DL-Lite ontology (which includes 875 disjointness axioms) and 20 queries of the CQAPri benchmark, together with the 18 datasets named \mn{uXcY} with $X\in\{1,5,20\}$ and $Y\in\{1,5,10,20,30,50\}$. Parameters $X$ and $Y$ are related to the size and the proportion of facts involved in some conflicts respectively (the higher the bigger), and the datasets are such that $\mn{uXcY}\subseteq \mn{uXcY'}$ for $Y\leq Y'$ and $\mn{uXcY}\subseteq \mn{uX'cY}$ for $X\leq X'$. 
Their sizes range from 75K to 2M facts and their proportions of facts involved in some conflict from 3\% to 46\%. 
The induced conflict graphs contain from 2K to 946K facts and from 2K to 3M conflicts. 

The Food Inspection dataset contains data about inspection of restaurants 
 in New York and Chicago \cite{Dataset1FoodNewYork,Dataset2FoodChicago}. 
We use the database schema and six queries proposed by \citeauthor{DBLP:conf/sat/DixitK19} \shortcite{DBLP:conf/sat/DixitK19}: there are 
four relations, each 
having a key constraint and one having a further 
FD. 
The dataset contains 523K facts, 
37\% of them belong to 
some conflict. 
The conflict graph contains 192K facts and 219K conflicts. 

We build the Physicians dataset from the National Downloadable File provided by the Centers for Medicare \& Medicaid Services \cite{Dataset3Physicians}. It contains information on medical professionals and their affiliations. 
We decompose it into seven relations, add four reasonable key constraints and two FDs, 
 and design six queries. 
In total, the dataset contains more than 8M facts and 2\% of them are in some conflict. 
The conflict graph contains 183K facts and 2.7M conflicts. 

\subsubsection*{Priority Relations} 
We build score-structured priority relations by randomly assigning each fact a score between $1$ and $n$. 
To construct a non-score-structured priority relation, 
we consider each conflict and assign a random direction to the corresponding   
edge in the conflict graph 
with a probability $p$, except if doing so creates a cycle,
and verify that the resulting priority is indeed not score-structured. 
On the Food Inspection and Physicians datasets, we build four priority relations: two score-structured with  
$n=2$ and $n=5$, and 
two non-score-structured with  
$p=0.5$ and $p=0.8$. 
For the CQAPri benchmark, we build 
two priority relations, one score-structured with $n=5$ and 
one non-score-structured with $p=0.8$, on our largest dataset (\mn{u20c50}), then propagate them to the other datasets. 

%
\section{Experimental Evaluation}

\begin{table}
\small{
\setlength{\tabcolsep}{0pt}
\begin{tabular*}{0.47\textwidth}{l @{\extracolsep{\fill}} r r r r r r }
\toprule
&& Triv.
& IAR$\setminus$Triv.
& AR$\setminus$IAR
& brave$\setminus$AR
& not brave
\\\midrule
S&&\multicolumn{2}{c}{1,655} & 7 & 77 & 0\\
$\{$P,C$\}$-2&& 1,668 & 0 & 7 & 48 & 16\\
$\{$P,C$\}$-5&& 1,679 & 1 & 7 & 29 & 23\\
P-0.5&& 1,671 & 1 & 7 & 45 & 15\\
C-0.5&& 1,671 & 23 & 0 & 30 & 15\\
P-0.8&& 1,680 & 5 & 7 & 25 & 22\\
C-0.8&& 1,680 & 24 & 0 & 13 & 22\\
\bottomrule
\end{tabular*}}
\caption{Number of answers of \mn{q1} over the Physicians dataset depending on the priority relation (none, score-structured with $n=2$ or $n=5$, and not score-structured with $p=0.5$ or $p=0.8$) and type of repairs (standard,  Pareto- or completion-optimal).}\label{tab:answers}
\end{table}

Our experimental evaluation aims at assessing (i) the impact of adopting 
different kinds of repairs, and (ii) the relative performances of alternative procedures for the same semantics. More precisely, we consider the following questions. 
\begin{itemize}
\item What is the impact in terms of number of answers of adopting optimal repairs rather than standard repairs, or completion-optimal repairs instead of Pareto-optimal repairs when the priority relation is not score-structured?
\item What is the impact of using one kind of repairs rather than another on the computation time?
\item Given a semantics and type of repair, what is the impact in terms of computation times of the choice of:
\begin{itemize}
\item  how to encode optimality (P$_1$ or P$_2$ for Pareto-optimal repairs, P$_1$, P$_2$ or C when $\succ$ is score-structured)?
\item how to encode contradictions ($\varphi_{\neg \Cmc}$ with \cqaprienc or \cavsatenc)?
\item the algorithm used to filter the non-trivial answers? 
\end{itemize}
\end{itemize}

In what follows, we 
summarize our main observations. Detailed results 
are given 
in the appendix. 


\subsubsection*{Comparing Semantics w.r.t.\ Number of Answers} 
Table \ref{tab:answers} shows the impact on the number of answers of the type of priority relation and chosen notion of optimal repairs for an example query. For each priority relation and repair type X, it gives the number of answers that: are trivially X-IAR (i.e.\ some cause contains only facts without outgoing edges in the directed conflict graph), hold under X-IAR but not trivially, hold under X-AR but not under X-IAR, hold under X-brave but not X-AR, and do not hold under X-brave semantics.

\definecolor{common}{HTML}{696969}
\definecolor{filterRemaining}{HTML}{BEBEBE}

\begin{figure}
\begin{tikzpicture}
\begin{axis}[ybar stacked,xtick=data, label style={font=\footnotesize},legend style={font=\footnotesize},label style={font=\footnotesize},bar width=1mm,width=0.5\textwidth,height=5cm, yticklabel style={rotate=80},xticklabel style={rotate=80},xticklabels={
AR, 
AR 2,
IAR 2,
brave 2,
AR 5,
IAR 5,
brave 5,
P-AR 0.5,
P-IAR 0.5,
P-brave 0.5,
C-AR 0.5,
C-IAR 0.5,
C-brave 0.5,
P-AR 0.8,
P-IAR 0.8,
P-brave 0.8,
C-AR 0.8,
C-IAR 0.8,
C-brave 0.8,
}, ymin=0]\addplot[common,fill=common] coordinates{
(0,87) 
(1,92) 
(2,103) 
(3,107) 
(4,66) 
(5,62) 
(6,62) 
(7,85) 
(8,76) 
(9,84)  
(10,0) 
(11,0) 
(12,0) 
(13,61) 
(14,61) 
(15,82) 
(16,0) 
(17,0) 
(18,0) 
};
\addplot[filterRemaining,fill=filterRemaining] coordinates{
(0,322) 
(1,789) 
(2,1905) 
(3,1092) 
(4,293) 
(5,621) 
(6,494) 
(7,1120) 
(8,2203) 
(9,1432) 
(10,0) 
(11,0) 
(12,0) 
(13,815) 
(14,1728) 
(15,1487) 
(16,0) 
(17,0) 
(18,0) 
};
\end{axis} 
\end{tikzpicture}
\caption{Best running times (in milliseconds) for each semantics and priority relation (none, score-structured with $n=2$ or $n=5$, not score-structured with $p=0.5$ or $p=0.8$) for query \mn{q2} over the Food Inspection dataset. An empty bar means that the query ran out of time / memory for all possible algorithms and encodings. The lower part of bars 
is the time to identify self-inconsistent facts and trivial answers, 
the upper part 
the time 
to filter non-trivial answers. 
}\label{fig:semantics-times-comparison}
\end{figure}

Priority relations leads to fewer  
edges in the directed conflict graph, which in turn makes more answers hold trivially. 
The more the priority relation sets preferences between facts (higher parameter $n$ or $p$ of the priority relation), the more trivially X-IAR answers we obtain. 
Adopting optimal repairs also significantly increases  
the number of potential answers that do not hold under X-brave semantics, which are very rare when using classical subset repairs. 

An interesting observation is that while, in the absence of a priority relation, many queries of the CQAPri benchmark do not have any AR answers that are not trivial, which makes trivial answers a good approximation of AR, 
this is no longer the case for optimal repairs. It hence seems even more important 
to actually compute the answers that hold under the desired semantics rather 
simply computing the polynomial lower bound given by the trivial answers. 

Regarding the impact of the choice between Pareto- and completion-optimal repairs, 
in many cases {\sc orbits} did not manage to compute the answers for completion-optimal repairs in our given time and memory limits.
When it does manage to compute them (for 5 queries on the Physicians dataset; none on the Food Inspection dataset; and from between 7 and 13 on \mn{uXc1} to less than 3 on \mn{uXc50}), we observe a difference with Pareto-answers in only two cases.

\begin{table}
\small{
\setlength{\tabcolsep}{3pt}
\begin{tabular*}{0.48\textwidth}{l l @{\extracolsep{\fill}} rrrrrr}
\toprule
&& \mn{q1}& \mn{q2}& \mn{q3}& \mn{q4}& \mn{q5}& \mn{q6}
\\\midrule
\multirow{3}{*}{Alg.\ 1}
&P$_1$& 417 & \cellcolor{Lgray}{141} & 350 & 12,799 & \cellcolor{Lgray}{224} & 4,009\\
&P$_2$ & 804 & \cellcolor{Lgray}{142} & 379 & 326,594 & \cellcolor{Lgray}{213} & 8,684\\
&C & 252,694 & \cellcolor{Lgray}{179} & 550 & oom & 284 & t.o\\
\midrule
\multirow{3}{*}{Alg.\ 2}
&P$_1$ & \cellcolor{Lgray}{\textcolor{red}{\textbf{268}}} & \cellcolor{Lgray}{166} & \cellcolor{Lgray}{326} & \cellcolor{Lgray}{\textcolor{red}{\textbf{1,730}}} & \cellcolor{Lgray}{214} & 11,263\\
&P$_2$ & 502 & \cellcolor{Lgray}{163} & 333 & 2,961 & \cellcolor{Lgray}{221} & 10,833\\
&C & oom & 632 & t.o & t.o & 551 & oom\\
\midrule
\multirow{3}{*}{Alg.\ 3}
&P$_1$  & \cellcolor{Lgray}{272} & \cellcolor{Lgray}{154} & \cellcolor{Lgray}{313} & t.o & \cellcolor{Lgray}{211} & 245,804\\
&P$_2$   & 466 & \cellcolor{Lgray}{146} & \cellcolor{Lgray}{\textcolor{red}{\textbf{281}}} & t.o & \cellcolor{Lgray}{\textcolor{red}{\textbf{201}}} & 241,030\\
&C   & oom & 624 & t.o & t.o & 550 & oom\\
\midrule
\multirow{3}{*}{Alg.\ 4}
&P$_1$  & 362 & \cellcolor{Lgray}{166} & 997 & 42,544 & 281 & 559,923\\
&P$_2$  & 566 & 193 & 972 & 36,923 & 304 & 546,199\\
&C & oom & 764 & t.o & t.o & 846 & oom\\
\midrule
\multirow{3}{*}{Alg.\ 5}
&P$_1$ & 383 & \cellcolor{Lgray}{\textcolor{red}{\textbf{135}}} & 335 & 8,192 & \cellcolor{Lgray}{211} & \cellcolor{Lgray}{\textcolor{red}{\textbf{3,419}}}\\
&P$_2$   & 565 & \cellcolor{Lgray}{157} & \cellcolor{Lgray}{309} & 225,170 & \cellcolor{Lgray}{207} & 5,963\\
&C & 192,429 & \cellcolor{Lgray}{164} & 544 & oom & \cellcolor{Lgray}{238} & t.o\\
\bottomrule
\end{tabular*}}
\caption{Query answer filtering time (in milliseconds, t.o:time out, oom:out of memory) under X-brave semantics (X~$\in\{$P,C$\}$), for each algorithm and encoding $\varphi_\mn{X\text{-}max}$, on Physicians dataset with 
 score-structured priority ($n=2$). Alg.\ 1: \algosat, Alg.\ 2: \algomaxsat, Alg.\ 3: \algomuses, Alg.\ 4: \algoassump, Alg.\ 5: \algocauses. 
Best time  in bold red and `close to best times' (\ie not exceeding the best 
by more than 50ms or 10\%) on 
grey. 
}\label{tab:times}
\end{table}

\subsubsection*{Comparing Semantics \wrt Computation Time} 
Figure \ref{fig:semantics-times-comparison} shows the best running times (across all algorithms and encoding variants) for each semantics and an example query. 
We did this comparison for all queries on Physicians and Food Inspection datasets and two CQAPri datasets. 

Given a kind of repair X, the relative difficulty of the X-AR, X-IAR and X-brave semantics depends on the query, dataset and sometimes the nature of the priority relation. 

Comparing S-AR, P-AR, and C-AR semantics, we observe that using optimal repairs may either increase or decrease 
the answer filtering time, intuitively because there are more trivial answers, but the encodings are more complex. 

Given a dataset, query and semantics, if we compare two priority relations of the same kind (score-structured or not), the one that sets preferences between more facts leads to lower running times. This can be explained by the increase of the number of trivial answers and maybe by the fact that the encodings involve less facts and encode more `forced choices' between facts so that there are less possibilities to explore. 
When we compare the `easiest' non score-structured priority relations ($p=0.8$) and the hardest score-structured ones ($n=2$), which lead to comparable sizes of directed conflict graphs, score-structured priority relations seem to be easier than non-score-structured ones.

\pgfplotstableread[col sep=comma]{
param,4,13,20,30,37,46
q1,1571,6989,20079,82349,253729,nan
q2,463,1455,3446,10017,28370,232386
q3,43,84,92,420,591,2622
q4,37391,233382,618392,nan,nan,nan
q5,42,73,103,102,112,846
q6,42,113,148,205,327,4772
q7,75,179,375,664,1180,2602
q8,57,141,158,185,408,713
q9,1037,3764,7085,19367,51791,370340
q10,64,226,379,493,983,1983
q11,131,306,516,1193,2519,10004
q12,356,1152,2160,5441,11383,71135
q13,238,527,923,2252,4726,21990
q14,144,318,507,974,1913,5822
q15,925,2721,5074,11378,23765,113593
q16,12944,63914,178995,567216,1491941,nan
q17,62,211,366,649,1026,2696
q18,389,907,2042,5376,12631,86121
q19,39,122,159,274,578,2460
q20,41,81,107,131,165,1762
}\Data
\pgfplotstabletranspose[colnames from=param]\TransposedData\Data
\begin{figure*}
\begin{subfigure}{0.33\textwidth}\centering\resizebox{\linewidth}{!}{
\begin{tikzpicture}
\begin{axis}[legend pos=outer north east,legend columns=2, xmax=50, ymax=1600000, ymode=log]
\addplot[blue,mark=square*,solid] table [y=q1, x=colnames]{\TransposedData};
\addlegendentry{q1}
\addplot[red,mark=triangle* ,solid] table [y=q2, x=colnames]{\TransposedData};
\addlegendentry{q2}
\addplot[brown,mark=otimes* ,solid] table [y=q3,  x=colnames]{\TransposedData};
\addlegendentry{q3}
\addplot[green,mark=diamond* ,solid] table [y=q4,  x=colnames]{\TransposedData};
\addlegendentry{q4}
\addplot[black,mark=pentagon* ,solid] table [y=q5,  x=colnames]{\TransposedData};
\addlegendentry{q5}
\addplot[orange,mark=* ,solid] table [y=q6,  x=colnames]{\TransposedData};
\addlegendentry{q6}
\addplot[violet, mark=star*,solid] table [y=q7,  x=colnames]{\TransposedData};
\addlegendentry{q7}
\addplot[blue,mark=star*,densely dotted] table [y=q8,  x=colnames]{\TransposedData};
\addlegendentry{q8}
\addplot[red,mark=square*,densely dotted] table [y=q9,  x=colnames]{\TransposedData};
\addlegendentry{q9}
\addplot[brown,mark=triangle*,densely dotted] table [y=q10,  x=colnames]{\TransposedData};
\addlegendentry{q10}
\addplot[green,mark=otimes*,densely dotted] table [y=q11,  x=colnames]{\TransposedData};
\addlegendentry{q11}
\addplot[black,mark=diamond*,densely dotted] table [y=q12,  x=colnames]{\TransposedData};
\addlegendentry{q12}
\addplot[orange,mark=pentagon*,densely dotted] table [y=q13,  x=colnames]{\TransposedData};
\addlegendentry{q13}
\addplot[violet,mark=*,densely dotted] table [y=q14,  x=colnames]{\TransposedData};
\addlegendentry{q14}
\addplot[blue,mark=*,densely dashed] table [y=q15,  x=colnames]{\TransposedData};
\addlegendentry{q15}
\addplot[red,mark=star*,densely dashed] table [y=q16,  x=colnames]{\TransposedData};
\addlegendentry{q16}
\addplot[brown,mark=square*,densely dashed] table [y=q17,  x=colnames]{\TransposedData};
\addlegendentry{q17}
\addplot[green,mark=triangle*,densely dashed] table [y=q18,  x=colnames]{\TransposedData};
\addlegendentry{q18}
\addplot[black,mark=otimes*,densely dashed] table [y=q19,  x=colnames]{\TransposedData};
\addlegendentry{q19}
\addplot[orange,mark=diamond*,densely dashed] table [y=q20,  x=colnames]{\TransposedData};
\addlegendentry{q20}
\end{axis}
\end{tikzpicture}}
\subcaption{\algosat}
\end{subfigure}
\pgfplotstableread[col sep=comma]{
param,4,13,20,30,37,46
q1,1221,3599,7972,26908,54701,135263
q2,464,1499,2837,8051,18973,84826
q3,23,111,115,204,246,456
q4,237134,nan,nan,nan,nan,nan
q5,23,54,70,85,99,1155
q6,25,129,157,159,198,434
q7,143,224,287,833,1052,2060
q8,85,111,150,145,207,369
q9,834,2755,6088,nan,nan,nan
q10,97,194,261,344,520,1043
q11,143,401,1131,nan,nan,nan
q12,1847,nan,nan,nan,nan,nan
q13,284,1120,41032,nan,nan,nan
q14,193,430,682,1497,3207,11765
q15,517,1441,2503,4454,6297,20432
q16,3965,26006,72169,115208,207914,180360
q17,107,273,390,774,1148,1942
q18,356,945,1675,4619,10020,23016
q19,22,152,188,365,681,2806
q20,24,115,123,137,145,332
}\Data
\pgfplotstabletranspose[colnames from=param]\TransposedData\Data
\begin{subfigure}{0.33\textwidth}\centering\resizebox{\linewidth}{!}{
\begin{tikzpicture}
\begin{axis}[legend pos=outer north east,legend columns=2, xmax=50, ymax=1600000, ymode=log]
\addplot[blue,mark=square*,solid] table [y=q1, x=colnames]{\TransposedData};
\addlegendentry{q1}
\addplot[red,mark=triangle* ,solid] table [y=q2, x=colnames]{\TransposedData};
\addlegendentry{q2}
\addplot[brown,mark=otimes* ,solid] table [y=q3,  x=colnames]{\TransposedData};
\addlegendentry{q3}
\addplot[green,mark=diamond* ,solid] table [y=q4,  x=colnames]{\TransposedData};
\addlegendentry{q4}
\addplot[black,mark=pentagon* ,solid] table [y=q5,  x=colnames]{\TransposedData};
\addlegendentry{q5}
\addplot[orange,mark=* ,solid] table [y=q6,  x=colnames]{\TransposedData};
\addlegendentry{q6}
\addplot[violet, mark=star*,solid] table [y=q7,  x=colnames]{\TransposedData};
\addlegendentry{q7}
\addplot[blue,mark=star*,densely dotted] table [y=q8,  x=colnames]{\TransposedData};
\addlegendentry{q8}
\addplot[red,mark=square*,densely dotted] table [y=q9,  x=colnames]{\TransposedData};
\addlegendentry{q9}
\addplot[brown,mark=triangle*,densely dotted] table [y=q10,  x=colnames]{\TransposedData};
\addlegendentry{q10}
\addplot[green,mark=otimes*,densely dotted] table [y=q11,  x=colnames]{\TransposedData};
\addlegendentry{q11}
\addplot[black,mark=diamond*,densely dotted] table [y=q12,  x=colnames]{\TransposedData};
\addlegendentry{q12}
\addplot[orange,mark=pentagon*,densely dotted] table [y=q13,  x=colnames]{\TransposedData};
\addlegendentry{q13}
\addplot[violet,mark=*,densely dotted] table [y=q14,  x=colnames]{\TransposedData};
\addlegendentry{q14}
\addplot[blue,mark=*,densely dashed] table [y=q15,  x=colnames]{\TransposedData};
\addlegendentry{q15}
\addplot[red,mark=star*,densely dashed] table [y=q16,  x=colnames]{\TransposedData};
\addlegendentry{q16}
\addplot[brown,mark=square*,densely dashed] table [y=q17,  x=colnames]{\TransposedData};
\addlegendentry{q17}
\addplot[green,mark=triangle*,densely dashed] table [y=q18,  x=colnames]{\TransposedData};
\addlegendentry{q18}
\addplot[black,mark=otimes*,densely dashed] table [y=q19,  x=colnames]{\TransposedData};
\addlegendentry{q19}
\addplot[orange,mark=diamond*,densely dashed] table [y=q20,  x=colnames]{\TransposedData};
\addlegendentry{q20}
\end{axis}
\end{tikzpicture}}
\subcaption{\algomaxsat}
\end{subfigure}
\pgfplotstableread[col sep=comma]{
param,4,13,20,30,37,46
q1,77235,nan,nan,nan,nan,nan
q2,714,5885,26245,188019,553947,nan
q3,62,117,129,249,277,473
q5,63,100,94,112,190,1009
q6,60,137,169,186,222,502
q7,144,217,322,814,994,2306
q8,90,122,131,136,189,374
q9,6587,nan,nan,nan,nan,nan
q10,114,185,234,320,483,1202
q11,139,3620,53085,nan,nan,nan
q12,176556,nan,nan,nan,nan,nan
q13,374,42531,nan,nan,nan,nan
q14,238,783,1568,3395,5552,12264
q15,2767,34044,346482,nan,nan,nan
q17,116,231,368,669,1127,2378
q18,504,3281,7578,72549,193728,nan
q19,67,122,166,330,757,2936
q20,62,109,136,157,194,369
}\Data
\pgfplotstabletranspose[colnames from=param]\TransposedData\Data
\begin{subfigure}{0.33\textwidth}\centering\resizebox{\linewidth}{!}{
\begin{tikzpicture}
\begin{axis}[legend pos=outer north east,legend columns=2, xmax=50, ymax=1600000, ymode=log]
\addplot[blue,mark=square*,solid] table [y=q1, x=colnames]{\TransposedData};
\addlegendentry{q1}
\addplot[red,mark=triangle* ,solid] table [y=q2, x=colnames]{\TransposedData};
\addlegendentry{q2}
\addplot[brown,mark=otimes* ,solid] table [y=q3,  x=colnames]{\TransposedData};
\addlegendentry{q3}
\addplot[black,mark=pentagon* ,solid] table [y=q5,  x=colnames]{\TransposedData};
\addlegendentry{q5}
\addplot[orange,mark=* ,solid] table [y=q6,  x=colnames]{\TransposedData};
\addlegendentry{q6}
\addplot[violet, mark=star*,solid] table [y=q7,  x=colnames]{\TransposedData};
\addlegendentry{q7}
\addplot[blue,mark=star*,densely dotted] table [y=q8,  x=colnames]{\TransposedData};
\addlegendentry{q8}
\addplot[red,mark=square*,densely dotted] table [y=q9,  x=colnames]{\TransposedData};
\addlegendentry{q9}
\addplot[brown,mark=triangle*,densely dotted] table [y=q10,  x=colnames]{\TransposedData};
\addlegendentry{q10}
\addplot[green,mark=otimes*,densely dotted] table [y=q11,  x=colnames]{\TransposedData};
\addlegendentry{q11}
\addplot[black,mark=diamond*,densely dotted] table [y=q12,  x=colnames]{\TransposedData};
\addlegendentry{q12}
\addplot[orange,mark=pentagon*,densely dotted] table [y=q13,  x=colnames]{\TransposedData};
\addlegendentry{q13}
\addplot[violet,mark=*,densely dotted] table [y=q14,  x=colnames]{\TransposedData};
\addlegendentry{q14}
\addplot[blue,mark=*,densely dashed] table [y=q15,  x=colnames]{\TransposedData};
\addlegendentry{q15}
\addplot[brown,mark=square*,densely dashed] table [y=q17,  x=colnames]{\TransposedData};
\addlegendentry{q17}
\addplot[green,mark=triangle*,densely dashed] table [y=q18,  x=colnames]{\TransposedData};
\addlegendentry{q18}
\addplot[black,mark=otimes*,densely dashed] table [y=q19,  x=colnames]{\TransposedData};
\addlegendentry{q19}
\addplot[orange,mark=diamond*,densely dashed] table [y=q20,  x=colnames]{\TransposedData};
\addlegendentry{q20}
\end{axis}
\end{tikzpicture}}
\subcaption{\algomuses}
\end{subfigure}
\caption{Time (in milliseconds, log. scale) to filter query answers under X-AR semantics (X~$\in\{$P,C$\}$) \wrt percentage of facts involved in some conflict for \mn{u20cY} with score-structured priority relation ($\varphi_\mn{P_{1}\text{-}max}$ and \cqaprienc encoding). Missing queries ran out of time or memory.}\label{fig:times-wrt-conf}
\end{figure*}

Finally, we conclude that our procedures do not perform well for completion-optimal repair-based semantics, which 
form most of the cases that fail due to lack of time or memory, 
and very often have higher running times than Pareto-optimal-based semantics with the same priority relation. 

\subsubsection*{Choice of Algorithm \& Encoding for Given Semantics} 
Table \ref{tab:times} presents the time needed to filter the candidate answers that hold under brave semantics based on optimal-repairs (score-structured case) for some example queries. It illustrates the huge impact that the choice of an algorithm and encoding can have. For example, \mn{q6} answers are filtered in 3.5s with algorithm \algocauses and $\varphi_\mn{P_{1}\text{-}max}$, but need at least 546s with the \algoassump algorithm, at least 5.9s with $\varphi_\mn{P_{2}\text{-}max}$, and cannot be filtered in our time and memory limits with encoding $\varphi_\mn{C\text{-}max}$. 
For X-AR and X-IAR semantics, we also observe sometimes huge variations when using the \cqaprienc or \cavsatenc version of $\varphi_{\neg\Cmc}$. For example, for X-AR semantics on \mn{u20c50} with a score-structured priority relation, the best times for queries \mn{q2} and \mn{q18} are both obtained with algorithm \algomaxsat and $\varphi_\mn{P_{1}\text{-}max}$, but with the \cavsatenc variant for \mn{q2} (50 seconds versus 85 with \cqaprienc) and \cqaprienc for \mn{q18} (23 seconds versus 47 with \cavsatenc). 
The comparison of the possible procedures for each semantics on the different datasets and queries shows that there is not a `best' method in general. 
However, we still gain some relevant insights.

The first one concerns the choice of $\varphi_\mn{X\text{-}max}$. 
The encoding $\varphi_\mn{P_{1}\text{-}max}$ is generally the best one for Pareto-optimal repairs, in the sense that it achieves `close to best times' (\cf Table \ref{tab:times}) much more often than the others. 
However, there are a few cases where $\varphi_\mn{P_{2}\text{-}max}$ performs significantly better, especially on the CQAPri datasets with fewer conflicts (\eg on \mn{u20c1} with a score-structured priority relation, the best time for filtering \mn{q9} answers under P-AR semantics is 480ms with $\varphi_\mn{P_{2}\text{-}max}$, while the best time with another encoding is 825ms). 
When the priority relation is score-structured, $\varphi_\mn{C\text{-}max}$ never significantly outperforms $\varphi_\mn{P_{1}\text{-}max}$ and $\varphi_\mn{P_{2}\text{-}max}$ and leads to much more time or memory failures. 

The second concerns the choice of algorithm for X-IAR semantics. For all kinds of repairs, algorithm \algoIARcauses is generally better than the others in terms of frequency of `close to best times'. It is sometimes outperformed by algorithms \algocauses or \algoIARfacts. The `generic' algorithms (\algosat, \algomaxsat, \algomuses, \algoassump) perform quite poorly, except on the simplest cases.

For the AR and brave semantics, it is more difficult to find an algorithm that is superior to the others. 
For non-score-structured priority relations and completion-optimal repairs, algorithm \algosat seems to be the best choice for both C-AR and C-brave semantics, but all algorithms fail in most cases. 
A related observation is that algorithms that consider answers individually and use smaller encodings seems to be often more robust in terms of time-out and out-of-memory, with a notable exception for P-AR and P-brave semantics on \mn{u20c50} with non-score-structured priority, where algorithms \algomaxsat and \algomuses are more robust.  
For S-AR, P-AR and P-brave, we observe different behaviours depending on the benchmark: 
 \algomaxsat and \algomuses tend to perform better for the CQAPri benchmark, while 
 \algosat tends to perform better for the Food Inspection dataset.

Finally, comparing \cqaprienc and \cavsatenc versions of $\varphi_{\neg\Cmc}$, we observe that the relative performance depends on the dataset, query, algorithm, and choice of $\varphi_\mn{X\text{-}max}$. However, we note that $\varphi_\mn{P_{2}\text{-}max}$ usually works better with \cqaprienc. Even if there is not a direct relationship between the encoding sizes and the running times, this is 
probably due to the fact that \cavsatenc enforces that both the facts that occur in the causes and their conflicts are part of the encoding, which significantly increases 
the size of $\varphi_\mn{P_{2}\text{-}max}$, but has little 
impact on $\varphi_\mn{P_{1}\text{-}max}$.  

Figure \ref{fig:times-wrt-conf} shows the evolution of the running times of three algorithms using the same encoding variants 
as the proportion of facts involved in some conflicts grows, on the \mn{u20cY} datasets with a score-structured priority for X-AR semantics (X~$\in\{$P,C$\}$). 
It illustrates the fact that the relative performance of the algorithms depends on the query and dataset (here, the proportion of facts involved in some conflict): For example, \algomaxsat is the best for \mn{q9} over the three first datasets, but runs out of time on the three last (more that 20\% of facts in conflict), while \algosat can handle them. 

%
\section{Conclusion} 
We have presented a comprehensive exploration of SAT-based approaches to querying inconsistent data using 
(optimal) repair-based inconsistency tolerant semantics, including the proposal of novel encodings and algorithms.
Our generic framework places existing approaches into a broader context and makes our results and 
system 
directly applicable to both the (pure) database and OMQA settings. 

Our experimental comparison of different SAT-based algorithms and encoding variants  
shows that the choice of algorithm and encoding may have huge impact on the computation time.  
While in some cases our results can be used to single out some approaches as more effective, 
more often there are no clear winner(s). 
This suggests that to minimize runtimes, 
it may make sense to  
launch multiple algorithms 
in parallel, and/or  
devise methods that can help predict which algorithm and encoding will perform 
best on a given dataset and query, e.g.\ using machine learning techniques. 


Our work lays 
important foundations  for 
the future development of mature systems for 
querying inconsistent data. We plan to investigate different
ways of improving the performance
for optimal repair-based semantics. For example, it would be interesting to explore alternative approaches for completion-optimal repairs 
based upon 
SAT modulo graph techniques \cite{DBLP:conf/jelia/GebserJR14}.
Another promising direction is to employ 
more refined polynomial 
lower approximations than the trivial answers, 
such as the grounded semantics \cite{DBLP:conf/kr/BienvenuB20}.


%
%
%
%

\section*{Acknowledgements}
This work was supported by the ANR AI Chair INTENDED (ANR-19-CHIA-0014).
\bibliographystyle{kr}
\bibliography{cqa-priority}

\newpage
\onecolumn
\appendix
\appendixpage
\startcontents[sections]
\printcontents[sections]{l}{1}{\setcounter{tocdepth}{2}}

\clearpage
\section{Proofs for Section \ref{sec:prelim}}

\begin{proposition}
Query entailment is  $\np$-hard under X-brave semantics (X $\in \{P,C\}$) and $\conp$-hard under X-IAR semantics (X $\in \{P,C\}$)
over KBs whose theory consists of a set of FDs.
\end{proposition}
\begin{proof}
We first give the $\np$-hardness proof for X-brave semantics (X $\in \{P,C\}$), by reduction 
from propositional 3CNF satisfiability. Consider a propositional CNF $\varphi= c_1 \ldots \wedge \ldots c_n$,
over variables $x_1, \ldots, x_k$. We consider the KB $\Kmc=(\Dmc_\varphi,\Tmc)$, where 
\begin{align*}
\Dmc_\varphi \quad =\quad &  \{R(x_i, x_i, b) \mid 1 \leq i \leq k, b \in \{0,1\}\} \cup \\
&  \{R(c_j,x_i, 1) \mid x_i \text { is a disjunct of } c_j\}  \cup  \{R(c_j,x_i, 0) \mid \neg x_i \text { is a disjunct of } c_j\}  \cup\\
&  \{R(c_j,u,u) \mid 1 \leq j \leq n\} \cup \{(\star, u, \star)\}
\end{align*}
and the theory $\Tmc$ consists of 
the FDs $R: 1 \rightarrow 3$ and $R:2 \rightarrow3$, which can be equivalently expressed by the FOL formulas:
$$\forall u,v,v'w,w' \neg (R(u,v,w) \wedge R(u,v',w') \wedge w\neq w') \qquad \forall u,u',v,w,w' \neg (R(u,v,w) \wedge R(u',v,w') \wedge w\neq w') $$
We let $\succ$ be the priority induced by the scoring function which assigns scores as follows: 
\begin{itemize}
\item Score 3: all facts of the form $R(x_i, x_i, b)$
\item Score 2: all facts of the form $R(c_j,\_,\_)$ (i.e. with some $c_j$ as first argument)
\item Score 1: the fact $(\star, u, \star)$
\end{itemize}
For example, if $\neg x_i \in c_j$, then $R(x_i,x_i,1) \succ R(c_j,x_i, 0)$, since $R(x_i,x_i,1)$ and $R(c_j,x_i, 0)$ violate the FD $R:2 \rightarrow3$,
and $R(x_i,x_i,1)$ has a higher score than $R(c_j,x_i, 0)$. We claim that: $\Kmc_\succ \bravemodels{P} (\star, u, \star)$ iff $\varphi$ is satisfiable. 
Note that since $\succ$ is score-structured, $\Kmc_\succ \bravemodels{P} (\star, u, \star)$ iff $\Kmc_\succ \bravemodels{C} (\star, u, \star)$, so proving the claim establishes the result for both types of repair. 

To prove the first direction of the claim,
suppose that $\Kmc_\succ \bravemodels{P} (\star, u, \star)$. Then there exists a Pareto-optimal repair $\Rmc$ of $\Kmc_\succ$
such that $(\star, u, \star) \in \Rmc$. As $\succ$ is score-structured, $\Rmc$ is also a $\subseteq_P$-repair  \cite{DBLP:conf/aaai/BienvenuBG14}, which means it can be built 
by taking a $\subseteq$-maximal $\Tmc$-consistent subset $S_3$ of the score-3 facts, then adding a $\subseteq$-maximal subset $S_2$ of the score-2 facts
such that $S_3 \cup S_2$ is $\Tmc$-consistent, followed by adding $(\star, u, \star)$ if $S_3 \cup S_2 \cup \{(\star, u, \star)\}$ is $\Tmc$-consistent. 
Clearly, $S_3$ will contain exactly one fact of the form $R(x_i, x_i, b)$ for each variable $x_i$, yielding a valuation $\nu_\Rmc$ of $\{x_1, \ldots, x_n\}$
that assigns truth value $b$ to $x_i$ if $R(x_i, x_i, b)\in \Rmc$. 
As $(\star, u, \star) \in \Rmc$ and $\Rmc$ is $\Tmc$-consistent (hence satisfies $R:2 \rightarrow3$), we know that 
$\Rmc$ does not contain any fact of the form $R(c_j,u,u)$. Indeed, if it did, then $S_3 \cup S_2 \cup \{(\star, u, \star)\}$ would not be $\Tmc$-consistent,
so  $(\star, u, \star)$ would not have been added. As $S_2$ is $\subseteq$-maximal subset of the score-2 facts
such that $S_3 \cup S_2$ is $\Tmc$-consistent, it must be the case that for every $1 \leq i \leq n$, there exists a fact $R(c_j,x_i,b) \in \Rmc$. 
However, due to $R:2 \rightarrow3$, we know that if $R(c_j,x_i,b) \in \Rmc$, then $R(x_i,x_i,b) \in \Rmc$. It follows that every clause is satisfied by 
the valuation $\nu_\Rmc$, so $\varphi$ is satisfiable. 

Now we prove the second direction. Suppose that $\varphi$ is satisfiable, and let $\mu$ be a satisfying valuation. Consider the subset $\Rmc \subseteq \Dmc_\varphi$ that contains:
\begin{itemize}
\item $R(x_i, x_i, b)$, if $\mu(x_i)=b$
\item A maximal \Tmc-consistent subset of $\{R(c_j, x_i, b) \mid R(c_j, x_i, b) \in \Dmc, \mu(x_i)=b\}$
\item $(\star, u, \star)$
\end{itemize}
We claim that $\Rmc$ is $\subseteq_P$-repair, which implies that it is also a Pareto- and completion-optimal repair, as $\succ$ is score-structured. 
First note that $\Rmc$ is $\Tmc$-consistent. Indeed, if $R(x_i,x_i,b)$ and $R(c_j, x_i, b')$ both appear, then by construction, we must have $b=b'$. 
Next note that we cannot add any further $R(x_i, x_i, b)$ facts without becoming $\Tmc$-inconsistent, 
which means that we have included a maximal subset of the score-3 facts. We now consider the score-2 facts. As every clause $c_j$ has at least one satisfied literal, we will include at least one fact of the form $R(c_j,x_i,b)$ per clause $c_j$, which will prevent us from adding any $R(c_j,u,u)$ facts. We also cannot add a fact $R(c_j,x_i,b)$ if $\mu(x_i) \neq b$, since then we would contradict a score-3 fact from $\Rmc$. Finally, the only score-1 fact, $(\star, u, \star)$, is already present in $\Rmc$. It follows that $\Rmc$ is indeed a $\subseteq_P$-repair, which implies that it is also a Pareto-optimal repair, since $\succ$ is score-structured. This shows that $\Kmc_\succ \bravemodels{P} (\star, u, \star)$. \medskip

We can use almost the same reduction to show the $\conp$-hardness proof for X-IAR semantics (X $\in \{P,C\}$). Indeed, it suffices to add to $\Dmc_\varphi$ an additional score-1 fact: $(\star, \odot, \odot)$. Using similar arguments to above, we can show that 
$\Kmc_\succ \iarmodels{P} (\star, \odot, \odot)$ iff $\varphi$ is unsatisfiable. Indeed, $\Kmc_\succ \iarmodels{P} (\star, \odot, \odot)$ iff  
$(\star, \odot, \odot)$ holds in every optimal repair iff $(\star, u, \star)$ is absent from every optimal repair. And we have shown above that the latter holds iff $\varphi$ is unsatisfiable. 
\end{proof}

\section{Proofs for Sections \ref{sec:encodings} and  \ref{sec:algos}}

\subsubsection*{Inconsistent or non-minimal `causes'} In this section and in all of the following proofs, we assume that $\causes{q(\ans),\Kmc}$ may actually be a superset of the causes such that every superfluous $\Bmc$ in it either (1) includes a real cause $\Cmc$ of $q(\ans)$ or (2) contains two distinct facts $\alpha$ and $\beta$ that form a conflict. Note that this only strengthens the results, which naturally hold also when $\causes{q(\ans),\Kmc}$ is the proper set of causes. 

\subsection{Basic Lemmas}\label{sec:basiclemmasencodings}

In Lemmas \ref{lem:notC} to \ref{lem:query}, $\Kmc_\succ$ is a prioritized KB with $\Kmc=(\Dmc,\Tmc)$, $\Rmc\in\xreps{\Kmc_\succ}$ (with X $\in \{P, C\}$), and $\nu$ is the valuation of $\{x_\alpha\mid \alpha\in \Dmc\}$ such that $\nu(x_\alpha)=\true$ iff $\alpha\in\Rmc$.

\begin{lemma}\label{lem:notC}
$\nu$ satisfies $ \varphi_{\neg \Cmc}$ iff $\Cmc\not\subseteq\Rmc$.
\end{lemma}
\begin{proof}
If $\nu$ satisfies $\varphi_{\neg \Cmc}$, there exists $\alpha\in\Cmc$ such that
\begin{itemize}
\item in case of \cavsatenc encoding: $\nu(x_\alpha)=\false$ and 
\item in both \cavsatenc and \cqaprienc encoding variants: there exists $\beta$ such that $\alpha\confof\beta$, $\alpha\not\succ\beta$ and $\nu(x_\beta)=\true$, \ie $\beta\in\Rmc$. 
\end{itemize}
Hence, since $\Rmc$ is \Tmc-consistent, we must have $\Cmc\not\subseteq\Rmc$. 

Conversely, if $\Cmc\not\subseteq\Rmc$, there exist $\alpha\in\Cmc$ and $\beta\in\Rmc$ such that $\alpha\confof\beta$ and $\alpha\not\succ\beta$. Otherwise $\Rmc\cup\{\alpha\}\setminus\{\beta\mid\alpha\succ\beta\}$ would be a Pareto-improvement of $\Rmc$, contradicting our assumption $\Rmc\in\xreps{\Kmc_\succ}$ for either X $\in \{P, C\}$ (recall that completion-optimal repairs are also Pareto-optimal repairs). Since $\nu(x_\alpha)=\false$ and $\nu(x_\beta)=\true$, $\nu$ satisfies $\varphi_{\neg \Cmc}$. 
\end{proof}

\begin{lemma}\label{lem:cause}
$\nu$ satisfies $ \varphi_{ \Cmc}$ iff $\Cmc\subseteq\Rmc$.
\end{lemma}
\begin{proof}
Straightforward.
\end{proof}

\begin{lemma}\label{lem:notQ}
$\nu$ satisfies $ \varphi_{\neg q(\ans)}$ iff $(\Rmc,\Tmc)\not\models q(\ans)$.
\end{lemma}
\begin{proof}
If $\nu$ satisfies $\varphi_{\neg q(\ans)}$, for every $\Cmc\in \causes{q(\ans),\Kmc}$, $\nu$ satisfies $\varphi_{\neg\Cmc}$, so by Lemma \ref{lem:notC}, $\Cmc\not\subseteq\Rmc$. Hence $(\Rmc,\Tmc)\not\models q(\ans)$. 
It follows that $\Kmc_\succ \not\armodels{X} q(\ans)$. 

If $(\Rmc,\Tmc)\not\models q(\ans)$, $\Rmc$ does not contain any cause for $q(\ans)$ (hence not any super-set of such cause either), and since $\Rmc$ is \Tmc-consistent, it does not contain any inconsistent subset. 
It follows that for every $\Bmc\in\causes{q(\ans),\Kmc}$ (even if $\Bmc$ is a super set of a cause or contains a pair of conflicting assertions), $\Bmc\not\subseteq\Rmc$, so by Lemma \ref{lem:notC}, $\nu$ satisfies $\varphi_{\neg\Bmc}$. 
Hence $\nu$ satisfies $\varphi_{\neg q(\ans)}$. 
\end{proof}

\begin{lemma}\label{lem:query}
$\nu$ can be extended to a valuation that satisfies $ \varphi_{q(\ans)}$ iff $(\Rmc,\Tmc)\models q(\ans)$.
\end{lemma}
\begin{proof}
If $(\Rmc,\Tmc)\models q(\ans)$, $\Rmc$ contains a real cause $\Cmc$ of $q(\ans)$. Let us extend $\nu$ with $\nu(x_\Cmc)=\true$, and $\nu(x_{\Bmc})=\false$ for every $\Bmc\in\causes{q(\ans),\Kmc_\succ}$ different from $\Cmc$. Then $\nu$ satisfies $\varphi_{q(\ans)}$. 

If $\nu$ can be extended to a valuation that satisfies $\varphi_{q(\ans)}$, there exists $\Bmc\in\causes{q(\ans),\Kmc_\succ}$ such that $\nu(x_\Bmc)=\true$ and for every $\alpha\in\Bmc$, $\nu(x_\alpha)=\true$, \ie $\Bmc\subseteq\Rmc$. Since $\Rmc$ is \Tmc-consistent, $\Bmc$ is \Tmc-consistent hence must include a real cause of $q(\ans)$. It follows that  $(\Rmc,\Tmc)\models q(\ans)$.
\end{proof}

The following lemma has already been shown for the particular case where $\succ$ is score structured. Indeed, it follows from the results that (i) when the priority relation is score-structured, (Pareto- or completion-)optimal repairs coincide with the $\subseteq_P$-repairs defined by \citeauthor{DBLP:conf/aaai/BienvenuBG14} \shortcite{DBLP:conf/aaai/BienvenuBG14} and (ii) the proof for the SAT encodings presented in \cite{DBLP:conf/aaai/BienvenuBG14} for the semantics based on $\subseteq_P$-repairs. 
However, it has not been observed before that this is also true when $\succ$ is not score-structured.

\begin{lemma}\label{lem:p1max}
Let $\Kmc_\succ$ be a prioritized KB with $\Kmc=(\Dmc,\Tmc)$, $F_1\subseteq \Dmc$, $F_2= F_1\cup \facts( \varphi_\mn{P_{1}\text{-}max}(F_1))$, and let $\nu$ be a valuation of $\{x_\alpha\mid \alpha\in F_2\}$.
\begin{itemize}
\item If $\nu$ satisfies $ \varphi_\mn{P_{1}\text{-}max}(F_1)\wedge \varphi_\mn{cons}(F_2)$ then $S_\nu=\{\alpha\mid \nu(x_\alpha)=\true\}$ can be extended to a Pareto-optimal repair of $\Kmc_\succ$. 
\item If there exists a Pareto-optimal repair $\Rmc$ of $\Kmc_\succ$ such that $\Rmc\cap F_2=S_\nu$, then $\nu$ satisfies $ \varphi_\mn{P_{1}\text{-}max}(F_1)\wedge \varphi_\mn{cons}(F_2)$.
\end{itemize}
\end{lemma}
\begin{proof}
\noindent$(\Rightarrow)$ Assume that $\nu$ satisfies $\varphi_\mn{P_{1}\text{-}max}(F_1)\wedge \varphi_\mn{cons}(F_2)$. Since $\nu$ satisfies $\varphi_\mn{cons}(F_2)$, it is easy to check that $S_\nu$ is \Tmc-consistent, so there exists some repair that extends it. 
Assume for a contradiction that all repairs that extend $S_\nu$ are not Pareto-optimal and take a repair $\Rmc_0\supseteq S_\nu$. 
\begin{itemize}
\item Since $\Rmc_0\notin\preps{\Kmc_\succ}$, there exists $\beta_0\in \Dmc\setminus\Rmc_0$ such that $S_0=\Rmc_0\cup\{\beta_0\}\setminus\{\alpha\mid \beta_0\succ\alpha\}$ is a Pareto-improvement of $\Rmc_0$. We first show that $S_\nu\subseteq S_0$. Assume for a contradiction that this is not the case: there is $\alpha\in S_\nu$ such that $\beta_0\succ\alpha$.
\begin{itemize}
\item Since $\alpha\in S_\nu\subseteq F_2=\reachr(F_1)$, then $\beta_0\in \reachr(F_1)$ and $\varphi_\mn{P_{1}\text{-}max}(F_1)$ contains the clause $x_{\beta_0} \vee \bigvee_{\beta_0\confof \gamma , \beta_0\not\succ\gamma } x_\gamma $. 
\item Since $\alpha\in S_\nu$ and $S_\nu$ is \Tmc-consistent, $\beta_0\notin S_\nu$ so $\nu(\beta_0)=\false$. Hence there exists $\gamma$ such that $\beta_0\confof\gamma$ and $\beta_0\not\succ\gamma$ such that $\nu(x_\gamma)=\true$, \ie $\gamma\in S_\nu\subseteq \Rmc_0$. 
\item Since $\beta_0\not\succ\gamma$ , by definition of $S_0$, it follows that $\gamma\in S_0$. This contradicts the \Tmc-consistency of $S_0$ hence the fact that $S_0$ is a Pareto-improvement of $\Rmc_0$.  
\end{itemize}
It follows that $S_\nu\subseteq S_0$.
\item Any repair $\Rmc_1\supseteq S_0$ is thus a Pareto-improvement of $\Rmc_0$ which contains $S_\nu$. 
\item Since we assume that all repairs that extend $S_\nu$ are not optimal, we can then construct an infinite sequence $\Rmc_0,\Rmc_1,\dots$ of repairs that all contain $S_\nu$ and such that $\Rmc_{i+1}$ is a Pareto-improvement of $\Rmc_i$. 
\item The number of repairs being finite, there must be a repetition in this sequence. 
By  Proposition 7 in \cite{DBLP:journals/amai/StaworkoCM12}, since $\succ$ is acyclic, so is the relation of Pareto-dominance between repairs, which contradicts the existence of such a cycle of repairs. 
\end{itemize}
Hence, there exists an optimal repair $\Rmc$ that extends $S_\nu$. 
\medskip

\noindent$(\Leftarrow)$  Assume that there exists $\Rmc\in \preps{\Kmc_\succ}$ such that $S_\nu=\Rmc\cap F_2$. 
Since $\Rmc$ is \Tmc-consistent, it is easy to check that $\nu$ satisfies $\varphi_\mn{cons}(F_2)$. 
Assume for a contradiction that $\nu$ does not satisfy $\varphi_\mn{P_{1}\text{-}max}(F_1)$.
\begin{itemize}
\item  There exists $\alpha\in F_2$ such that $\nu(x_\alpha)=\false$ and for every $\beta\confof\alpha$ such that $\alpha\not\succ\beta$ (note that such $\beta$ is in $\reachr(F_1)$), $\nu(x_\beta)=\false$. 
\item Since $S_\nu=\Rmc\cap F_2$, then $\alpha\notin\Rmc$ and for every $\beta\confof\alpha$ such that $\alpha\not\succ\beta$, $\beta\notin\Rmc$. 
\item Hence $\Rmc'=\Rmc\cup\{\alpha\}\setminus\{\beta\mid\beta\confof\alpha\}$ is a Pareto-improvement of $\Rmc$, contradicting $\Rmc\in \preps{\Kmc_\succ}$.
\end{itemize}
Hence $\nu$ satisfies $\varphi_\mn{P_{1}\text{-}max}(F_1)$.
\end{proof}

\begin{lemma}\label{lem:p2max}
Let $\Kmc_\succ$ be a prioritized KB with $\Kmc=(\Dmc,\Tmc)$, $F_1\subseteq \Dmc$, $F_2= F_1\cup \facts( \varphi_\mn{P_{2}\text{-}max}(F_1))$, and let $\nu$ be a valuation of $\{x_\alpha\mid \alpha\in F_2\}$.
\begin{itemize}
\item If $\nu$ satisfies $ \varphi_\mn{P_{2}\text{-}max}(F_1)\wedge \varphi_\mn{cons}(F_2)$ then $S_\nu=\{\alpha\mid \nu(x_\alpha)=\true\}$ can be extended to a Pareto-optimal repair of $\Kmc_\succ$. 
\item If there exists a Pareto-optimal repair $\Rmc$ of $\Kmc_\succ$ such that $\Rmc\cap F_2=S_\nu$, then $\nu$ satisfies $ \varphi_\mn{P_{2}\text{-}max}(F_1)\wedge \varphi_\mn{cons}(F_2)$.
\end{itemize}\end{lemma}
\begin{proof}
\noindent$(\Rightarrow)$ Assume that $\nu$ satisfies $\varphi_\mn{P_{2}\text{-}max}(F_1)\wedge \varphi_\mn{cons}(F_2)$. Since $\nu$ satisfies $\varphi_\mn{cons}(F_2)$, it is easy to check that $S_\nu$ is \Tmc-consistent, so there exists some repair that extends it. 
Assume for a contradiction that all repairs that extend $S_\nu$ are not Pareto-optimal and take a repair $\Rmc_0\supseteq S_\nu$. 
\begin{itemize}
\item Since $\Rmc_0\notin\preps{\Kmc_\succ}$, there exists $\beta_0\in \Dmc\setminus\Rmc_0$ such that $S_0=\Rmc_0\cup\{\beta_0\}\setminus\{\alpha\mid \beta_0\succ\alpha\}$ is a Pareto-improvement of $\Rmc_0$. We first show that $S_\nu\subseteq S_0$. Assume for a contradiction that this is not the case: there is $\alpha\in S_\nu$ such that $\beta_0\succ\alpha$. 
\begin{itemize}
\item Since $\alpha\in S_\nu\subseteq F_2$, $\alpha$ is in $F_1$ or in $\facts( \varphi_\mn{P_{2}\text{-}max}(F_1))=\reachr^-(F_1)$. In both cases $\varphi_\mn{P_{2}\text{-}max}(F_1)$ contains the following clause: $\neg x_\alpha\vee \bigvee_{\beta_0\confof\gamma, \beta_0\not\succ\gamma} x_\gamma$ since $\beta_0\succ\alpha$. 
\item Since $\nu(x_\alpha)=\true$, there must be $\gamma$ such that $\beta_0\confof\gamma$ and $\beta_0\not\succ\gamma$ such that $\nu(x_\gamma)=\true$, \ie $\gamma\in S_\nu\subseteq \Rmc_0$. 
\item Since $\beta_0\not\succ\gamma$ , by definition of $S_0$, it follows that $\gamma\in S_0$. This contradicts the \Tmc-consistency of $S_0$ hence the fact that $S_0$ is a Pareto-improvement of $\Rmc_0$.  
\end{itemize}
It follows that $S_\nu\subseteq S_0$.
\item Any repair $\Rmc_1\supseteq S_0$ is thus a Pareto-improvement of $\Rmc_0$ which contains $S_\nu$. 
\item Since we assume that all repairs that extend $S_\nu$ are not Pareto-optimal, we can then construct an infinite sequence $\Rmc_0,\Rmc_1,\dots$ of repairs that all contain $S_\nu$ and such that $\Rmc_{i+1}$ is a Pareto-improvement of $\Rmc_i$. 
\item The number of repairs being finite, there must be a repetition in this sequence. 
By  Proposition 7 in \cite{DBLP:journals/amai/StaworkoCM12}, since $\succ$ is acyclic, so is the relation of Pareto-dominance between repairs, which contradicts the existence of such a cycle of repairs. 
\end{itemize}
Hence, there exists a Pareto-optimal repair $\Rmc$ that extends $S_\nu$. 
\medskip

\noindent$(\Leftarrow)$ Assume that there exists $\Rmc\in \preps{\Kmc_\succ}$ such that $S_\nu=\Rmc\cap F_2$. 
Since $\Rmc$ is \Tmc-consistent, it is easy to check that $\nu$ satisfies $\varphi_\mn{cons}(F_2)$. 
Assume for a contradiction that $\nu$ does not satisfy $\varphi_\mn{P_{2}\text{-}max}(F_1)$.
\begin{itemize}
\item  There exists $\alpha\in F_2$ such that $\nu$ does not satisfy $
\bigwedge_{\beta\succ\alpha} (\neg x_\alpha\vee \bigvee_{\beta \confof \gamma, \beta\not\succ\gamma} x_\gamma )$, \ie (i) $\nu(x_\alpha)=\true$ and (ii) there exists $\beta\succ\alpha$ and for every $\gamma$ such that $\beta\confof\gamma$ and $\beta\not\succ\gamma$, $\nu(x_\gamma)=\false$. 
\item Since $S_\nu=\Rmc\cap F_2$, it follows that  $\alpha\in\Rmc$ and for every $\gamma$ such that $\beta\confof\gamma$ and $\beta\not\succ\gamma$, $\gamma\notin\Rmc$. 
\item Let $\Rmc'=\Rmc\cup\{\beta\}\setminus\{\gamma\mid\beta\confof\gamma\}$. Since for every $\gamma\in\Rmc$ such that $\beta\confof\gamma$, it holds that $\beta\succ\gamma$, it follows that $\Rmc'$ is a Pareto-improvement of $\Rmc$, contradicting $\Rmc\in \preps{\Kmc_\succ}$. 
\end{itemize}
Hence $\nu$ satisfies $\varphi_\mn{P_{2}\text{-}max}(F_1)$.
\end{proof}

\begin{lemma}\label{lem:cmax}
Let $\Kmc_\succ$ be a prioritized KB with $\Kmc=(\Dmc,\Tmc)$, $F_1\subseteq \Dmc$, $F_2= F_1\cup \facts( \varphi_\mn{C\text{-}max}(F_1))$, and let $\nu$ be a valuation of $\{x_\alpha\mid \alpha\in F_2\}$.
\begin{itemize}
\item If $\nu$ can be extended to a valuation that satisfies $ \varphi_\mn{C\text{-}max}(F_1)\wedge \varphi_\mn{cons}(F_2)$ then $S_\nu=\{\alpha\mid \nu(x_\alpha)=\true\}$ can be extended to a completion-optimal repair of $\Kmc_\succ$.
\item If there exists a completion-optimal repair $\Rmc$ of $\Kmc_\succ$ such that $\Rmc\cap F_2=S_\nu$, then $\nu$ can be extended to a valuation of the variables of $ \varphi_\mn{C\text{-}max}(F_1)\wedge \varphi_\mn{cons}(F_2)$ that satisfies $ \varphi_\mn{C\text{-}max}(F_1)\wedge \varphi_\mn{cons}(F_2)$. 
\end{itemize}
\end{lemma}
\begin{proof}
\noindent$(\Rightarrow)$ Assume that $\nu$ has been extended to a valuation of the variables of $\varphi_\mn{C\text{-}max}(F_1)\wedge \varphi_\mn{cons}(F_2)$ that satisfies $\varphi_\mn{C\text{-}max}(F_1)\wedge \varphi_\mn{cons}(F_2)$. Let  $\succ'$ be the binary relation over $F_2$ such that $\alpha\succ'\beta$ iff $\nu(x_{\alpha\succ'\beta})=\true$. 
\begin{itemize}
\item Since $\nu$ satisfies $\varphi_\mn{compl}$, $\succ'$ is such that for all $\alpha,\beta\in F_2$ such that $\alpha\confof\beta$, (i) $\alpha\succ'\beta$ or $\beta\succ'\alpha$ , (ii) it is not the case that $\alpha\succ'\beta$ and $\beta\succ'\alpha$ and (iii) $\succ'$ extends $\succ$. 
\item Moreover, since $\nu$ satisfies $\varphi_\mn{acyc}$, $\succ'$ is acyclic: if there was a cycle $\alpha_1\succ'\dots\succ'\alpha_n\succ'\alpha_1$, then by construction of $\succ'$ we would have $\nu(x_{\alpha_1\succ'\alpha_2})=\dots
=\nu(x_{\alpha_n\succ'\alpha_1})=\true$ and $\nu(t_{\alpha_1,\alpha_2})=\dots
=\nu(t_{\alpha_1,\alpha_n})=\true$, which contradicts the clause $\neg x_{\alpha_n\succ'\alpha_1}\vee \neg t_{\alpha_1,\alpha_n}$. 
\end{itemize}
Hence $\succ'$ is a completion of $\succ$ on $F_2$. 
Let us complete $\succ'$ into an arbitrary completion of $\succ$ on $\Kmc$.  
Since $\succ'$ is total, $\Kmc_{\succ'}$ has a unique Pareto-optimal repair $\Rmc_{\succ'}$ \cite{DBLP:journals/amai/StaworkoCM12}, which is a completion-optimal repair.  
We next show that $S_\nu\subseteq \Rmc_{\succ'}$. 
\begin{itemize}
\item The optimal repair $\Rmc_{\succ'}$ can be obtained as follows (\cf Algorithm \mn{CCategoricity} by \citeauthor{DBLP:conf/icdt/KimelfeldLP17} \shortcite{DBLP:conf/icdt/KimelfeldLP17}, which is here adapted to the case where the total priority relation guarantees the uniqueness of the optimal repair): Start with $\Rmc_{\succ'}=\emptyset$, then add to $\Rmc_{\succ'}$ the set $P=\{\alpha \mid \forall\beta\in\Kmc, \beta\not\succ'\alpha\}$, remove from $\Dmc$ the sets $P$ and $N=\{\alpha \mid \tup{\Tmc,P\cup\{\alpha\}}\models \bot\}$, and repeat this process until $\Dmc$ is empty. 
\item We show by induction that every $\alpha\in P\cap F_2$ is such that $\nu(x_\alpha)=\true$ and every $\alpha\in N\cap F_2$ is such that $\nu(x_\alpha)=\false$.  
\begin{itemize}
\item In the first step, $P$ contains facts that are not dominated in $\succ'$ and $N$ contains their conflicting facts. 
\begin{itemize}
\item Let $\alpha\in P\cap F_2$ and assume for a contradiction that $\nu(x_\alpha)=\false$. 
Since $\nu$ satisfies $\varphi_\mn{pref}$, there exists $\beta\in F_2$ such that $\alpha\confof\beta$ and $\nu(x_\beta)=\true$ and $\nu(x_{\beta\succ'\alpha})=\true$, \ie $\beta\succ'\alpha$, which contradicts the assumption on $\alpha$. 
\item Now assume that $\alpha\in N\cap F_2$, \ie $\alpha$ is in conflict with some $\beta\in P$. Since $\beta\succ'\alpha$ and $\succ'$ extends $\succ$, it follows that $\beta\in F_2$ by construction of $F_2=\reachr(F_1)$. Hence $\nu(x_\beta)=\true$, and since $\nu$ satisfies $\varphi_\mn{cons}(F_2)$, it follows that $\nu(x_\alpha)=\false$. 
\end{itemize}
\item Assume that at steps $1,\dots, i$ of the algorithm, $\nu(x_\alpha)=\true$ if $\alpha\in P\cap F_2$ and $\nu(x_\alpha)=\false$ if $\alpha\in N\cap F_2$, and consider step $i+1$. 
\begin{itemize}
\item Let $\alpha\in P\cap F_2$ and assume for a contradiction that $\nu(x_\alpha)=\false$. 
As before, since $\nu$ satisfies $\varphi_\mn{pref}$, there exists $\beta\in F_2$ such that $\alpha\confof\beta$ and $\nu(x_\beta)=\true$ and $\nu(x_{\beta\succ'\alpha})=\true$, \ie $\beta\succ'\alpha$. However, by definition of $P$, $\beta\succ'\alpha$ implies that $\beta$ has already be removed from $\Dmc$, so $\nu(x_\beta)=\true$ implies that $\beta$ belongs to $\Rmc_{\succ'}$. This contradicts the assumption $\alpha\in P$ by $\Tmc$-consistency of $\Rmc_{\succ'}$.
\item The fact that $\nu(x_\alpha)=\false$ if $\alpha\in N\cap F_2$ follows as in the base case. 
\end{itemize}
\end{itemize}
\end{itemize}
By definition of $S_\nu$, we thus have $S_\nu= \Rmc_{\succ'}\cap F_2$. 
\medskip

\noindent$(\Leftarrow)$ Assume that there exists $\Rmc\in \creps{\Kmc_\succ}$ such that $S_\nu=\Rmc\cap F_2$. Since $\Rmc$ is \Tmc-consistent, it is easy to check that $\nu$ satisfies $\varphi_\mn{cons}(F_2)$. 
Moreover, there exists a completion $\succ'$ of $\succ$ such that $\Rmc$ is the unique Pareto-optimal repair of $\Kmc_{\succ'}$. 
Extend $\nu$ by setting $\nu(x_{\alpha\succ'\beta})=\true$ iff $\alpha\succ'\beta$, $\nu(t_{\alpha,\beta})=\true$ iff there is a path $\alpha\succ'\dots\succ'\beta$, and $\nu(x_{\beta\rightarrow\alpha})=\true$ iff $\beta\succ'\alpha$ and $\beta\in S_\nu$. 
\begin{itemize}
\item Since $\succ'$ is a total and anti-symmetric binary relation that extends $\succ$, it is easy to check that $\nu$ satisfies $\varphi_\mn{compl}$. 
\item Moreover, since $\succ'$ is acyclic, $\nu$ satisfies $\varphi_\mn{acyc}$: for every $\alpha,\beta\in F_2$, $\nu $ satisfies $\neg x_{\alpha\succ'\beta} \vee t_{\alpha,\beta}$ by definition of $\nu(x_{\alpha\succ'\beta})$ and $\nu(t_{\alpha,\beta})$; it satisfies $ \neg x_{\alpha\succ'\beta} \vee \neg t_{\beta,\alpha}$ because if $\alpha\succ'\beta$ then there does not exist any path $\beta\succ'\dots\succ'\alpha$; and if there is a path $\alpha\succ'\dots\succ'\beta$ and $\beta\succ'\gamma$, there is a path $\alpha\succ'\dots\succ'\gamma$ so $\nu$ satisfies  
$\neg t_{\alpha,\beta} \vee \neg x_{\beta\succ'\gamma} \vee t_{\alpha,\gamma}$. 
\item Finally, we show that $\nu$ satisfies $\varphi_\mn{pref}$. Assume for a contradiction that this is not the case: either (i) there exists $\alpha\in F_2$ such that $\nu(x_\alpha)=\false$ and for every conflicting $\beta\in F_2$, $\nu(x_{\beta\rightarrow\alpha})=\false$ or (ii) there exists conflicting $\alpha,\beta\in F_2$ such that $\nu(x_{\beta\rightarrow\alpha})=\true$ and $\nu(x_\beta)=\false$ or $\nu(x_{\beta\succ'\alpha})=\false$. Case (ii) contradicts the definition of $\nu$ so we are in case (i). Hence since $S_\nu=\Rmc\cap F_2$, $\alpha\notin\Rmc$ and for every conflicting $\beta\in F_2$, either $\beta\not\succ'\alpha$ or $\beta\notin\Rmc$. Since $\Rmc\in\preps{\Kmc_{\succ'}}$, 
$\alpha\notin\Rmc$ implies that there exists $\beta\in\Dmc$ such that  $\beta\succ'\alpha$ and $\beta\in\Rmc$. Moreover, since $\succ'$ extends $\succ$, $\beta\succ'\alpha$ implies that $\beta\in F_2$. We conclude that $\nu$ satisfies $\varphi_\mn{pref}$.
\end{itemize}
Hence $\nu$ satisfies $\varphi_\mn{C\text{-}max}(F_1)$. 
\end{proof}

 \subsection{Propositional Encodings for X-AR, X-Brave and X-IAR semantics}
  
\begin{proposition}\label{prop:simpleARencoding}
$\Kmc_\succ \armodels{X} q(\ans)$ if and only if $\Phi_{X\text{-}AR}(q(\ans))$ is unsatisfiable.
\end{proposition}
\begin{proof}
Recall that $\Phi_{X\text{-}AR}(q(\ans))=\varphi_{\neg q(\ans)} \wedge \varphi_\mn{X\text{-}max}(F_1)\wedge \varphi_\mn{cons}(F_2)$ where  $F_1=\facts(\varphi_{\neg q(\ans)})$ and $F_2= F_1 \cup  \facts(\varphi_\mn{X\text{-}max}(F_1))$.

\noindent$(\Rightarrow)$ Assume that $\Phi_{X\text{-}AR}(q(\ans))$ is satisfiable and let $S_\nu=\{\alpha\mid \nu(x_\alpha)=\true\}$ for some $\nu$ that satisfies $\Phi_{X\text{-}AR}(q(\ans))$. 
Since $\nu$ satisfies $\varphi_\mn{X\text{-}max}(F_1)\wedge\varphi_\mn{cons}(F_2)$, by Lemma \ref{lem:p1max}, \ref{lem:p2max}, or \ref{lem:cmax}, $S_\nu$ can be extended to a X-optimal repair $\Rmc$ of $\Kmc_\succ$. 
Since $\nu$ satisfies $\varphi_{\neg q(\ans)}$, by Lemma \ref{lem:notQ}, $(\Rmc,\Tmc)\not\models q(\ans)$. 
It follows that $\Kmc_\succ \not\armodels{X} q(\ans)$.

\noindent$(\Leftarrow)$ Assume that $\Kmc_\succ \not\armodels{X} q(\ans)$: There exists $\Rmc\in\xreps{\Kmc_\succ}$ such that $(\Rmc,\Tmc)\not\models q(\ans)$. Let $\nu$ be the valuation of $\{x_\alpha\mid \alpha\in F_2\}$ defined by $\nu(x_\alpha)=\true$ iff $\alpha\in\Rmc$. 
By Lemma \ref{lem:p1max}, \ref{lem:p2max}, or \ref{lem:cmax}, $\nu$ can be extended to a valuation of the variables of $\varphi_\mn{X\text{-}max}(F_1)\wedge\varphi_\mn{cons}(F_2)$ that satisfies $\varphi_\mn{X\text{-}max}(F_1)\wedge\varphi_\mn{cons}(F_2)$. 
Since $(\Rmc,\Tmc)\not\models q(\ans)$, by Lemma \ref{lem:notQ},  $\nu$ satisfies $\varphi_{\neg q(\ans)}$. 
It follows that $\Phi_{X\text{-}AR}(q(\ans))$ is satisfiable. 
\end{proof}

\begin{proposition}\label{prop:multiARencoding}
$\Kmc_\succ \armodels{X} q(\ans)$ if and only if $x_{\ans}$ is false in every satisfying assignment of $\Psi_{X\text{-}AR}(\mi{PotAns})$
\end{proposition}
\begin{proof}
Recall that $\Psi_{X\text{-}AR}(\mi{PotAns})= \bigwedge_{\ans\in\mi{PotAns}} \varphi'_{\neg q(\ans)} \wedge 
\varphi_\mn{X\text{-}max}(F'_1) \wedge \varphi_\mn{cons}(F'_2)$ where $F'_1 = \facts(\bigwedge_{\ans\in\mi{PotAns}} \varphi'_{\neg q(\ans)})$ and 
$F'_2= F'_1 \cup \facts(\varphi_\mn{X\text{-}max}(F'_1))$.

\noindent$(\Rightarrow)$ Assume that there exists a satisfying assignment $\nu$ of $\Psi_{X\text{-}AR}(\mi{PotAns})$ such that $\nu(x_{\ans})=\true$, and 
let $S_\nu=\{\alpha\mid \nu(x_\alpha)=\true\}$. 
Since $\nu$ satisfies $\varphi_\mn{X\text{-}max}(F'_1)\wedge\varphi_\mn{cons}(F'_2)$, by Lemma \ref{lem:p1max}, \ref{lem:p2max}, or \ref{lem:cmax}, $S_\nu$ can be extended to a X-optimal repair $\Rmc$ of $\Kmc_\succ$. 
Since $\nu$ satisfies $\varphi'_{\neg q(\ans)}$ and $\nu(x_{\ans})=\true$, then $\nu$ must satisfy $\varphi_{\neg q(\ans)}$. 
It follows by Lemma \ref{lem:notQ} that $(\Rmc,\Tmc)\not\models q(\ans)$. Hence $\Kmc_\succ \not\armodels{X} q(\ans)$.

\noindent$(\Leftarrow)$ Assume that $\Kmc_\succ \not\armodels{X} q(\ans)$: There exists $\Rmc\in\xreps{\Kmc_\succ}$ such that $(\Rmc,\Tmc)\not\models q(\ans)$. 
\begin{itemize}
\item As in the proof of Proposition \ref{prop:simpleARencoding}, we can find a valuation $\nu$ of the variables of $\varphi_\mn{X\text{-}max}(F'_1) \wedge \varphi_\mn{cons}(F'_2)$ that satisfies $\varphi_\mn{X\text{-}max}(F'_1) \wedge \varphi_\mn{cons}(F'_2)$ and satisfies $\varphi_{\neg q(\ans)}$. 
\item Let us extend $\nu$ by setting $\nu(x_{\ans})=\true$  and $\nu(x_{\ans'})=\false$ for every $\ans'\in\mi{PotAns}$ such that $\ans\neq\ans'$. 
\item Since $\nu$ satisfies $\varphi_{\neg q(\ans)}$, it also satisfies $\varphi'_{\neg q(\ans)}$. 
\item 
Let $\ans'\in\mi{PotAns}$, $\ans'\neq\ans$. Assume for a contradiction that $\nu$ does not satisfy $\varphi'_{\neg q(\ans')}$: There is a cause $\Cmc$ of $q(\ans')$ such that $\nu$ does not satisfy $\varphi'_{\neg \Cmc}(x_{\ans'})$. In \cqaprienc case, there is a contradiction because $\varphi'_{\neg \Cmc}(x_{\ans'})$ is a single clause that contains $\neg x_{\ans'}$. In \cavsatenc case, this implies that there is a fact $\alpha\in\Cmc$ such that $\nu(x_\alpha)=\false$, \ie $\alpha\notin\Rmc$, and for every $\beta\confof\alpha$ such that $\alpha\not\succ\beta$, $\nu(x_\beta)=\false$, \ie $\beta\notin\Rmc$. This contradicts the fact that $\Rmc\in \xreps{\Kmc_\succ}$ as $\Rmc\cup\{\alpha\}\setminus\{\beta\mid\alpha\succ\beta\}$ would be an improvement of $\Rmc$. Hence $\nu$ satisfies $\varphi'_{\neg q(\ans')}$.
\end{itemize}
It follows that $\nu$ satisfies $\Psi_{X\text{-}AR}(\mi{PotAns})$.
\end{proof}

\begin{proposition}\label{prop:simpleBraveencoding}
$\Kmc_\succ \bravemodels{X} q(\ans)$ if and only if $\Phi_{X\text{-}brave}(q(\ans))$ is satisfiable.
\end{proposition}
\begin{proof}
Recall that $\Phi_{X\text{-}brave}(q(\ans))=\varphi_{q(\ans)} \wedge \varphi_\mn{X\text{-}max}(F_1)\wedge \varphi_\mn{cons}(F_2)$ where  $F_1=\facts(\varphi_{q(\ans)})$ and $F_2= F_1 \cup  \facts(\varphi_\mn{X\text{-}max}(F_1))$.

\noindent$(\Rightarrow)$ Assume that $\Kmc_\succ \bravemodels{X} q(\ans)$: There exists $\Rmc\in\xreps{\Kmc_\succ}$ such that $(\Rmc,\Tmc)\models q(\ans)$. Let $\nu$ be the valuation of $\{x_\alpha\mid \alpha\in F_2\}$ defined by $\nu(x_\alpha)=\true$ iff $\alpha\in\Rmc$. 
By Lemma \ref{lem:p1max}, \ref{lem:p2max}, or \ref{lem:cmax}, $\nu$ can be extended to a valuation of the variables of $\varphi_\mn{X\text{-}max}(F_1)\wedge\varphi_\mn{cons}(F_2)$ that satisfies $\varphi_\mn{X\text{-}max}(F_1)\wedge\varphi_\mn{cons}(F_2)$. 
Since $(\Rmc,\Tmc)\models q(\ans)$, by Lemma \ref{lem:query}, $\nu$ can be extended to a valuation that satisfies $\varphi_{q(\ans)}$. 
It follows that $\Phi_{X\text{-}brave}(q(\ans))$ is satisfiable. 

\noindent$(\Leftarrow)$ Assume that $\Phi_{X\text{-}brave}(q(\ans))$ is satisfiable and let $S_\nu=\{\alpha\mid \nu(x_\alpha)=\true\}$ for some $\nu$ that satisfies $\Phi_{X\text{-}brave}(q(\ans))$. 
Since $\nu$ satisfies $\varphi_\mn{X\text{-}max}(F_1)\wedge\varphi_\mn{cons}(F_2)$, by Lemma \ref{lem:p1max}, \ref{lem:p2max}, or \ref{lem:cmax}, $S_\nu$ can be extended to a X-optimal repair $\Rmc$ of $\Kmc_\succ$. 
Since $\nu$ satisfies $\varphi_{q(\ans)}$, by Lemma \ref{lem:query}, $(\Rmc,\Tmc)\models q(\ans)$. Hence $\Kmc_\succ \bravemodels{X} q(\ans)$.
\end{proof}

\begin{proposition}
$\Kmc_\succ \bravemodels{X} q(\ans)$ if and only if $\Phi_{X\text{-}brave}(\Cmc)$ is satisfiable for some $\Cmc\in\causes{q(\ans),\Kmc}$.
\end{proposition}
\begin{proof}
Recall that $\Phi_{X\text{-}brave}(\Cmc)=\varphi_{\Cmc} \wedge \varphi_\mn{X\text{-}max}(F_1)\wedge \varphi_\mn{cons}(F_2)$ where  $F_1=\facts(\varphi_{\Cmc})$ and $F_2= F_1 \cup  \facts(\varphi_\mn{X\text{-}max}(F_1))$. 
The proof follows from Lemmas \ref{lem:p1max}, \ref{lem:p2max}, or \ref{lem:cmax} and \ref{lem:cause}, and is similar to the proof for Proposition \ref{prop:simpleBraveencoding}.
\end{proof}

\begin{proposition}
$\Kmc_\succ \bravemodels{X} q(\ans)$ if and only if $x_{\ans}$ is true in some satisfying assignment of $\Psi_{X\text{-}brave}(\mi{PotAns})$.
\end{proposition}
\begin{proof}
Recall that $\Psi_{X\text{-}brave}(\mi{PotAns})= \bigwedge_{\ans\in\mi{PotAns}} \varphi'_{q(\ans)} \wedge 
\varphi_\mn{X\text{-}max}(F'_1) \wedge \varphi_\mn{cons}(F'_2)$ where $F'_1 = \facts(\bigwedge_{\ans\in\mi{PotAns}} \varphi'_{ q(\ans)})$ and 
$F'_2= F'_1 \cup \facts(\varphi_\mn{X\text{-}max}(F'_1))$.

\noindent$(\Rightarrow)$ Assume that $\Kmc_\succ \bravemodels{X} q(\ans)$: There exists $\Rmc\in\xreps{\Kmc_\succ}$ such that $(\Rmc,\Tmc)\models q(\ans)$. 
\begin{enumerate}
\item As in the proof of Proposition \ref{prop:simpleBraveencoding}, we can find a valuation $\nu$ of the variables of $\varphi_\mn{X\text{-}max}(F'_1) \wedge \varphi_\mn{cons}(F'_2)$ that satisfies $\varphi_\mn{X\text{-}max}(F'_1) \wedge \varphi_\mn{cons}(F'_2)$ and satisfies $\varphi_{q(\ans)}$. 

\item Let us extend $\nu$ by setting $\nu(x_{\ans})=\true$, and for every $\ans'\in\mi{PotAns}$ such that $\ans\neq\ans'$, $\nu(x_{\ans'})=\false$ and $\nu(x_\Cmc)=\false$ for every $\Cmc\in\causes{q(\ans'),\Kmc}\setminus\causes{q(\ans),\Kmc}$ (\ie the value of $\nu(x_\Cmc)$ has not been fixed in step (1)). 

\item Since $\nu$ satisfies $\varphi_{q(\ans)}$ and $\nu(x_{\ans})=\true$, it also satisfies $\varphi'_{q(\ans)}$. 

\item For every $\ans'\in\mi{PotAns}$ such that $\ans'\neq\ans$, $\nu$ satisfies $\varphi'_{q(\ans')}$ because (i) $\nu(x_{\ans'})=\false$ and $\neg x_{\ans'}$ is a disjunct of the first clause, and (ii) every other clause of $\varphi'_{q(\ans')}$ either contains some $\neg x_\Cmc$ with $\Cmc\in\causes{q(\ans'),\Kmc}\setminus\causes{q(\ans),\Kmc}$ and $\nu(x_\Cmc)=\false$, or is a clause shared with $\varphi_{q(\ans)}$ and thus is satisfied by $\nu$. 
\end{enumerate}
It follows that $\nu$ satisfies $\Psi_{X\text{-}brave}(\mi{PotAns})$. 

\noindent$(\Leftarrow)$ Assume that there exists a satisfying assignment $\nu$ of $\Psi_{X\text{-}brave}(\mi{PotAns})$ such that $\nu(x_{\ans})=\true$, and 
let $S_\nu=\{\alpha\mid \nu(x_\alpha)=\true\}$. 
Since $\nu$ satisfies $\varphi_\mn{X\text{-}max}(F'_1)\wedge\varphi_\mn{cons}(F'_2)$, by Lemma \ref{lem:p1max}, \ref{lem:p2max}, or \ref{lem:cmax}, $S_\nu$ can be extended to a X-optimal repair $\Rmc$ of $\Kmc_\succ$. 
Since $\nu$ satisfies $\varphi'_{q(\ans)}$ and $\nu(x_{\ans})=\true$, then $\nu$ must satisfy $\varphi_{q(\ans)}$. 
It follows by Lemma \ref{lem:query} that $(\Rmc,\Tmc)\models q(\ans)$. Hence $\Kmc_\succ \bravemodels{X} q(\ans)$.
\end{proof}

\begin{proposition}\label{prop:simpleIARencoding}
$\Kmc_\succ \iarmodels{X} q(\ans)$ if and only if $\Phi_{X\text{-}IAR}(q(\ans))$ is unsatisfiable.
\end{proposition}
\begin{proof}
Recall that $\Phi_{X\text{-}IAR}(q(\ans))= 
\bigwedge_{\Cmc\in \causes{q(\ans),\Kmc}} \Big (\varphi^\Cmc_{\neg \Cmc} \wedge \varphi^\Cmc_\mn{X\text{-}max}(F_1^\Cmc)\wedge \varphi^\Cmc_\mn{cons}(F_2^\Cmc) \Big)
$ where $F_1^\Cmc=\facts(\varphi^\Cmc_{\neg \Cmc})$, and $F_2^\Cmc=F_1^\Cmc \cup \facts(\varphi^\Cmc_\mn{X\text{-}max}(F_1^\Cmc))$.

\noindent$(\Rightarrow)$ Assume that $\Phi_{X\text{-}IAR}(q(\ans))$ is satisfiable and let $\nu$ be a valuation that satisfies $\Phi_{X\text{-}IAR}(q(\ans))$. 
\begin{itemize}
\item Let $\Cmc\in \causes{q(\ans),\Kmc}$ and $S_\nu^\Cmc=\{\alpha\mid \nu(x^\Cmc_\alpha)=\true, \alpha\in F_2^\Cmc\}$. 
\item Since $\nu$ satisfies $\varphi^\Cmc_\mn{X\text{-}max}(F_1^\Cmc)\wedge \varphi^\Cmc_\mn{cons}(F_2^\Cmc)$, by Lemma \ref{lem:p1max}, \ref{lem:p2max}, or \ref{lem:cmax}, $S_\nu^\Cmc$ can be extended to a X-optimal repair $\Rmc^\Cmc$ of $\Kmc_\succ$. 
\item Since $\nu$ satisfies $\varphi^\Cmc_{\neg \Cmc}$, by Lemma \ref{lem:notC}, $\Cmc\not\subseteq\Rmc^\Cmc$. 
\end{itemize}
Since for every cause $\Cmc$ we can find $\Rmc^\Cmc\in\xreps{\Kmc_\succ}$ that does not include $\Cmc$, it follows that $\Kmc_\succ \not\iarmodels{X} q(\ans)$.

\noindent$(\Leftarrow)$ Assume that $\Kmc_\succ \not\iarmodels{X} q(\ans)$ and let $\Bmc \in\causes{q(\ans),\Kmc}$. 
\begin{itemize}
\item There exists $\Rmc^\Bmc\in\xreps{\Kmc_\succ}$ such that $\Bmc\not\subseteq\Rmc^\Bmc$ (if $\Bmc$ contains a real cause of $q(\ans)$ this comes from $\Kmc_\succ \not\iarmodels{X} q(\ans)$, and if $\Bmc$ contains a pair of conflicting assertions it is not included in any repair). 
\item Let $\nu^\Bmc$ be the valuation of $\{x^\Bmc_\alpha\mid \alpha\in F_2^\Bmc\}$ defined by $\nu(x^\Bmc_\alpha)=\true$ iff $\alpha\in\Rmc^\Bmc$. 
\item By Lemma \ref{lem:p1max}, \ref{lem:p2max}, or \ref{lem:cmax}, $\nu^\Bmc$ can be extended to a valuation of the variables of $\varphi^\Bmc_\mn{X\text{-}max}(F_1^\Bmc)\wedge\varphi^\Bmc_\mn{cons}(F_2^\Bmc)$ that satisfies $\varphi^\Bmc_\mn{X\text{-}max}(F_1^\Bmc)\wedge\varphi^\Bmc_\mn{cons}(F_2^\Bmc)$. 
\item Since $\Bmc\not\subseteq\Rmc^\Bmc$,  by Lemma \ref{lem:notC}, $\nu^\Bmc$ satisfies $\varphi^\Bmc_{\neg \Bmc}$. 
\end{itemize}
The valuation $\nu$ defined by $\nu(x_\alpha^\Cmc)=\nu^\Cmc(x_\alpha^\Cmc)$ for every $\Cmc\in \causes{q(\ans),\Kmc}$ and $x_\alpha^\Cmc$ such that $\alpha\in F_2^\Cmc$ thus satisfies $\Phi_{X\text{-}IAR}(q(\ans))$.
\end{proof}

\begin{proposition}
$\Kmc_\succ \iarmodels{X} q(\ans)$ if and only if $\Phi_{X\text{-}IAR}(\Cmc)$ is unsatisfiable for some $\Cmc\in\causes{q(\ans),\Kmc}$.
\end{proposition}
\begin{proof}
Recall that $\Phi_{X\text{-}IAR}(\Cmc)=\varphi_{\neg \Cmc} \wedge \varphi_\mn{X\text{-}max}(F_1)\wedge \varphi_\mn{cons}(F_2)$ where $F_1=\facts(\varphi_{\neg \Cmc})$, and $F_2= F_1 \cup \facts(\varphi_\mn{X\text{-}max}(F_1))$. 
The proof follows from Lemmas \ref{lem:p1max}, \ref{lem:p2max}, or \ref{lem:cmax} and \ref{lem:notC}.
\end{proof}

\begin{proposition}
$\Kmc_\succ \iarmodels{X} q(\ans)$ if and only if $x_{\ans}$ is false in every satisfying assignment of $\Psi_{X\text{-}IAR}(\mi{PotAns})$.
\end{proposition}
\begin{proof}
$\Psi_{X\text{-}IAR}(\mi{PotAns})= \bigwedge_{\ans\in\mi{PotAns}} \Big ( \bigwedge_{\Cmc\in \causes{q(\ans),\Kmc}} (\varphi'^\Cmc_{\neg \Cmc}(x_{\ans})  \wedge \varphi^\Cmc_\mn{X\text{-}max}(F^{\Cmc}_{1}) \wedge \varphi^\Cmc_\mn{cons}(F^{\Cmc}_2)) \Big)$
where 
$F^{\Cmc}_1 = \facts(\varphi'^\Cmc_{\neg \Cmc}(x_{\ans}))$, and $F^\Cmc_2 = F^\Cmc_1 \cup \facts(\varphi^\Cmc_\mn{X\text{-}max}(F^\Cmc_{1}))$

\noindent$(\Rightarrow)$ Assume that there exists a satisfying assignment $\nu$ of $\Psi_{X\text{-}IAR}(\mi{PotAns})$ such that $\nu(x_{\ans})=\true$. 
As in the proof of Proposition \ref{prop:simpleIARencoding} , for every $\Cmc\in \causes{q(\ans),\Kmc}$, one can find $\Rmc^\Cmc$ that extends $S_\nu^\Cmc=\{\alpha\mid \nu(x^\Cmc_\alpha)=\true, \alpha\in F_2^\Cmc\}$. 
Since $\nu$ satisfies $\varphi'^\Cmc_{\neg \Cmc}(x_{\ans})$, and $\nu(x_{\ans})=\true$, then $\nu$ must satisfy $\varphi^\Cmc_{\neg \Cmc}$. 
It follows by Lemma \ref{lem:notC} that $\Cmc\not\subseteq\Rmc^\Cmc$. 
Hence $\Kmc_\succ \not\iarmodels{X} q(\ans)$.

\noindent$(\Leftarrow)$ Assume that $\Kmc_\succ \not\iarmodels{X} q(\ans)$.  
\begin{itemize}
\item As in the proof of Proposition \ref{prop:simpleIARencoding}, for each $\Bmc \in\causes{q(\ans),\Kmc}$, we can find a valuation $\nu^\Bmc$ of the variables of $\varphi^\Bmc_\mn{X\text{-}max}(F_1^\Bmc)\wedge\varphi^\Bmc_\mn{cons}(F_2^\Bmc)$ that satisfies $\varphi^\Bmc_\mn{X\text{-}max}(F_1^\Bmc)\wedge\varphi^\Bmc_\mn{cons}(F_2^\Bmc)$ and $\varphi^\Bmc_{\neg \Bmc}$.
\item Let $\nu$ be the valuation defined by 
\begin{itemize}
\item $\nu(x_\alpha^\Bmc)=\nu^\Bmc(x_\alpha^\Bmc)$ for every $\Bmc\in \causes{q(\ans),\Kmc}$ and $x_\alpha^\Bmc$ such that $\alpha\in F_2^\Bmc$;
\item $\nu(x_{\ans})=\true$; 
\item for every $\ans'\in\mi{PotAns}$ such that $\ans\neq\ans'$, $\nu(x_{\ans'})=\false$ and for every $\Cmc\in\causes{q(\ans'),\Kmc}\setminus \causes{q(\ans),\Kmc}$:
\begin{itemize} 
\item we find a valuation $\nu^\Cmc$ of the variables of $\varphi^\Cmc_\mn{X\text{-}max}(F_1^\Cmc)\wedge\varphi^\Cmc_\mn{cons}(F_2^\Cmc)$ that satisfies $\varphi^\Cmc_\mn{X\text{-}max}(F_1^\Cmc)\wedge\varphi^\Cmc_\mn{cons}(F_2^\Cmc)$ and is such that for every $\alpha\in\Cmc$, either $\nu^\Cmc(x_\alpha^\Cmc)=\true$ or there exists $\beta\confof\alpha$ such that $\alpha\not\succ\beta$ and $\nu^\Cmc(x_\beta^\Cmc)=\true$. This is doable by Lemma \ref{lem:p1max}, \ref{lem:p2max}, or \ref{lem:cmax}, since for every fact $\alpha$, any X-optimal repair contains either $\alpha$ or some $\beta\confof\alpha$ such that $\alpha\not\succ\beta$. 
\item We define $\nu(x^\Cmc_\alpha)=\nu^\Cmc(x_\alpha^\Cmc)$ for every $x_\alpha^\Cmc$ such that $\alpha\in F_2^\Cmc$. 
\end{itemize}
\end{itemize}
\item For each $\Bmc \in\causes{q(\ans),\Kmc}$, since $\nu$ satisfies $\varphi^\Bmc_{\neg \Bmc}$, it also satisfies $\varphi'^\Bmc_{\neg \Bmc}(x_{\ans})$. 
\item Let $\ans'\in\mi{PotAns}$, $\ans'\neq\ans$, and $\Cmc\in\causes{q(\ans'),\Kmc}$. 
If $\Cmc\in\causes{q(\ans),\Kmc}$, then $\nu$ satisfies $\varphi^\Cmc_{\neg \Cmc}$ hence also satisfies $\varphi'^\Cmc_{\neg \Cmc}(x_{\ans'})$. 
Otherwise, $\Cmc\in\causes{q(\ans'),\Kmc}\setminus \causes{q(\ans),\Kmc}$ and 
by definition of $\varphi'^\Cmc_{\neg \Cmc}(x_{\ans'})$ which consists either of a single clause that contains $\neg x_{\ans'}$ (case \cqaprienc) or whose clauses contain either $\neg x_{\ans'}$ or the disjunction of some $x_\alpha^\Cmc$ with $\alpha\in\Cmc$ and all $x_\beta^\Cmc$ such that $\alpha\confof\beta$ and $\alpha\not\succ\beta$ (case \cavsatenc), it follows that $\nu$ satisfies $\varphi'^\Cmc_{\neg \Cmc}(x_{\ans'})$.
\end{itemize}
It follows that $\nu$ satisfies $\Psi_{X\text{-}IAR}(\mi{PotAns})$.
\end{proof}

\begin{proposition}\label{prop:simpleIARfact}
For every $\alpha\in\Dmc$, $\Kmc_\succ \iarmodels{X} \alpha$ if and only if $\Phi_{X\text{-}IAR}(\alpha)$ is unsatisfiable.
\end{proposition}
\begin{proof}
Recall that 
$\Phi_{X\text{-}IAR}(\alpha)=\Phi_{X\text{-}IAR}(\{\alpha\})=\varphi_{\neg \{\alpha\}} \wedge \varphi_\mn{X\text{-}max}(F_1)\wedge \varphi_\mn{cons}(F_2)$ where $F_1=\facts(\varphi_{\neg \alpha})$, and $F_2= F_1 \cup \facts(\varphi_\mn{X\text{-}max}(F_1))$. 

\noindent$(\Rightarrow)$ Assume that $\Phi_{X\text{-}IAR}(\alpha)$ is satisfiable and let $\nu$ be a valuation that satisfies $\Phi_{X\text{-}IAR}(\alpha)$. 
 Let $S_\nu=\{\beta\mid \nu(x_\beta)=\true, \beta\in F_2^\Cmc\}$. 
 Since $\nu$ satisfies $\varphi_\mn{X\text{-}max}(F_1)\wedge \varphi_\mn{cons}(F_2)$, by Lemma \ref{lem:p1max}, \ref{lem:p2max}, or \ref{lem:cmax}, $S_\nu$ can be extended to a X-optimal repair $\Rmc$ of $\Kmc_\succ$. 
 Since $\nu$ satisfies $\varphi_{\neg \{\alpha\}}$, by Lemma \ref{lem:notC}, $\alpha\notin\Rmc$. 
It follows that $\Kmc_\succ \not\iarmodels{X} \alpha$.

\noindent$(\Leftarrow)$ Assume that $\Kmc_\succ \not\iarmodels{X} \alpha$. 
 There exists $\Rmc\in\xreps{\Kmc_\succ}$ such that $\alpha\notin\Rmc$. 
 Let $\nu$ be the valuation of $\{x_\beta\mid \beta\in F_2\}$ defined by $\nu(x_\beta)=\true$ iff $\beta\in\Rmc$. 
 By Lemma \ref{lem:p1max}, \ref{lem:p2max}, or \ref{lem:cmax}, $\nu$ can be extended to a valuation of the variables of $\varphi_\mn{X\text{-}max}(F_1)\wedge\varphi_\mn{cons}(F_2)$ that satisfies $\varphi_\mn{X\text{-}max}(F_1)\wedge\varphi_\mn{cons}(F_2)$. 
Since $\alpha\notin\Rmc$, by Lemma \ref{lem:notC}, $\nu$ satisfies $\varphi_{\neg\{\alpha\}}$. 
Hence $\nu$ satisfies $\Phi_{X\text{-}IAR}(\alpha)$.
\end{proof}

\begin{proposition}
For every $\alpha\in\Dmc$, $\Kmc_\succ \iarmodels{X} \alpha$ if and only if $\alpha\in\mi{Rel}$ implies that $y_{\alpha}$ is false in every satisfying assignment of $\Psi_{X\text{-}IAR}(\mi{Rel})$.
\end{proposition}
\begin{proof}
$\Psi_{X\text{-}IAR}(\mi{Rel})= \bigwedge_{\alpha\in\mi{Rel}} \varphi'_{\neg \{\alpha\}}(y_\alpha)  \wedge \varphi_\mn{X\text{-}max}(F_{1}) \wedge \varphi_\mn{cons}(F_2)$ where $F'_1=\facts(\bigwedge_{\alpha\in\mi{Rel}} \varphi'_{\neg \{\alpha\}}(y_\alpha) )$, and $F'_2= F_1 \cup \facts(\varphi_\mn{X\text{-}max}(F'_1))$. 

\noindent$(\Rightarrow)$ Assume that there exists a satisfying assignment $\nu$ of $\Psi_{X\text{-}IAR}(\mi{Rel})$ such that $\nu(y_\alpha)=\true$, and 
let $S_\nu=\{\beta\mid \nu(x_\beta)=\true\}$. 
Since $\nu$ satisfies $\varphi_\mn{X\text{-}max}(F'_1)\wedge\varphi_\mn{cons}(F'_2)$, by Lemma \ref{lem:p1max}, \ref{lem:p2max}, or \ref{lem:cmax}, $S_\nu$ can be extended to a X-optimal repair $\Rmc$ of $\Kmc_\succ$. 
Since $\nu$ satisfies $\varphi'_{\neg \{\alpha\}}$ and $\nu(y_\alpha)=\true$, then $\nu$ must satisfy $\varphi_{\neg \{\alpha\}}$. 
It follows by Lemma \ref{lem:notC} that $\alpha\notin\Rmc$. Hence $\Kmc_\succ \not\iarmodels{X} \alpha$.

\noindent$(\Leftarrow)$ Assume that $\Kmc_\succ \not\armodels{X} \alpha$: There exists $\Rmc\in\xreps{\Kmc_\succ}$ such that $\alpha\notin\Rmc$. 
\begin{itemize}
\item As in the proof of Proposition \ref{prop:simpleIARfact}, we can find a valuation $\nu$ of the variables of $\varphi_\mn{X\text{-}max}(F'_1) \wedge \varphi_\mn{cons}(F'_2)$ that satisfies $\varphi_\mn{X\text{-}max}(F'_1) \wedge \varphi_\mn{cons}(F'_2)$ and satisfies $\varphi_{\neg \{\alpha\}}$. 
\item Let us extend $\nu$ by setting $\nu(y_\alpha)=\true$  and $\nu(y_{\alpha'})=\false$ for every $\alpha'\in\mi{Rel}$ such that $\alpha\neq\alpha'$. 
\item Since $\nu$ satisfies $\varphi_{\neg\{\alpha\}}$, it also satisfies $\varphi'_{\neg \{\alpha\}}(y_{\alpha})$. 
\item Let $\alpha'\in\mi{Rel}$, $\alpha'\neq\alpha$. 
Assume for a contradiction that $\nu$ does not satisfy $\varphi'_{\neg  \{\alpha'\}}(y_{\alpha'})$. In \cqaprienc case, there is a contradiction because $\varphi'_{\neg  \{\alpha'\}}(y_{\alpha'})$ is a single clause that contains $\neg y_{\alpha'}$. In \cavsatenc case, this implies that $\nu(x_{\alpha'})=\false$, \ie $\alpha'\notin\Rmc$, and for every $\beta\confof\alpha'$ such that $\alpha'\not\succ\beta$, $\nu(x_\beta)=\false$, \ie $\beta\notin\Rmc$. This contradicts the fact that $\Rmc\in \xreps{\Kmc_\succ}$ as $\Rmc\cup\{\alpha'\}\setminus\{\beta\mid\alpha'\succ\beta\}$ would be an improvement of $\Rmc$. Hence $\nu$ satisfies $\varphi'_{\neg  \{\alpha'\}}(y_{\alpha'})$. 
\end{itemize}
It follows that $\nu$ satisfies $\Psi_{X\text{-}IAR}(\mi{Rel})$. 
\end{proof}

\subsection{Non-Binary Conflicts}
To define encodings in the case where conflicts are not binary, we will use the following notation:
\begin{itemize}
\item $\alpha\confof\Bmc$ denotes $\Bmc\cup\{\alpha\}\in\conflicts{\Kmc}$ and $\alpha\notin\Bmc$;
\item $\Bmc\rightsquigarrow\alpha$ denotes $\alpha\confof\Bmc$ and for every $\beta\in\Bmc$, $\alpha\not\succ\beta$. 
\end{itemize}

We redefine $\mathcal{G}_{\Kmc_\succ}$ as a directed hypergraph whose hyperedges are $(\{\alpha\}, \Bmc)$ for every $\Bmc\rightsquigarrow\alpha$
and use hypergraph reachability to define $\reachr(F)$: $\beta$ is reachable from $\alpha$ in $\mathcal{G}_{\Kmc_\succ}$ if $\beta=\alpha$ or there exists a hyperedge $(\{\gamma\}, \Bmc)$ such that $\beta\in\Bmc$ and $\gamma$ is reachable from $\alpha$. 
We next redefine the basic building blocks of our propositional encodings ($\varphi_\Cmc$ needs no modification):

$$\varphi_{\neg \Cmc} =  (\bigvee_{\alpha\in\Cmc}\bigvee_{\Bmc\rightsquigarrow\alpha} x_{\Cmc,\Bmc} )
 \wedge \bigwedge_{\substack{\alpha\in\Cmc, \Bmc\rightsquigarrow\alpha,\\\beta\in\Bmc}} \neg x_{\Cmc,\Bmc} \vee x_\beta$$

$$\varphi_\mn{cons}(F) =  \bigwedge_{\substack{\Cmc\subseteq F,\\ \Cmc\in\conflicts{\Kmc}}}\ \bigvee_{\gamma\in\Cmc} \neg x_\gamma$$

$$\varphi_\mn{P\text{-}max}(F) = 
\bigwedge_{\alpha\in \reachr(F)} (x_\alpha \vee \bigvee_{\Bmc\rightsquigarrow\alpha} x_{\alpha,\Bmc}) 
\wedge  
\bigwedge_{\substack{\alpha\in \reachr(F),\\\Bmc\rightsquigarrow\alpha,\\\beta\in\Bmc}}\neg x_{\alpha,\Bmc} \vee x_\beta$$

$$\varphi_\mn{C\text{-}max}(F) =\varphi_\mn{pref}\wedge\varphi_\mn{compl}\wedge\varphi_\mn{acyc}$$ where $\varphi_\mn{compl}$ and $\varphi_\mn{acyc}$ remain as in the binary case, except that $\alpha\confof\beta$ is replaced by $\{\alpha,\beta\}\subseteq\Cmc\in\conflicts{\Kmc}$ and 
$$\varphi_\mn{pref}=\bigwedge_{\alpha\in \reachr(F)} (x_\alpha \vee \bigvee_{\Bmc\rightsquigarrow\alpha} x_{\Bmc\rightarrow\alpha}) \wedge \bigwedge_{\substack{\alpha\in \reachr(F),\\\Bmc\rightsquigarrow\alpha,\\\beta\in\Bmc}} \Big ( (\neg x_{\Bmc\rightarrow\alpha} \vee x_\beta) \wedge (\neg x_{\Bmc\rightarrow\alpha} \vee x_{\beta\succ'\alpha}) \Big ).$$

We show the analogues of the basic lemmas of Section \ref{sec:basiclemmasencodings} for the formulas that differ in the non-binary case. As in Section~\ref{sec:basiclemmasencodings}, in the lemmas below, $\Kmc_\succ$ is a prioritized KB with $\Kmc=(\Dmc,\Tmc)$, $\Rmc\in\xreps{\Kmc_\succ}$ (with $X\in \{P, C\}$), and $\nu$ is the valuation of $\{x_\alpha\mid \alpha\in \Dmc\}$ such that $\nu(x_\alpha)=\true$ iff $\alpha\in\Rmc$.\smallskip

\begin{lemma}
Let $\Kmc_\succ$ be a prioritized KB with $\Kmc=(\Dmc,\Tmc)$, $\Rmc\in\xreps{\Kmc_\succ}$, and $\nu$ be the valuation of $\{x_\alpha\mid \alpha\in \Dmc\}$ such that $\nu(x_\alpha)=\true$ iff $\alpha\in\Rmc$. 
Then $\nu$ can be extended to a valuation that satisfies $ \varphi_{\neg \Cmc}$ iff $\Cmc\not\subseteq\Rmc$.
\end{lemma}
\begin{proof}
Assume that $\nu$ has been extended to satisfy $\varphi_{\neg \Cmc}$. There exists $\alpha\in\Cmc$ and $\Bmc\rightsquigarrow\alpha$ such that $\nu(x_{\Cmc,\Bmc})=\true$ and for every $\beta\in\Bmc$, $\nu(x_\beta)=\true$, \ie $\beta\in\Rmc$. Since $\Rmc$ is \Tmc-consistent, $\Bmc\cup\{\alpha\}\not\subseteq\Rmc$ so $\alpha\notin\Rmc$. Hence $\Cmc\not\subseteq\Rmc$. 

Assume that $\Cmc\not\subseteq\Rmc$, \ie there exists $\alpha\in\Cmc$ such that $\alpha\notin\Rmc$. There must exist $\Bmc\rightsquigarrow\alpha$ such that $\Bmc\subseteq\Rmc$. Otherwise $\Rmc\cup\{\alpha\}\setminus\{\beta\mid\alpha\succ\beta\}$ would be a Pareto-improvement of $\Rmc$, contradicting $\Rmc\in\xreps{\Kmc_\succ}$   (since $X\in \{P, C\}$ and $\creps{\Kmc_\succ}\subseteq\preps{\Kmc_\succ}$). 
Extend $\nu$ by setting $\nu(x_{\Cmc,\Bmc})=\true$. Since $\nu(x_\beta)=\true$ for every $\beta\in\Bmc$ , it follows that $\nu$ satisfies $\varphi_{\neg \Cmc}$. 
\end{proof}

\begin{lemma}
Let $\Kmc_\succ$ be a prioritized KB with $\Kmc=(\Dmc,\Tmc)$, $F_1\subseteq \Dmc$, $F_2= F_1\cup \facts( \varphi_\mn{P\text{-}max}(F_1))$, and let $\nu$ be a valuation of $\{x_\alpha\mid \alpha\in F_2\}$.
\begin{itemize}
\item If $\nu$ can be extended to a valuation that satisfies $ \varphi_\mn{P\text{-}max}(F_1)\wedge \varphi_\mn{cons}(F_2)$ then $S_\nu=\{\alpha\mid \nu(x_\alpha)=\true\}$ can be extended to a Pareto-optimal repair of $\Kmc_\succ$. 
\item If there exists a Pareto-optimal repair $\Rmc$ of $\Kmc_\succ$ such that $\Rmc\cap F_2=S_\nu$, then $\nu$ can be extended to a valuation that satisfies $ \varphi_\mn{P\text{-}max}(F_1)\wedge \varphi_\mn{cons}(F_2)$.
\end{itemize}
\end{lemma}
\begin{proof}
\noindent$(\Rightarrow)$ Assume that $\nu$ has been extended to satisfy $\varphi_\mn{P\text{-}max}(F_1)\wedge \varphi_\mn{cons}(F_2)$. Since $\nu$ satisfies $\varphi_\mn{cons}(F_2)$, it is easy to check that $S_\nu$ is \Tmc-consistent. 
Let $\Rmc'=S_\nu$, $\Dmc'=\Dmc$ and repeat the following steps until $\Dmc'$ is empty (this is possible since there is always at least one non-dominated $\alpha$ in $\Dmc'$, as $\succ$ is acyclic and $\Dmc'$ is finite): 
\begin{itemize}
\item choose $\alpha\in\Dmc'$ such that for every $\beta\in\Dmc'$, $\beta\not\succ\alpha$;
\item if $\Rmc'\cup\{\alpha\}$ is $\Tmc$-consistent, add $\alpha$ to $\Rmc'$;
\item remove $\alpha$ from $\Dmc'$.
\end{itemize}
Let $\Rmc$ be the final $\Rmc'$. By construction, $S_\nu\subseteq\Rmc$ and we show that $\Rmc\in\preps{\Kmc_\succ}$. Assume for a contradiction that this is not the case: there exists $\beta_0\in\Dmc\setminus\Rmc$ such that $S=\Rmc\cup\{\beta_0\}\setminus\{\alpha\mid \beta_0\succ\alpha\}$ is $\Tmc$-consistent. 
Since $\beta_0\notin\Rmc$, by construction of $\Rmc$, there exists $\Cmc\in\conflicts{\Kmc}$ such that $\beta_0\in\Cmc$ and $\Cmc\setminus\{\beta_0\}\subseteq\Rmc'$ when $\beta_0$ was chosen in the construction of $\Rmc$ (otherwise $\beta_0$ would have been added to $\Rmc'$).
\begin{itemize}
\item 
Let $Min(\Cmc)=\{\gamma \mid \gamma\in\Cmc,\forall\delta\in\Cmc, \gamma\not\succ\delta\}$. Assume for a contradiction that there exists $\gamma\in Min(\Cmc)\cap S_\nu$.  Since $\Cmc\setminus\{\gamma\}\rightsquigarrow\gamma$ (by definition of $Min(\Cmc)$) and $\gamma\in S_\nu\subseteq F_2=R(F_1)$, it follows that $\Cmc\subseteq R(F_1)$. In particular, $\beta_0\in R(F_1)$ so $\nu$ satisfies $(x_{\beta_0} \vee \bigvee_{\Bmc\rightsquigarrow{\beta_0}} x_{{\beta_0},\Bmc})$ 
and  
$\bigwedge_{\substack{\Bmc\rightsquigarrow{\beta_0},\\\beta\in\Bmc}}\neg x_{{\beta_0},\Bmc} \vee x_\beta $. Thus either (i) $\nu(\beta_0)=\true$, \ie $\beta_0\in S_\nu\subseteq\Rmc$, contradicting our assumption on $\beta_0$, or (ii) there exists $\Bmc\rightsquigarrow{\beta_0}$ such that $\Bmc\subseteq S_\nu\subseteq\Rmc$ and since for every $\beta\in\Bmc$, $\beta_0\not\succ\beta$, $\Bmc\subseteq S$, contradicting the $\Tmc$-consistency of $S$. 
Hence $Min(\Cmc)\cap S_\nu=\emptyset$ ($\dagger$). 
\item
Since $S$ is $\Tmc$-consistent, $\Cmc\not\subseteq S$ so there exists $\alpha_0\in \Cmc$ such that $\beta_0\succ\alpha_0$. Hence $\alpha_0$ has been chosen after $\beta_0$ in the construction of $\Rmc$. 
\item 
Since $\alpha_0$ has been chosen after $\beta_0$ in the construction of $\Rmc$ and was already in $\Rmc'$ when $\beta_0$ was chosen (by assumption on $\Cmc$), it must be the case that $\alpha_0\in S_\nu$. Hence, by ($\dagger$), $\alpha_0\notin Min(\Cmc)$ so there exists $\alpha_1\in\Cmc$ such that $\alpha_0\succ\alpha_1$. 
\item Since $\beta_0\succ\alpha_0\succ\alpha_1$, so that $\alpha_1$ was chosen after $\beta_0$ in the construction of $\Rmc$ and was already in $\Rmc'$ when $\beta_0$ was chosen, it must again be the case that $\alpha_1\in S_\nu$. We can iterate this argument to create an infinite chain $\beta_0\succ\alpha_0\succ\alpha_1\succ\dots$, since $\succ$ is acyclic. This yields a contradiction since $\Cmc$ is finite.
\end{itemize}
Hence $\Rmc\in\preps{\Kmc_\succ}$.
\medskip

\noindent$(\Leftarrow)$  Assume that there exists $\Rmc\in \preps{\Kmc_\succ}$ such that $S_\nu=\Rmc\cap F_2$. 
Since $\Rmc$ is \Tmc-consistent, it is easy to check that $\nu$ satisfies $\varphi_\mn{cons}(F_2)$. 
Assume for a contradiction that $\nu$ cannot be extended to satisfy $\varphi_\mn{P\text{-}max}(F_1)$.
\begin{itemize}
\item  There exists $\alpha\in \reachr(F_1) \subseteq F_2$ such that $\nu(x_\alpha)=\false$ and for every $\Bmc\rightsquigarrow\alpha$, $\nu(x_\beta)=\false$ for some $\beta\in\Bmc$.  Indeed, otherwise, setting $\nu(x_{\alpha,\Bmc})=\true$ for $\Bmc$ with $\Bmc\rightsquigarrow\alpha$, $\nu(x_\beta)=\true$ for all $\beta\in\Bmc$, and $\nu(x_{\alpha,\Bmc'})=\false$ otherwise will satisfy $\varphi_\mn{P\text{-}max}(F_1)$. 
\item Since $S_\nu=\Rmc\cap F_2$, it follows that $\alpha\notin\Rmc$ and for every $\Bmc\rightsquigarrow\alpha$, there is some $\beta\in\Bmc$ such that $\beta\notin\Rmc$. 
\item Hence, for every $\Bmc\subseteq\Rmc$ such that $\alpha\confof\Bmc$, we have $\Bmc\not\rightsquigarrow\alpha$, i.e.\ there exists $\beta\in\Bmc$ such that $\alpha\succ\beta$.
\item It follows that $\Rmc'=\Rmc\cup\{\alpha\}\setminus \{\beta\mid \alpha\succ\beta\}$ is a Pareto-improvement of $\Rmc$, contradicting $\Rmc\in \preps{\Kmc_\succ}$.
\end{itemize}
Hence $\nu$ satisfies $\varphi_\mn{P\text{-}max}(F_1)$.
\end{proof}

\begin{lemma}
Let $\Kmc_\succ$ be a prioritized KB with $\Kmc=(\Dmc,\Tmc)$, $F_1\subseteq \Dmc$, $F_2= F_1\cup \facts( \varphi_\mn{C\text{-}max}(F_1))$, and let $\nu$ be a valuation of $\{x_\alpha\mid \alpha\in F_2\}$.
\begin{itemize}
\item If $\nu$ can be extended to a valuation that satisfies $ \varphi_\mn{C\text{-}max}(F_1)\wedge \varphi_\mn{cons}(F_2)$ then $S_\nu=\{\alpha\mid \nu(x_\alpha)=\true\}$ can be extended to a completion-optimal repair of $\Kmc_\succ$.
\item If there exists a completion-optimal repair $\Rmc$ of $\Kmc_\succ$ such that $\Rmc\cap F_2=S_\nu$, then $\nu$ can be extended to a valuation of the variables of $ \varphi_\mn{C\text{-}max}(F_1)\wedge \varphi_\mn{cons}(F_2)$ that satisfies $ \varphi_\mn{C\text{-}max}(F_1)\wedge \varphi_\mn{cons}(F_2)$. 
\end{itemize}
\end{lemma}
\begin{proof}
\noindent$(\Rightarrow)$ Assume that $\nu$ satisfies $\varphi_\mn{C\text{-}max}(F_1)\wedge \varphi_\mn{cons}(F_2)$. Let  $\succ'$ be the binary relation over $F_2$ such that $\alpha\succ'\beta$ iff $\nu(x_{\alpha\succ'\beta})=\true$. 
As in the proof of Lemma \ref{lem:cmax}, since $\nu$ satisfies $\varphi_\mn{compl}$ and $\varphi_\mn{acyc}$, $\succ'$ is a completion of $\succ$ on $F_2$ that we complete into an arbitrary completion on $\Kmc$, and there is a unique Pareto-optimal repair $\Rmc_{\succ'}$, which is a completion-optimal repair.  
We next show that $S_\nu\subseteq \Rmc_{\succ'}$. 
\begin{itemize}
\item Recall from proof of Lemma \ref{lem:cmax} that $\Rmc_{\succ'}$ can be obtained as follows: Start with $\Rmc_{\succ'}=\emptyset$, then add to $\Rmc_{\succ'}$ the set $P=\{\alpha \mid \forall\beta\in\Kmc, \beta\not\succ'\alpha\}$, remove from $\Dmc$ the sets $P$ and $N=\{\alpha \mid \tup{\Tmc,P\cup\{\alpha\}}\models \bot\}$, and repeat this process until $\Dmc$ is empty. 
\item We show by induction that every $\alpha\in P\cap F_2$ is such that $\nu(x_\alpha)=\true$ and every $\alpha\in N\cap F_2$ is such that $\nu(x_\alpha)=\false$.  
\begin{itemize}
\item In the first step, $P$ contains facts that have no facts of greater priority and $N$ contains facts that belong to some conflict that contains only facts from $P$ except them. 
\begin{itemize}
\item Let $\alpha\in P\cap F_2$ and assume for a contradiction that $\nu(x_\alpha)=\false$. 
Since $\nu$ satisfies $\varphi_\mn{pref}$, there exists $\Bmc$ such that $\Bmc\rightsquigarrow\alpha$ 
and for every $\beta\in\Bmc$, $\nu(x_\beta)=\true$ and $\nu(x_{\beta\succ'\alpha})=\true$, \ie $\beta\succ'\alpha$, which contradicts the assumption on $\alpha$. 
\item Now assume that $\alpha\in N\cap F_2$, \ie there is some $\Bmc\subseteq P$ such that $\alpha\confof\Bmc$. Since for every $\beta\in\Bmc$, $\beta\succ'\alpha$ and $\succ'$ extends $\succ$, then for every  $\beta\in\Bmc$, $\alpha\not\succ\beta$, so $\Bmc\rightsquigarrow\alpha$. 
It follows that 
$\Bmc \subseteq F_2$
by construction of $F_2=\reachr(F_1)$. 
Hence for every $\beta\in\Bmc$, $\beta\in P\cap F_2$ so $\nu(x_\beta)=\true$. Since $\nu$ satisfies $\varphi_\mn{cons}(F_2)$, it follows that $\nu(x_\alpha)=\false$. 
\end{itemize}
\item Assume that at steps $1,\dots, i$ of the algorithm, $\nu(x_\alpha)=\true$ if $\alpha\in P\cap F_2$ and $\nu(x_\alpha)=\false$ if $\alpha\in N\cap F_2$, and consider step $i+1$. 
\begin{itemize}
\item Let $\alpha\in P\cap F_2$ and assume for a contradiction that $\nu(x_\alpha)=\false$. 
As before, since $\nu$ satisfies $\varphi_\mn{pref}$, there exists $\Bmc$ such that $\Bmc\rightsquigarrow\alpha$ 
and for every $\beta\in\Bmc$, $\nu(x_\beta)=\true$ and $\nu(x_{\beta\succ'\alpha})=\true$, \ie $\beta\succ'\alpha$. However, by definition of $P$, $\alpha\in P$ and $\beta\succ'\alpha$ imply that $\beta$ has already be removed from $\Dmc$, so belonged to $P$ or $N$ at step $j\leq i$. It follows that $\nu(x_\beta)=\true$ implies that $\beta$ belongs to $\Rmc_{\succ'}$. Hence $\Bmc\subseteq \Rmc_{\succ'}$, which contradicts the assumption $\alpha\in P$ by $\Tmc$-consistency of $\Rmc_{\succ'}$. 
\item The fact that $\nu(x_\alpha)=\false$ if $\alpha\in N\cap F_2$ follows as in the base case. 
\end{itemize}
\end{itemize}
\end{itemize}
By definition of $S_\nu$, we thus have $S_\nu= \Rmc_{\succ'}\cap F_2$. 
\medskip

\noindent$(\Leftarrow)$ Assume that there exists $\Rmc\in \creps{\Kmc_\succ}$ such that $S_\nu=\Rmc\cap F_2$. Since $\Rmc$ is \Tmc-consistent, it is easy to check that $\nu$ satisfies $\varphi_\mn{cons}(F_2)$. 
Moreover, there exists a completion $\succ'$ of $\succ$ such that $\Rmc$ is the unique Pareto-optimal repair of $\Kmc_{\succ'}$. 
Extend $\nu$ by setting $\nu(x_{\alpha\succ'\beta})=\true$ iff $\alpha\succ'\beta$, $\nu(t_{\alpha,\beta})=\true$ iff there is a path $\alpha\succ'\dots\succ'\beta$, and $\nu(x_{\Bmc\rightarrow\alpha})=\true$ iff for every $\beta\in\Bmc$, $\beta\succ'\alpha$ and $\beta\in S_\nu$. 
As in the proof of Lemma \ref{lem:cmax}, we can show that $\nu$ satisfies $\varphi_\mn{compl}$ and $\varphi_\mn{acyc}$. 
We show that $\nu$ satisfies $\varphi_\mn{pref}$. 

Assume for a contradiction that this is not the case: either (i) there exists $\alpha \in \reachr(F_1) \subseteq  F_2$ such that $\nu(x_\alpha)=\false$ and for every $\Bmc$ such that $\Bmc\rightsquigarrow\alpha$, 
$\nu(x_{\Bmc\rightarrow\alpha})=\false$ or (ii) there exists $\alpha$ and $\Bmc$ such that $\Bmc\rightsquigarrow\alpha$, 
$\nu(x_{\Bmc\rightarrow\alpha})=\true$ and there exists $\beta\in\Bmc$ such that $\nu(x_\beta)=\false$ or $\nu(x_{\beta\succ'\alpha})=\false$. 
Case (ii) contradicts the definition of $\nu$ (since $\nu(x_{\Bmc\rightarrow\alpha})=\true$ iff for every $\beta\in\Bmc$, $\beta\succ'\alpha$ and $\beta\in S_\nu$) 
 so we are in case (i).
 \begin{itemize}
\item Since $S_\nu=\Rmc\cap F_2$, $\nu(x_\alpha)=\false$ implies that $\alpha\notin\Rmc$.
\item For every $\Bmc$ such that $\Bmc\rightsquigarrow\alpha$, 
by definition of $\nu$, $\nu(x_{\Bmc\rightarrow\alpha})=\false$ implies that there exists $\beta\in\Bmc$ such that $\beta\not\succ'\alpha$ or $\beta\notin\Rmc$. 
\item However, since $\Rmc\in\preps{\Kmc_{\succ'}}$ and  $\succ'$ is total, 
$\alpha\notin\Rmc$ implies that there exists $\Bmc\subseteq\Rmc$ such that $\alpha\confof\Bmc$ and for every $\beta\in\Bmc$, $\beta\succ'\alpha$. Indeed, otherwise, $\Rmc'=\Rmc\cup\{\alpha\}\setminus\{\beta\mid \alpha\succ'\beta\}$ would be a Pareto-improvement of $\Rmc$ \wrt $\succ'$ since every conflict $\Cmc\subseteq \Rmc\cup\{\alpha\}$ would contain some $\beta\in\Rmc$ such that $\beta\not\succ'\alpha$, \ie such that $\alpha\succ'\beta$. Moreover, $\Bmc\rightsquigarrow\alpha$, since if there was $\beta\in\Bmc$ such that $\alpha\succ\beta$, we would have $\alpha\succ'\beta$ and hence $\beta\not\succ'\alpha$. This means that we have found a subset $\Bmc \subseteq \Rmc$ with $\Bmc\rightsquigarrow\alpha$ such that $\beta\succ'\alpha$ for every $\beta\in\Bmc$, which contradicts the statement in the previous item.
\end{itemize}
We conclude that $\nu$ satisfies $\varphi_\mn{pref}$. 
Hence $\nu$ satisfies $\varphi_\mn{C\text{-}max}(F_1)$. 
\end{proof}

\section{Pseudocode for Algorithms}
Algorithm \ref{general-algo} is our high-level procedure: it removes the self-inconsistent facts with the procedure \mn{RemoveSelfIncons} and finds the trivial answers with the procedure \mn{HandleTriviallyIAR}. Then it filters the remaining answers with the \mn{FilterRemainingAnswers} procedure which is one of the algorithms presented in the next subsections.

\begin{algorithm}
\caption{\mn{General\ Algorithm}}\label{general-algo}
\textbf{Input:} semantics $\sem$, repair type X, directed conflict graph $\mathcal{G}_{\Kmc_\succ}$, set of potential answers $\mi{PotAns}$ and their causes $\{\causes{q(\ans),\Kmc}\mid \ans\in\mi{PotAns}\}$\\
\textbf{Output:} set $\mi{Ans}$ of answers to $q$ over $\Kmc_\succ$ under X-$\sem$
\begin{algorithmic}
\STATE $\mathcal{G}_{\Kmc_\succ}, \mi{PotAns}, \{\causes{q(\ans),\Kmc}\mid \ans\in\mi{PotAns}\}\leftarrow \mn{RemoveSelfIncons}(\mathcal{G}_{\Kmc_\succ},\{\causes{q(\ans),\Kmc}\mid \ans\in\mi{PotAns}\})$
\STATE $\mi{Ans}, \causes{q(\ans),\Kmc}\leftarrow\mn{HandleTriviallyIAR}(\mathcal{G}_{\Kmc_\succ},\{\causes{q(\ans),\Kmc}\mid \ans\in\mi{PotAns}\})$
\STATE $\mi{Ans} \leftarrow \mi{Ans} \cup \mn{FilterRemainingAnswers}(\sem, X,\mathcal{G}_{\Kmc_\succ},\{\causes{q(\ans),\Kmc}\mid \ans\in\mi{PotAns}\})$
\STATE Output $\mi{Ans}$
\end{algorithmic}
\end{algorithm}

\begin{algorithm}
\caption{\mn{RemoveSelfIncons}}
\textbf{Input:} directed conflict graph $\mathcal{G}_{\Kmc_\succ}$, set of potential answers $\mi{PotAns}$ and their causes $\{\causes{q(\ans),\Kmc}\mid \ans\in\mi{PotAns}\}$\\
\textbf{Output:} updated $\mathcal{G}_{\Kmc_\succ}$, $\mi{PotAns}$ and $\{\causes{q(\ans),\Kmc}\mid \ans\in\mi{PotAns}\}$ without self-inconsistent facts
\begin{algorithmic}
\FORALL{$\{\alpha\}\in \conflicts{\Kmc}$}
	\FORALL{$\ans\in\mi{PotAns}$ }
		\FORALL{$\Cmc\in  \causes{q(\ans),\Kmc}$ }
		\IF{$\alpha\in\Cmc$}
			\STATE $\causes{q(\ans),\Kmc}\leftarrow \causes{q(\ans),\Kmc}\setminus\{\Cmc\}$
		\ENDIF
		\ENDFOR
		\IF{$\causes{q(\ans),\Kmc}=\emptyset$}
			\STATE $\mi{PotAns}\leftarrow\mi{PotAns}\setminus\{\ans\}$
		\ENDIF
	\ENDFOR
	\FORALL{$\{\alpha,\beta\}\in \conflicts{\Kmc}$}
	\STATE $\conflicts{\Kmc}\leftarrow \conflicts{\Kmc}\setminus \{\alpha,\beta\}$
	\ENDFOR
\ENDFOR
\STATE Output $\mathcal{G}_{\Kmc_\succ}$, $\mi{PotAns}$, $\{\causes{q(\ans),\Kmc}\mid \ans\in\mi{PotAns}\}$
\end{algorithmic}
\end{algorithm}

\begin{algorithm}
\caption{\mn{HandleTriviallyIAR}}
\textbf{Input:} directed conflict graph $\mathcal{G}_{\Kmc_\succ}$, set of potential answers $\mi{PotAns}$ and their causes $\{\causes{q(\ans),\Kmc}\mid \ans\in\mi{PotAns}\}$\\
\textbf{Output:} set $\mi{Ans}$ of trivial answers, updated $\{\causes{q(\ans),\Kmc}\mid \ans\in\mi{PotAns}\}$ without trivial answers and facts
\begin{algorithmic}
\STATE $\mi{Ans}\leftarrow\emptyset$
\FORALL{$\ans\in\mi{PotAns}$ }
\FORALL{$\Cmc\in  \causes{q(\ans),\Kmc}$, if $\ans\notin \mi{Ans}$ }
	\STATE $\Cmc\leftarrow\Cmc\setminus\{\alpha\mid \{\alpha,\beta\}\in\conflicts{\Kmc} \Rightarrow \alpha\succ\beta\}$
	\IF{$\Cmc=\emptyset$}
		\STATE $\mi{Ans}\leftarrow\mi{Ans}\cup\{\ans\}$
	\ENDIF
\ENDFOR
\ENDFOR
\STATE Output $\mi{Ans}$, $\{\causes{q(\ans),\Kmc}\mid \ans\in\mi{PotAns}\}$
\end{algorithmic}
\end{algorithm}

\subsection{Generic Algorithms Applicable to X-AR, X-IAR and X-brave}
The algorithms \algosat (Algorithm \ref{single-ans-filter}), \algomaxsat (Algorithm \ref{global-filter}), \algomuses (Algorithm \ref{muses-filter}), and \algoassump (Algorithm \ref{assumptions-filter}) can be used for any semantics.
\begin{itemize}
\item \algosat filters answers one by one, constructing a SAT encoding for each answer and using a standard SAT solver. 
\item \algomaxsat constructs a single encoding with hard clauses and soft clauses that correspond to the answers to check and uses a weighted MaxSAT solver to filter as many answers as possible before iterating to filter remaining answers. Note that in the X-IAR case, we actually don't need to iterate since each cause of each answer has its own variables.
\item \algomuses uses the same encoding for all answers with hard and soft clauses and computes all minimal unsatisfiable subsets of soft clauses \wrt the hard clauses: those that contain a single variable correspond to the true (resp.\ false) answers under X-AR or X-IAR (resp.\ X-brave) semantics. 
\item \algoassump takes the hard clauses of this global encoding as the SAT encoding and uses the soft clauses as assumptions to check for each answer whether it holds or not.
\end{itemize}

\noindent Let $\mn{val}_\sem=\mn{UNSAT}$ for $\sem=$ AR and $\sem=$ IAR, and $\mn{val}_\sem=\mn{SAT}$ for $\sem=$ brave.

\begin{algorithm}
\caption{\algosat}\label{single-ans-filter}
\textbf{Input:} semantics $\sem$, repair type X, directed conflict graph $\mathcal{G}_{\Kmc_\succ}$, set of potential answers $\mi{PotAns}$ and their causes $\{\causes{q(\ans),\Kmc}\mid \ans\in\mi{PotAns}\}$ \\
\textbf{Output:} set $\mi{Ans}$ of answers to $q$ over $\Kmc_\succ$ under X-$\sem$
\begin{algorithmic}
\STATE $\mi{Ans}\leftarrow\emptyset$
\FORALL{$\ans\in\mi{PotAns}$ }
	\STATE $\varphi\leftarrow \Phi_{X\text{-}\sem}(q(\ans))$
	\IF{$\mn{SAT}(\varphi)=\mn{val}_\sem$}
		\STATE $\mi{Ans}\leftarrow\mi{Ans}\cup\{\ans\}$
	\ENDIF
\ENDFOR
\STATE Output $\mi{Ans}$
\end{algorithmic}
\end{algorithm}

\begin{algorithm}
\caption{\algomaxsat}\label{global-filter}
\textbf{Input:} semantics $\sem$, repair type X, directed conflict graph $\mathcal{G}_{\Kmc_\succ}$, set of potential answers $\mi{PotAns}$ and their causes $\{\causes{q(\ans),\Kmc}\mid \ans\in\mi{PotAns}\}$\\
\textbf{Output:} set $\mi{Ans}$ of answers to $q$ over $\Kmc_\succ$ under X-$\sem$
\begin{algorithmic}
\STATE $\mi{Ans}\leftarrow\emptyset$
\STATE $ \varphi\leftarrow \Psi_{X\text{-}\sem}(\mi{PotAns})\wedge \bigwedge_{\ans\in\mi{Potans}}x_{\ans}$
\STATE $\mi{moreAnswers}\leftarrow\true$
\STATE $\mi{assumedLitterals}\leftarrow\emptyset$	
\STATE $\mi{filteredAns}\leftarrow\emptyset$
\WHILE{$\mi{moreAnswers}=\true$ and  $\mi{filteredAns}\neq \mi{PotAns}$}	
\STATE $\mi{moreAnswers}\leftarrow\false$
\STATE $\mi{optModel}\leftarrow \mn{MaxSAT}(\varphi, \mi{assumedLitterals})$
\FORALL{$\ans\in\mi{PotAns}$ }
	\IF{$\mi{optModel}\models x_{\ans}$  }
		\STATE $\mi{moreAnswers}\leftarrow\true$	
		\STATE $\mi{filteredAns}\leftarrow\mi{filteredAns}\cup\{\ans\}$
		\STATE $\mi{assumedLitterals}\leftarrow \mi{assumedLitterals}\cup\{\neg x_{\ans}\}$
	\ENDIF
\ENDFOR
\ENDWHILE
\IF{$\mn{val}_\sem=\mn{UNSAT}$}
	\STATE $\mi{Ans}\leftarrow \mi{PotAns}\setminus\mi{filteredAns}$
\ELSE
	\STATE $\mi{Ans}\leftarrow \mi{Ans}\cup \mi{filteredAns}$	
\ENDIF
\STATE Output $\mi{Ans}$
\end{algorithmic}
\end{algorithm}

\begin{algorithm}
\caption{\algomuses}\label{muses-filter}
\textbf{Input:} semantics $\sem$, repair type X, directed conflict graph $\mathcal{G}_{\Kmc_\succ}$, set of potential answers $\mi{PotAns}$ and their causes $\{\causes{q(\ans),\Kmc}\mid \ans\in\mi{PotAns}\}$\\
\textbf{Output:} set $\mi{Ans}$ of answers to $q$ over $\Kmc_\succ$ under X-$\sem$
\begin{algorithmic}
\STATE $\mi{Ans}\leftarrow\emptyset$
\STATE $ \varphi\leftarrow \Psi_{X\text{-}\sem}(\mi{PotAns})\wedge \bigwedge_{\ans\in\mi{Potans}}x_{\ans}$
\STATE $\mi{MUSes}\leftarrow \mn{computeAllMUSes}(\varphi)$
\IF{$\mn{val}_\sem=\mn{UNSAT}$}
	\STATE $\mi{Ans}\leftarrow \mi{Ans}\cup \{\ans\mid \{x_{\ans}\}\in\mi{MUSes}\}$
\ELSE
	\STATE $\mi{Ans}\leftarrow \mi{PotAns}\setminus \{\ans\mid \{x_{\ans}\}\in\mi{MUSes}\}$	
\ENDIF
\STATE Output $\mi{Ans}$
\end{algorithmic}
\end{algorithm}

\begin{algorithm}
\caption{\algoassump}\label{assumptions-filter}
\textbf{Input:} semantics $\sem$, repair type X, directed conflict graph $\mathcal{G}_{\Kmc_\succ}$, set of potential answers $\mi{PotAns}$ and their causes $\{\causes{q(\ans),\Kmc}\mid \ans\in\mi{PotAns}\}$\\
\textbf{Output:} set $\mi{Ans}$ of answers to $q$ over $\Kmc_\succ$ under X-$\sem$
\begin{algorithmic}
\STATE $\mi{Ans}\leftarrow\emptyset$
\STATE $ \varphi\leftarrow \Psi_{X\text{-}\sem}(\mi{PotAns})$
\FORALL{$\ans\in\mi{PotAns}$ }
\IF{$\mn{SAT}(\varphi[x_{\ans}])=\mn{val}_\sem$}
	\STATE $\mi{Ans}\leftarrow \mi{Ans}\cup \{\ans\}$
\ENDIF
\ENDFOR
\STATE Output $\mi{Ans}$
\end{algorithmic}
\end{algorithm}

\subsection{Algorithms Specific to X-IAR or X-brave}

The algorithm \algocauses (Algorithm \ref{cause-by-cause-filter}) checks whether each cause is in some (resp.\ all) optimal repair(s): if an answer has such a cause, then it holds under X-brave (resp.\ X-IAR) semantics.  

\noindent The algorithms \algoIARcauses (Algorithm \ref{IAR-cause-filter}) and \algoIARfacts (Algorithm \ref{IAR-assertions-filter}) are specific to X-IAR semantics. 
\begin{itemize}
\item \algoIARcauses computes X-IAR and non-X-IAR facts as it goes: it checks the causes one by one, by first checking whether the cause contains some known non-X-IAR facts or only known X-IAR facts, and otherwise checking the status of each remaining fact, stoping at the first non-X-IAR one. 
\item \algoIARfacts also maintain two sets of X-IAR and non-X-IAR facts. For each answer, it checks whether the facts in the union of the causes are X-IAR or not with a weighted MaxSAT solver, then it checks whether some answer cause contains only such facts.
\end{itemize}

\begin{algorithm}
\caption{\algocauses}\label{cause-by-cause-filter}
\textbf{Input:} semantics $\sem\in\{$brave, IAR$\}$, repair type X, directed conflict graph $\mathcal{G}_{\Kmc_\succ}$, set of potential answers $\mi{PotAns}$ and their causes $\{\causes{q(\ans),\Kmc}\mid \ans\in\mi{PotAns}\}$\\
\textbf{Output:} set $\mi{Ans}$ of answers to $q$ over $\Kmc_\succ$ under X-$\sem$
\begin{algorithmic}
\STATE $\mi{Ans}\leftarrow\emptyset$
\FORALL{$\ans\in\mi{PotAns}$ }
	\FORALL{$\Cmc\in \causes{q(\ans),\Kmc}$, if $\ans\notin \mi{Ans}$}
		\STATE $\varphi\leftarrow\Phi_{X\text{-}\sem}(\Cmc)$
		\IF{$\mn{SAT}(\varphi)=\mn{val}_\sem$}
			\STATE $\mi{Ans}\leftarrow\mi{Ans}\cup\{\ans\}$
		\ENDIF
	\ENDFOR
\ENDFOR
\STATE Output $\mi{Ans}$
\end{algorithmic}
\end{algorithm}

\begin{algorithm}
\caption{\algoIARcauses}\label{IAR-cause-filter}
\textbf{Input:}  repair type X, $\mathcal{G}_{\Kmc_\succ}$, set of potential answers $\mi{PotAns}$ and their causes $\{\causes{q(\ans),\Kmc}\mid \ans\in\mi{PotAns}\}$\\
\textbf{Output:} set $\mi{Ans}$ of answers to $q$ over $\Kmc_\succ$ under X-IAR semantics
\begin{algorithmic}
\STATE $\mi{Ans}\leftarrow\emptyset$
\STATE $\mi{IARfacts}\leftarrow\emptyset$
\STATE $\mi{nonIARfacts}\leftarrow\emptyset$
\FORALL{$\ans\in\mi{PotAns}$ }
	\FORALL{$\Cmc\in \causes{q(\ans),\Kmc}$, if $\ans\notin \mi{Ans}$ and $\Cmc\cap\mi{nonIARfacts}=\emptyset $}
		\STATE $\Cmc\leftarrow \Cmc\setminus\mi{IARfacts}$
		\IF{$\Cmc=\emptyset$}
			\STATE $\mi{Ans}\leftarrow\mi{Ans}\cup\{\ans\}$
		\ELSE
			\STATE $\mi{IARCause}\leftarrow \true$
			\FORALL{$\alpha\in\Cmc$, if $\mi{IARCause}= \true$}
			\STATE $\varphi\leftarrow \Phi_{X\text{-}IAR}(\alpha)$
			\IF{$\mn{SAT}(\varphi)=\mn{UNSAT}$}
				\STATE $\mi{IARfacts}\leftarrow\mi{IARfacts}\cup\{\alpha\}$
			\ELSE
				\STATE $\mi{nonIARfacts}\leftarrow\mi{nonIARfacts}\cup\{\alpha\}$
				\STATE $\mi{IARCause}\leftarrow \false$
			\ENDIF
			\ENDFOR
			\IF{$\mi{IARCause}= \true$}
				\STATE $\mi{Ans}\leftarrow\mi{Ans}\cup\{\ans\}$
			\ENDIF
		\ENDIF
	\ENDFOR
\ENDFOR
\STATE Output $\mi{Ans}$
\end{algorithmic}
\end{algorithm}

\begin{algorithm}
\caption{\algoIARfacts}\label{IAR-assertions-filter}
\textbf{Input:}  repair type X, $\mathcal{G}_{\Kmc_\succ}$, set of potential answers $\mi{PotAns}$ and their causes $\{\causes{q(\ans),\Kmc}\mid \ans\in\mi{PotAns}\}$\\
\textbf{Output:} set $\mi{Ans}$ of answers to $q$ over $\Kmc_\succ$ under X-IAR semantics
\begin{algorithmic}
\STATE $\mi{Ans}\leftarrow\emptyset$
\STATE $\mi{IARfacts}\leftarrow\emptyset$
\STATE $\mi{nonIARfacts}\leftarrow\emptyset$
\FORALL{$\ans\in\mi{PotAns}$ }
	\STATE $\mi{RelevantFacts}\leftarrow \bigcup_{\Cmc\in\causes{q(\ans),\Kmc}}\Cmc \setminus(\mi{IARfacts}\cup\mi{nonIARfacts})$
	\STATE $\varphi\leftarrow \Psi_{X\text{-}IAR}(\mi{RelevantFacts}) \wedge \bigwedge_{\alpha\in\mi{RelevantFacts} } y_\alpha$
	\STATE $\mi{moreNonIAR}\leftarrow\true$
\STATE $\mi{assumedLitterals}\leftarrow\emptyset$	
\STATE $\mi{newNonIAR}\leftarrow\emptyset$
\WHILE{$\mi{moreNonIAR}=\true$ and  $\mi{newNonIAR}\neq \mi{RelevantFacts}$}	
\STATE $\mi{moreNonIAR}\leftarrow\false$
\STATE $\mi{optModel}\leftarrow \mn{MaxSAT}(\varphi, \mi{assumedLitterals})$
\FORALL{$\alpha\in\mi{RelevantFacts}$ }
	\IF{$\mi{optModel}\models y_{\alpha}$  }
		\STATE $\mi{moreNonIAR}\leftarrow\true$	
		\STATE $\mi{newNonIAR}\leftarrow\mi{newNonIAR}\cup\{\alpha\}$
		\STATE $\mi{assumedLitterals}\leftarrow \mi{assumedLitterals}\cup\{\neg y_{\alpha}\}$
	\ENDIF
\ENDFOR
\ENDWHILE
	\STATE $\mi{IARfacts}\leftarrow\mi{IARfacts}\cup\mi{RelevantFacts}\setminus\mi{newNonIAR}$
	\STATE $\mi{nonIARfacts}\leftarrow\mi{nonIARfacts}\cup\mi{newNonIAR}$
	\FORALL{$\Cmc\in \causes{q(\ans),\Kmc}$, if $\ans\notin \mi{Ans}$}
		\STATE $\Cmc\leftarrow \Cmc\setminus\mi{IARfacts}$
		\IF{$\Cmc=\emptyset$}
			\STATE $\mi{Ans}\leftarrow\mi{Ans}\cup\{\ans\}$
		\ENDIF
	\ENDFOR
\ENDFOR
\STATE Output $\mi{Ans}$
\end{algorithmic}
\end{algorithm}

\clearpage
\section{Details on Experimental Setting}
\subsection{KBs Description}
\subsubsection*{CQAPri benchmark} 
We refer to section 3.1.1 of \cite{DBLP:phd/hal/Bourgaux16} for the detailed description of the CQAPri benchmark DL-Lite ontology, datasets and queries. Note that while the queries are given as CQs, they correspond to UCQs when rewritten according to the ontology. 
We use the 18 datasets named \mn{uXcY} with $X\in\{1,5,20\}$ and $Y\in\{1,5,10,20,30,50\}$. Parameters $X$ and $Y$ are related to the size and the proportion of assertions involved in some conflicts respectively (the higher the bigger), and the datasets are such that $\mn{uXcY}\subseteq \mn{uXcY'}$ for $Y\leq Y'$ and $\mn{uXcY}\subseteq \mn{uX'cY}$ for $X\leq X'$. 
Their sizes range from 75K to 2M facts and their proportions of facts involved in some conflict are between 3\% and 46\%. 
The conflicts graphs contain from 2,373 to 946,819 facts and from 2,314 to 3,130,377 conflicts. 
In the smallest conflict graph (\mn{u1c1}), each of the facts is in conflict with between 1 and 614 facts with an average of 2. In the largest one (\mn{u20c50}), each of the facts is in conflict with between 1 and 744 facts with an average of 6.6.

\subsubsection*{Food Inspection dataset} 
The Food Inspection dataset contains data about inspection of food establishments in New York and Chicago taken from \cite{Dataset1FoodNewYork} and \cite{Dataset2FoodChicago}. 
The original data is in two csv files of respectively 399,961 and 213,606 rows and 26 and 17 columns. 
We use the database schema and queries proposed in \cite{DBLP:conf/sat/DixitK19}: the data is decomposed into four relations with some key constraints and one functional dependency. All details are provided in Figure \ref{food-benchmark-description}. We modified query $\mn{q5}$ because the value $Fail$ is not used in the New York dataset, so we keep only the first CQ of the original UCQ; and we don't report experiments using $\mn{q1}$ (which is the Boolean version of $\mn{q2}$) because it trivially hold under all semantics as it has causes without any conflict, and is thus not interesting in the scope of this paper. 
In total, the dataset contains more than 523K facts. Compared to the setting of \cite{DBLP:conf/sat/DixitK19}, our dataset is a bit larger because the original datasets have been extended with new entries since then. The conflict graph contains 192,028 facts (37\% of the facts are in some conflict) and 219,854 conflicts. Each of the conflict graph facts is in conflict with between 1 and 566 facts with an average of 2.3.

\begin{figure}[h]
\begin{tabular*}{\textwidth}{l @{\extracolsep{\fill}}  r}
\hline
Relation & \#facts \\
\hline
NY\_Insp(LicenseNo,Risk,InspDate,InspType,Result) & 241,419\\
CH\_Insp(LicenseNo,Risk,InspDate,InspType,Result) & 212,907\\
NY\_Rest(Name, LicenseNo, Cuisine, Address, Zip) & 28,878\\
CH\_Rest(Name, LicenseNo, Facility, Address, Zip) & 40,375\\
\hline
\end{tabular*}
\medskip

\begin{tabular*}{\textwidth}{l @{\extracolsep{\fill}}  l r}
\hline
Constraint & Type &\#facts in violations \\
\hline
 NY\_Insp(LicenseNo,InspDate,InspType $\rightarrow$ Risk, Result) & Key & 182,510 \\
 CH\_Insp(LicenseNo,InspDate,InspType $\rightarrow$ Risk, Result)& Key &1,155\\
 NY\_Rest(LicenseNo$\rightarrow$ Name, Cuisine, Address, Zip)& Key &0\\
 CH\_Rest(LicenseNo $\rightarrow$ Name, Facility, Address, Zip)& Key & 2,390 \\
CH\_Rest(Name $\rightarrow$ Zip) & FD & 6,294 \\
\hline
\end{tabular*}

\begin{align*}
\mn{q2}(x)=&
\text{NY\_Rest}(x,\_,\_,\_,\_) \wedge \text{CH\_Rest}(x,\_,\_,\_,\_)\\
\mn{q3}(x)=&
\text{NY\_Rest}(x,y,\_,\_,\_) \wedge \text{CH\_Rest}(x,y',\_,\_,\_)
\wedge 
\text{NY\_Insp}(y, \_, z, \_, \_) \wedge \text{CH\_Insp}(y', \_, z, \_, \_)\\
\mn{q4}(x,y)=&
\text{CH\_Rest}(x,y,\_,\_,\_) \wedge \text{CH\_Insp}(y,\_,\_,\_,Pass)\\
\mn{q5}(x)=&
\text{CH\_Rest}(x,y,\_,\_,\_) \wedge \text{CH\_Insp}(y,\_,\_,\_,Fail)\\
\mn{q6}(x,y)=&
\text{CH\_Rest}(x,\_,\_,\_,y) \wedge \text{NY\_Rest}(x,z,\_,\_,\_)
\wedge 
\text{NY\_Insp}(z, NotCritical, \_, \_,\_)
\end{align*}
\caption{Food Inspection dataset: database schema, constraints and queries. In the queries, $\_$ denotes existentially quantified variables that are not shared among atoms and free variables are given in query heads.}\label{food-benchmark-description}
\end{figure}

\subsubsection*{Physicians dataset} 
We build the Physicians dataset from the National Downloadable File provided by the Centers for Medicare \& Medicaid Services (CMS) \cite{Dataset3Physicians}. It contains information on medical professionals and the organizations they are associated with. 
The original data is a csv file of 2,214,220 rows and 40 columns. 
We decomposed it into seven relations, added some reasonable key constraints and functional dependencies and design some queries. All details are provided in Figure \ref{physicians-benchmark-description}.  
In total, the dataset contains more than 8,254K facts. 
The conflict graph contains 183,387 facts (2\% of the facts are in some conflict) and 2,708,718 conflicts. Each of the conflict graph facts is in conflict with between 1 and 2,069  facts with an average of 29.5. 

\begin{figure}[h]
\begin{tabular*}{\textwidth}{l @{\extracolsep{\fill}}r}
\hline
Relation & \#facts\\
\hline
Physicians(npi, ind\_pac\_id, ind\_enrl\_id, lst\_nm, frst\_nm, mid\_nm, suff, gndr) & 1,194,300\\
PrimarySpecialty(npi, pri\_spec) & 1,150,513\\
SecondarySpecialty(npi, sec\_spec) &154,570 \\
Credentials(npi, cred , med\_sch,  grd\_yr) & 1,146,083\\
GroupPractice(npi,org\_nm, org\_pac\_id, num\_org\_mem, adr\_ln\_1, adr\_ln\_2,  cty, st, zip) & 2,214,129\\
HospAffiliation(npi, hosp\_afl, hosp\_afl\_lbn) &1,248,351\\
Assignment(npi, ind\_assgn) & 1,146,086\\
\hline
\end{tabular*}
\medskip

\begin{tabular*}{\textwidth}{l @{\extracolsep{\fill}} l r}
\hline
Constraint & Type &\#facts in violations \\
\hline
Physicians(npi$\rightarrow$ ind\_pac\_id, ind\_enrl\_id, lst\_nm, frst\_nm, mid\_nm, suff, gndr) & Key& 91,408 \\
PrimarySpecialty(npi$\rightarrow$ pri\_spec)& Key& 8,800\\
Credentials(npi $\rightarrow$ cred , med\_sch,  grd\_yr)& Key& 0\\
GroupPractice(zip$\rightarrow$ cty, st ) & FD& 83,179\\
HospAffiliation(hosp\_afl $\rightarrow$ hosp\_afl\_lbn) & FD& 0\\
Assignment(npi $\rightarrow$ ind\_assgn) &Key& 0\\
\hline
\end{tabular*}

\begin{align*}
\mn{q1}(x)=&
\text{PrimarySpecialty}\mi{(x, INTERNAL\ MEDICINE)}\wedge\text{Assignment}(x, Y)
\wedge
\text{GroupPractice}(x,\_, \_, \_, \_, \_,  \_, AZ, \_)
\\
\mn{q2}(x)=&
\text{Physicians}(x, \_, \_, \_, \_, \_, \_, M)
\wedge \text{Credentials}(x, MD, \_,  \_)\\&
\wedge \text{PrimarySpecialty}\mi{(x, HAND\ SURGERY)}
\wedge\text{SecondarySpecialty}\mi{(x, ORTHOPEDIC\ SURGERY)}
\\
\mn{q3}(x,y)=&
\text{Physicians}\mi{(z, \_, \_, y, x, \_, \_, \_)}
\wedge 
\text{GroupPractice}\mi{(z,\_, \_, \_, \_, \_,  NEW\ YORK, \_, \_)}\\
\mn{q4}(x,x')=&
\text{GroupPractice}\mi{(x, \_, \_, \_, \_, \_,  \_, TX, \_)}\wedge
\text{GroupPractice}\mi{(x', \_, \_ \_, \_, \_,  \_, TX, \_)}\wedge\\&
\text{Credentials}\mi{(x, MD , y,  \_)}\wedge
\text{Credentials}\mi{(x', MD , y,  \_)}\wedge
\text{HospAffiliation}\mi{(x, z, \_)}\wedge 
\text{HospAffiliation}\mi{(x', z, \_)}
\\
\mn{q5}(x,x')=&
\text{PrimarySpecialty}\mi{(x, y)}\wedge
\text{PrimarySpecialty}\mi{(x', y)}\wedge\\&
\text{Credentials}\mi{(x, \_ , COLUMBIA\ U.C.\ PHYSICIANS\ AND\ SURGEONS,  z)}\wedge\\&
\text{Credentials}\mi{(x', \_ , COLUMBIA\ U.C.\ PHYSICIANS\ AND\ SURGEONS,  z)}\wedge\\&
\text{Physicians}\mi{(x, \_, \_,\_, \_, \_, \_, F)}\wedge 
\text{Physicians}\mi{(x',\_, \_, \_, \_, \_, \_, F)}
\\
\mn{q6}(x,x')=&\text{Physicians}\mi{(x, \_, \_, y, \_, \_, \_, \_)}
\wedge
\text{Physicians}\mi{(x', \_, \_, y, \_, \_, \_, \_)}
\wedge\\&
\text{GroupPractice}\mi{(x, \_, \_, \_, \_, \_,  \_, WA, \_)}
\wedge
\text{GroupPractice}\mi{(x', \_, \_, \_, \_, \_,  \_, OR, \_)}
\end{align*}
\caption{Physicians dataset: database schema, constraints and queries. In the queries, $\_$ denotes existentially quantified variables that are not shared among atoms and free variables are given in query heads.}\label{physicians-benchmark-description}
\end{figure}

\subsection{Construction of Input Files} 
To obtain the potential answers and their causes, we put the data in relational databases that we query with \textsf{ProvSQL}, a \textsf{PostgreSQL} extension which provides us with the answer causes \cite{DBLP:journals/pvldb/SenellartJMR18}. For the CQAPri benchmark, we first rewrite the conjunctive queries \wrt the ontology into union of conjunctive queries with \textsf{Rapid} \cite{CTS-CADE-11}. We refer to section 3.1.1 of \cite{DBLP:phd/hal/Bourgaux16} for details on the rewritten queries. 

We build the conflict graphs in a similar fashion, using `conflict queries' that ask whether some constraint is not satisfied by the dataset. The cause of this queries are precisely the conflicts of the KB.

\begin{remark}
We did not measure the time needed to compute the conflict graph or the causes of a query since our focus is on the comparison of the answer filtering algorithms. However such times are reported for the CQAPri benchmark in \cite{DBLP:phd/hal/Bourgaux16}. In particular, the time for the construction of conflict graphs ranged from 2 to 40 seconds for the databases we consider here. Note that the computation of the conflict graph and causes for answers depends on the logic / constraints considered, contrary to the filtering step we consider here. 
\end{remark}

\subsection{Sizes of Query Answers and Causes Files} 
Table \ref{tab:size-causes} gives the size of the files that contain the potential answers and their causes for the Food Inspection and Physicians datasets and four example datasets of the CQAPri benchmark, with minimal and maximal sizes and minimal and maximal proportion of facts in conflicts: \mn{u1c1}, \mn{u1c50}, \mn{u20c1}, \mn{u20c50}.

\begin{table}
\begin{center}
{\small
\begin{tabular*}{0.8\textwidth}{l @{\extracolsep{\fill}}  r  r }
\begin{tabular*}{0.5\textwidth}{l @{\extracolsep{\fill}}  r r r r  }
\toprule
& \mn{u1c1} &  \mn{u1c50} & \mn{u20c1} & \mn{u20c50}\\
\midrule
\mn{q1} & 2,106 & 2,120 & 57,498 & 57,857\\
\mn{q2} & 1,673 & 1,768 & 44,465 & 47,158 \\
\mn{q3} & 19 & 19 & 19 & 19 \\
\mn{q4} & 10,427 & 10,664 & 259,050 & 264,861\\
\mn{q5} & 27 & 32& 27 & 32\\
\mn{q6} & 36 & 51 & 36 & 51 \\
\mn{q7} & 5 & 6 & 5 & 6 \\
\mn{q8} & 0 & 0 & 7 &  7\\
\mn{q9} & 173 & 217 & 4,683 & 5,791\\
\mn{q10} & 0.2 & 0.5 & 10 & 11\\
\mn{q11} & 69 & 73 & 1,871 & 2,000\\
\mn{q12} & 162 & 174 & 3,823 & 4,123 \\
\mn{q13} & 87 & 97 & 2,368 & 2,635 \\
\mn{q14} & 17 & 17 & 440 & 440 \\
\mn{q15} & 16 & 16 & 439 & 447\\
\mn{q16} & 4,171 & 4,295 & 114,364 & 121,492\\
\mn{q17} & 1 & 1 & 35 & 40 \\
\mn{q18} & 1,615 & 1,664 & 44,242 & 45,286 \\
\mn{q19} & 0 & 0 & 14 & 301 \\
\mn{q20} & 3 & 3& 3 & 3\\
\bottomrule
\end{tabular*}
&
&
\begin{tabular*}{0.2\textwidth}{r }
\begin{tabular*}{0.2\textwidth}{l @{\extracolsep{\fill}}  r  }
\toprule
& Food Inspection  \\
\midrule
\mn{q2} & 3,692\\
\mn{q3} &  244  \\
\mn{q4} &7,016 \\
\mn{q5} & 3,007\\
\mn{q6} & 22,258  \\
\bottomrule
\end{tabular*}
\\
\\
\begin{tabular*}{0.2\textwidth}{l @{\extracolsep{\fill}}  r  }
\toprule
& Physicians  \\
\midrule
\mn{q1} & 234 \\
\mn{q2} & 38  \\
\mn{q3} & 1,692 \\
\mn{q4} & 351,060\\
\mn{q5} & 459 \\
\mn{q6} & 66,196  \\
\bottomrule
\end{tabular*}
\end{tabular*}
\end{tabular*}
}
\end{center}
\caption{Answers and causes file size in kB. }\label{tab:size-causes}
\end{table}

\subsection{Priority Relations and Sizes of Directed Conflict Graphs} 
We build score-structured priority relations by randomly assigning to each fact a score between $1$ and $n$. 
To construct a priority relation that is not score-structured, we consider each conflict and assign a random direction to the corresponding edge in the conflict graph with a probability $p$, except if doing so creates a cycle. We finally check if we got a non score-structured preference relation. 

On the Food Inspection and Physicians datasets, we build four priority relations: two score-structured with $n{=}2$ and $n{=}5$ respectively, and two non score-structured with $p{=}0.5$ and $p{=}0.8$ respectively. 
In the case of the CQAPri benchmark, we build only two priority relations (score-strutured with $n{=}5$ and non score-structured with $p{=}0.8$) on our biggest dataset (\mn{u20c50}), then propagate them to the other datasets. 
We check that the non score-structured relation is not score-structured already on \mn{u1c30} and on \mn{u5c1}. 
For datasets \mn{u1cY} with $Y<30$, there is no cycles in the conflict graph so all possible priority relations are actually score-structured.

The size of the directed conflict graph (that contains both edges $(\alpha,\beta)$ and $(\beta,\alpha)$ if $\{\alpha,\beta\}\in\conflicts{\Kmc}$ and $\alpha\not\succ\beta$ and $\beta\not\succ\alpha$, and contains a single edge $(\alpha,\beta)$ if $\{\alpha,\beta\}\in\conflicts{\Kmc}$ and $\beta\succ\alpha$) is roughly equal to $\frac{n+1}{2n}$ of the original conflict graph for score-structured priority with $n$ levels (since the probability of having $\alpha\succ\beta$ or $\beta\succ\alpha$, hence keeping only one of the edges $(\alpha,\beta)$ and $(\beta,\alpha)$ is $\frac{n-1}{n}$, so that $\frac{n-1}{2n}$ of the edges are removed). 
In the case of non-structured priority, because of the acyclicity condition, the proportion of edges removed is a bit less than $\frac{p}{2}$, so the size of the directed conflict graph is greater than $\frac{2-p}{2}$. 
We chose our two values of $n$ and $p$ so that $\frac{n+1}{2n}=\frac{2-p}{2}$. Table \ref{tab:conf-graph-size} reports the actual sizes of the directed conflict graphs.

\begin{table}
\begin{tabular*}{\textwidth}{l @{\extracolsep{\fill}} r r r r r }
\toprule
& no priority & score $n=2$ & score $n=5$ & non score $p=0.5$ & non score $p=0.8$\\
\midrule
Food Inspection & 439,708 & 329,909 & 263,868 & 356,446 & 308,856\\
Physicians & 5,417,436 & 4,063,007 & 3,250,863 & 4,621,295 & 4,186,468\\
\bottomrule
\end{tabular*}
\smallskip

\begin{tabular*}{\textwidth}{l @{\extracolsep{\fill}} r r r r }
\toprule
& \mn{u1c1} &\mn{u1c50}& \mn{u20c1} &\mn{u20c50} \\
\midrule
no priority  & 4,628 & 162,608 & 146,424 & 6,260,754
\\
score $n=5$ & 2,810 & 97,677 & 87,760 & 3,756,160
\\
non score $p=0.8$ & 3,140 & 109,292 &102,025 & 4,326,325
\\
\bottomrule
\end{tabular*}
\caption{Number of edges in directed conflict graphs, depending on the priority relation.}\label{tab:conf-graph-size}
\end{table}

\clearpage
\section{Details on Experimental Evaluation}

\subsection{Number of Answers Under the Different Semantics}  

Table \ref{cqapri-nb-answers-noprio} shows the number of answers that hold under the different semantics we consider in four example datasets \mn{u1c1}, \mn{u1c50}, \mn{u20c1} and \mn{u20c50} of the CQAPri benchmark in the case of classical repairs (no priority relation): (plain) IAR semantics, AR semantics but not IAR semantics, brave semantics but not AR semantics, and potential answers that would hold in the dataset without any constraints but do not hold under brave semantics. 

Tables \ref{cqapri-nb-answers-score}, \ref{cqapri-nb-answers-nonscore-pareto} and \ref{cqapri-nb-answers-nonscore-completion} show the number of answers that hold under the different semantics we consider in the case of optimal repairs with the score-structured and non score-structured priority relations respectively:  answers that are trivially X-IAR (some cause contains only facts that do not have outgoing edges in the directed conflict graph), X-IAR answers that are not trivially X-IAR, X-AR answers that are not X-IAR, X-brave answers that are not X-AR, and potential answers that do not hold under X-brave semantics. 

Tables \ref{food-nb-answers-noprio}, \ref{food-nb-answers-score}, \ref{food-nb-answers-nonscore} and  \ref{physicians-nb-answers-noprio}, \ref{physicians-nb-answers-score}, \ref{physicians-nb-answers-nonscore} show the number of answers in the different classes for the different priority relations and repair types on the Food Inspection dataset and  Physicians dataset respectively.
The empty cells correspond to cases where we were not able to compute the answers under the required semantics in our time and memory limits.

\paragraph{Classical v.s.\ optimal repairs} 
In the CQAPri benchmark, without priority relation, a lot of queries do not have any AR answers that are not trivial, making trivial answers a good approximation (computable in polynomial time) of AR in many cases (this is not the case for the Food Inspection and Physicians datasets). 
This is not true when using optimal repairs: in big datasets or datasets with a lot of conflicts, most queries have (a lot of) answers that hold under X-AR/X-IAR while not being trivial. Hence, even if there are more answers that are trivially true when using priority relations than without, it seems even more useful to actually compute these answers rather than going for the polynomial lower bound given by the trivial answers. 

While there are very few potential answers that are not brave under the classical semantics, there are very common with optimal repairs. 

\paragraph{Pareto- v.s.\ completion-optimal repairs} We were not able to compute answers for completion-optimal repairs on the Food Inspection dataset and most of the queries of the CQAPri benchmark. For the few queries of the CQAPri benchmark for which we manage to compute them, they coincide with those obtained with Pareto-optimal repairs. On the Physicians dataset, we manage to compute the answers with completion-optimal repairs for five queries out of six (only under C-IAR semantics for one of them), and two of them differ from those obtained with Pareto-optimal repairs (\mn{q1} and \mn{q6}).

\begin{table}
\caption{Number of answers in the different classes. CQAPri benchmark without priority relation. }\label{cqapri-nb-answers-noprio}
{\small
\begin{tabular*}{\textwidth}{l @{\extracolsep{\fill}}  r r r r | r r r r  }
\toprule
& \multicolumn{4}{c}{\mn{u1c1}}& \multicolumn{4}{c}{\mn{u20c1}}\\
& IAR & AR$\backslash$IAR & brave$\backslash$AR & not brave& IAR & AR$\backslash$IAR & brave$\backslash$AR & not brave\\
\midrule
\mn{q1} & 20,029 & 0 & 380 & 0& 532,331 & 0 & 12,776 & 0\\
\mn{q2} & 7,215 & 20 & 12 & 0& 189,186 & 228 & 1,019 & 0\\
\mn{q3} & 85 & 0 & 0 & 0& 85 & 0 & 0 & 0\\
\mn{q4} & 78,101 & 0 & 5,636 & 0& 1,123,422 & 0 & 944,364 & 0\\
\mn{q5} & 10 & 0 & 0 & 0& 10 & 0 & 0 & 0\\
\mn{q6} & 235 & 0 & 0 & 0& 235 & 0 & 0 & 0\\
\bottomrule
\end{tabular*}
}
\end{table}

\begin{table}
\caption{Number of answers in the different classes. CQAPri benchmark with optimal repairs for a score-structured priority relation. }\label{cqapri-nb-answers-score}
{\small
\begin{tabular*}{\textwidth}{l @{\extracolsep{\fill}} r r r r r | r r r r r }
\toprule
& \multicolumn{5}{c}{\mn{u1c1}}& \multicolumn{5}{c}{\mn{u20c1}}\\
& Triv.& IAR$\backslash$Triv. & AR$\backslash$IAR & brave$\backslash$AR & not brave& Triv.& IAR$\backslash$Triv. & AR$\backslash$IAR & brave$\backslash$AR & not brave\\
\midrule
\mn{q1} & 20,159 & 103 & 0 & 147 & 0& 537,616 & 4,842 & 0 & 2,597 & 52\\
\mn{q2} & 7,237 & 0 & 0 & 4 & 6 & 189,730 & 271 & 35 & 192 & 205\\
\mn{q3} & 85 & 0 & 0 & 0 & 0 & 85 & 0 & 0 & 0 & 0\\
\mn{q4} & 81,691 & 1,087 & 0 & 735 & 224& 1,458,368 & 411,269 & 0 & 195,919 & 2,230\\
\mn{q5} & 10 & 0 & 0 & 0 & 0 & 10 & 0 & 0 & 0 & 0\\
\mn{q6} & 235 & 0 & 0 & 0 & 0& 235 & 0 & 0 & 0 & 0\\
\bottomrule
\end{tabular*}
}

\end{table}

\begin{table}
\caption{Number of answers in the different classes. CQAPri benchmark with a non score-structured priority relation. Pareto-optimal repairs.}\label{cqapri-nb-answers-nonscore-pareto}
{\small
\begin{tabular*}{\textwidth}{l @{\extracolsep{\fill}} r r r r r | r r r r r }
\toprule
& \multicolumn{5}{c}{\mn{u1c1}}& \multicolumn{5}{c}{\mn{u20c1}}\\
& Triv.& IAR$\backslash$Triv. & AR$\backslash$IAR & brave$\backslash$AR & not brave& Triv.& IAR$\backslash$Triv. & AR$\backslash$IAR & brave$\backslash$AR & not brave\\
\midrule
\mn{q1} & 20,187 & 222 & 0 & 0 & 0 & 537,105 & 7,905 & 0 & 89 & 8\\
\mn{q2} & 7,242 & 0 & 0 & 0 & 5 & 189,733 & 446 & 16 & 51 & 187\\
\mn{q3} & 85 & 0 & 0 & 0 & 0& 85 & 0 & 0 & 0 & 0\\
\mn{q4} & 78,477 & 3,282 & 0 & 1,978 & 0& 1,298,399 & 766,010 & 0 & 2,861 & 516\\
\mn{q5} & 10 & 0 & 0 & 0 & 0& 10 & 0 & 0 & 0 & 0\\
\mn{q6} & 235 & 0 & 0 & 0 & 0 & 235 & 0 & 0 & 0 & 0\\
\bottomrule
\end{tabular*}
}
\end{table}
\begin{table}
\caption{Number of answers in the different classes. CQAPri benchmark with a non score-structured priority relation. Completion-optimal repairs.}\label{cqapri-nb-answers-nonscore-completion}
{\small
\begin{tabular*}{\textwidth}{l @{\extracolsep{\fill}} r r r r r | r r r r r }
\toprule
& \multicolumn{5}{c}{\mn{u1c1}}& \multicolumn{5}{c}{\mn{u20c1}}\\
& Triv.& IAR$\backslash$Triv. & AR$\backslash$IAR & brave$\backslash$AR & not brave& Triv.& IAR$\backslash$Triv. & AR$\backslash$IAR & brave$\backslash$AR & not brave\\
\midrule
\mn{q2} & 7,242 & 0 & 0 & 0 & 5&  &  &  &  & \\
\mn{q3} & 85 & 0 & 0 & 0 & 0& 85 & 0 & 0 & 0 & 0\\
\mn{q5} & 10 & 0 & 0 & 0 & 0& 10 & 0 & 0 & 0 & 0\\
\mn{q6} & 235 & 0 & 0 & 0 & 0& 235 & 0 & 0 & 0 & 0\\
\mn{q7} & 137 & 0 & 0 & 0 & 0&  &  &  &  & \\
\mn{q8} & 0 & 0 & 0 & 0 & 0& 10 & 21 & 0 & 0 & 0\\
\bottomrule
\end{tabular*}
}
\end{table}

\begin{table}
\caption{Number of answers in the different classes. Food Inspection dataset without priority relation. }\label{food-nb-answers-noprio}
{\small
\begin{tabular*}{\textwidth}{l @{\extracolsep{\fill}}  r r r r }
\toprule
& IAR & AR$\backslash$IAR & brave$\backslash$AR & not brave \\\midrule
\mn{q2} & 306 & 126 & 18 & 0\\
\mn{q3} & 4 & 7 & 41 & 0\\
\mn{q4} & 24,842 & 109 & 7,046 & 0\\
\mn{q5} & 14,371 & 478 & 2,077 & 0\\
\mn{q6} & 177 & 0 & 885 & 0\\
\bottomrule
\end{tabular*}
}
\end{table}

\begin{table}
\caption{Number of answers in the different classes. Food Inspection dataset with score-structured priorities. }\label{food-nb-answers-score}
\subcaption{Score-structured priority relation ($n=2$). Optimal repairs.}
{\small
\begin{tabular*}{\textwidth}{l @{\extracolsep{\fill}} r r r r r}
\toprule
& Triv.& IAR$\backslash$Triv. & AR$\backslash$IAR & brave$\backslash$AR & not brave\\\midrule
\mn{q2} & 343 & 1 & 92 & 8 & 6\\
\mn{q3} & 8 & 0 & 7 & 30 & 7\\
\mn{q4} & 25,723 & 142 & 60 & 4,256 & 1,816\\
\mn{q5} & 14,944 & 36 & 317 & 1,101 & 528\\
\mn{q6} & 280 & 1 & 0 & 502 & 279\\
\bottomrule
\end{tabular*}
}

\subcaption{Score-structured priority relation ($n=5$). Optimal repairs.}
{\small
\begin{tabular*}{\textwidth}{l @{\extracolsep{\fill}} r r r r r}
\toprule
& Triv.& IAR$\backslash$Triv. & AR$\backslash$IAR & brave$\backslash$AR & not brave\\\midrule
\mn{q2} & 383 & 2 & 51 & 4 & 10\\
\mn{q3} & 15 & 0 & 5 & 18 & 14\\
\mn{q4} & 26,367 & 239 & 27 & 2,098 & 3,266\\
\mn{q5} & 15,376 & 58 & 182 & 472 & 838\\
\mn{q6} & 331 & 1 & 0 & 233 & 497\\
\bottomrule
\end{tabular*}
}
\end{table}

\begin{table}
\caption{Number of answers in the different classes. Food Inspection dataset with non score-structured priorities. In the case of completion-optimal repairs (both for $p=0.5$ and $p=0.8$), all queries run out of time or memory. }\label{food-nb-answers-nonscore}

\subcaption{Priority relation not score-structured ($p=0.5$). Pareto-optimal repairs. }
{\small
\begin{tabular*}{\textwidth}{l @{\extracolsep{\fill}} r r r r r}
\toprule
& Triv.& IAR$\backslash$Triv. & AR$\backslash$IAR & brave$\backslash$AR & not brave\\\midrule
\mn{q2} & 335 & 14 & 86 & 11 & 4\\
\mn{q3} & 6 & 1 & 11 & 23 & 11\\
\mn{q4} & 25,649 & 353 & 60 & 3,517 & 2,418\\
\mn{q5} & 14,919 & 131 & 340 & 864 & 672\\
\mn{q6} & 274 & 13 & 0 & 326 & 449\\
\bottomrule
\end{tabular*}
}

\subcaption{Priority relation not score-structured ($p=0.8$). Pareto-optimal repairs. }
{\small
\begin{tabular*}{\textwidth}{l @{\extracolsep{\fill}} r r r r r}
\toprule
& Triv.& IAR$\backslash$Triv. & AR$\backslash$IAR & brave$\backslash$AR & not brave\\\midrule
\mn{q2} & 357 & 38 & 46 & 3 & 6\\
\mn{q3} & 11 & 8 & 6 & 10 & 17\\
\mn{q4} & 26,170 & 570 & 23 & 1,868 & 3,366\\
\mn{q5} & 15,262 & 231 & 167 & 388 & 878\\
\mn{q6} & 306 & 31 & 0 & 172 & 553\\
\bottomrule
\end{tabular*}
}
\end{table}

\begin{table}
\caption{Number of answers in the different classes. Physicians dataset without priority relation. }\label{physicians-nb-answers-noprio}
{\small
\begin{tabular*}{\textwidth}{l @{\extracolsep{\fill}}  r r r r}
\toprule
& IAR & AR$\backslash$IAR & brave$\backslash$AR & not brave\\
\midrule
\mn{q1} & 1,655 & 7 & 77 & 0\\
\mn{q2} & 290 & 20 & 7 & 0\\
\mn{q3} & 16,827 & 854 & 16 & 0\\
\mn{q4} & 326,021 & 9 & 6,371 & 12\\
\mn{q5} & 3,056 & 114 & 18 & 0\\
\mn{q6} & 144,731 & 24,345 & 946 & 0\\
\bottomrule
\end{tabular*}
}
\end{table}

\begin{table}
\caption{Number of answers in the different classes. Physicians dataset with score-structured priorities. }\label{physicians-nb-answers-score}
\subcaption{Score-structured priority relation ($n=2$). Optimal repairs.}
{\small
\begin{tabular*}{\textwidth}{l @{\extracolsep{\fill}} r r r r r}
\toprule
& Triv.& IAR$\backslash$Triv. & AR$\backslash$IAR & brave$\backslash$AR & not brave\\\midrule
\mn{q1} & 1,668 & 0 & 7 & 48 & 16\\
\mn{q2} & 300 & 0 & 11 & 5 & 1\\
\mn{q3} & 17,239 & 2 & 445 & 9 & 2\\
\mn{q4} & 326,380 & 494 & 85 & 4,900 & 554\\
\mn{q5} & 3,109 & 0 & 65 & 12 & 2\\
\mn{q6} & 156,267 & 2 & 12,877 & 876 & 0\\
\bottomrule
\end{tabular*}
}

\subcaption{Score-structured priority relation ($n=5$). Optimal repairs.}
{\small
\begin{tabular*}{\textwidth}{l @{\extracolsep{\fill}} r r r r r}
\toprule
& Triv.& IAR$\backslash$Triv. & AR$\backslash$IAR & brave$\backslash$AR & not brave\\
\midrule
\mn{q1} & 1,679 & 1 & 7 & 29 & 23\\
\mn{q2} & 311 & 0 & 3 & 3 & 0\\
\mn{q3} & 17,506 & 2 & 181 & 5 & 3\\
\mn{q4} & 327,641 & 861 & 85 & 2,630 & 1,196\\
\mn{q5} & 3,145 & 0 & 39 & 0 & 4\\
\mn{q6} & 163,722 & 113 & 5,545 & 498 & 144\\
\bottomrule
\end{tabular*}
}
\end{table}

\begin{table}
\caption{Number of answers in the different classes. Physicians dataset with non score-structured priorities. }\label{physicians-nb-answers-nonscore}

\subcaption{Priority relation not score-structured ($p=0.5$). Pareto-optimal repairs. }
{\small
\begin{tabular*}{\textwidth}{l @{\extracolsep{\fill}} r r r r r}
\toprule
& Triv.& IAR$\backslash$Triv. & AR$\backslash$IAR & brave$\backslash$AR & not brave\\\midrule
\mn{q1} & 1,671 & 1 & 7 & 45 & 15\\
\mn{q2} & 303 & 0 & 10 & 1 & 3\\
\mn{q3} & 17,257 & 19 & 412 & 5 & 4\\
\mn{q4} & 326,973 & 1,134 & 85 & 3,643 & 578\\
\mn{q5} & 3,111 & 0 & 59 & 10 & 8\\
\mn{q6} & 156,422 & 797 & 12,013 & 671 & 119\\
\bottomrule
\end{tabular*}
}

\subcaption{Priority relation not score-structured ($p=0.5$). Completion-optimal repairs. }
{\small
\begin{tabular*}{\textwidth}{l @{\extracolsep{\fill}} r r r r r}
\toprule
& Triv.& IAR$\backslash$Triv. & AR$\backslash$IAR & brave$\backslash$AR & not brave\\\midrule
\mn{q1} & 1,671 & 23 & 0 & 30 & 15\\
\mn{q2} & 303 & 0 & 10 & 1 & 3\\
\mn{q3} & 17,257 & 19 & 412 & 5 & 4\\
\mn{q4} & 326,973 &  &  &  & \\
\mn{q5} & 3,111 & 0 & 59 & 10 & 8\\
\mn{q6} & 156,422 & 1,277 &  &  & \\
\bottomrule
\end{tabular*}
}

\subcaption{Priority relation not score-structured ($p=0.8$). Pareto-optimal repairs. }
{\small
\begin{tabular*}{\textwidth}{l @{\extracolsep{\fill}} r r r r r}
\toprule
& Triv.& IAR$\backslash$Triv.  & AR$\backslash$IAR & brave$\backslash$AR & not brave\\
\midrule
\mn{q1} & 1,680 & 5 & 7 & 25 & 22\\
\mn{q2} & 305 & 0 & 7 & 1 & 4\\
\mn{q3} & 17,503 & 32 & 158 & 2 & 2\\
\mn{q4} & 326,823 & 874 & 85 & 3,505 & 1,126\\
\mn{q5} & 3,154 & 1 & 27 & 4 & 2\\
\mn{q6} & 163,563 & 1,009 & 4,778 & 661 & 11\\
\bottomrule
\end{tabular*}
}

\subcaption{Priority relation not score-structured ($p=0.8$). Completion-optimal repairs. }
{\small
\begin{tabular*}{\textwidth}{l @{\extracolsep{\fill}} r r r r r}
\toprule
& Triv.& IAR$\backslash$Triv.  & AR$\backslash$IAR & brave$\backslash$AR & not brave\\
\midrule
\mn{q1} & 1,680 & 24 & 0 & 13 & 22\\
\mn{q2} & 305 & 0 & 7 & 1 & 4\\
\mn{q3} & 17,503 & 32 & 158 & 2 & 2\\
\mn{q4} & 326,823 &  &  &  & \\
\mn{q5} & 3,154 & 1 & 27 & 4 & 2\\
\mn{q6} & 163,563 & 1,358 &  &  & \\
\bottomrule
\end{tabular*}
}
\end{table}

\clearpage
\subsection{Running Times of All Algorithms and Encodings for Each Semantics}

We present for all semantics the total query answers filtering times for each possible combination of algorithm and encoding. 
These times include the time needed to handle the self-inconsistent facts, the answers that are trivially IAR, and filter the remaining answers with the considered algorithm and encoding. 
For each dataset and query, the best time is written in bold red and the `close to best times', \ie times that are not longer than the best time by more than 50ms or by more than 10\% of the best time, are highlighted with a grey background. 

For the CQAPri benchmark, we present the results for the four datasets \mn{u1c1}, \mn{u1c50}, \mn{u20c1} and \mn{u20c50}, which have the smallest and biggest sizes and smallest and highest proportion of facts involved in some conflict.

\input{appendix-expe-time-tables}

\clearpage
\subsection{Semantics Comparison w.r.t. Computation Time}
The figures below compare the best running times (among all possible algorithms and encodings) for each semantics and priority relation (none, score-structured with $n=2$ or $n=5$, not score-structured with $p=0.5$ or $p=0.8$) on two datasets of the CQAPri benchmark as well as on the Food Inspection and Physicians datasets. 
The lower parts of bars correspond to the times for handling self-inconsistent facts and trivial answers. The upper parts  correspond to the remaining times (\ie filtering answers that are not trivially true). An empty bar means that the query run out of time or memory for all possible algorithms and encodings.

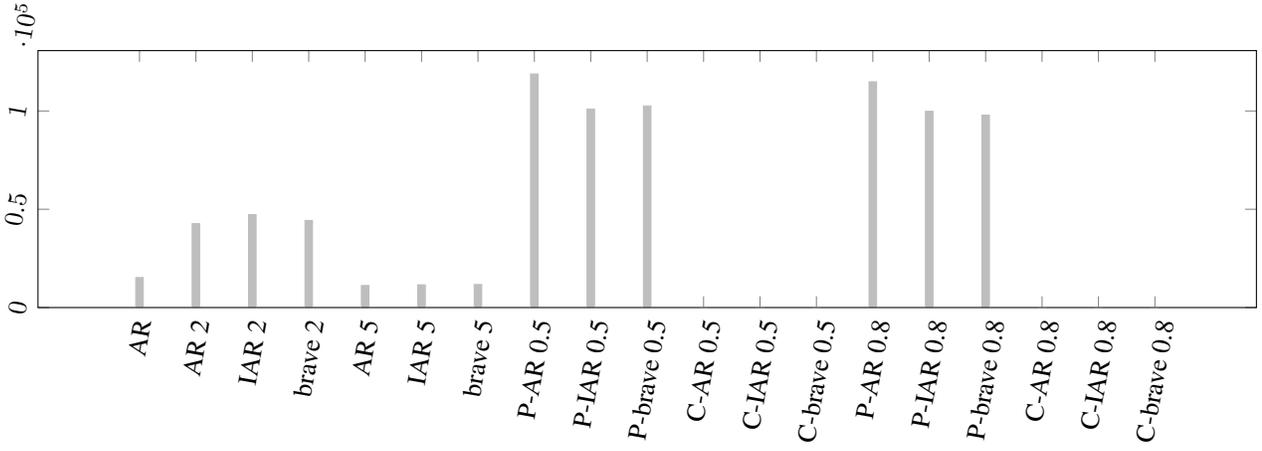
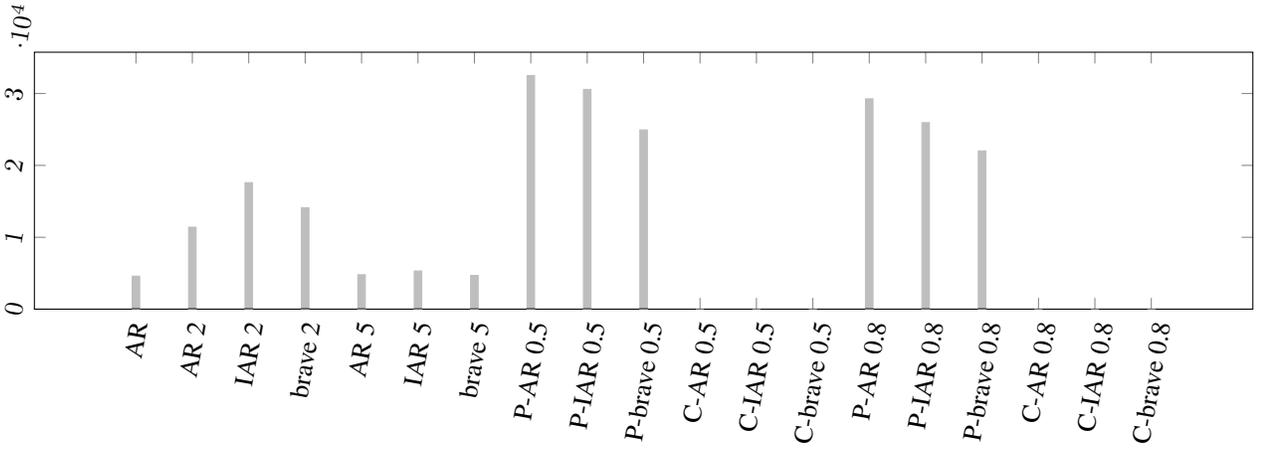
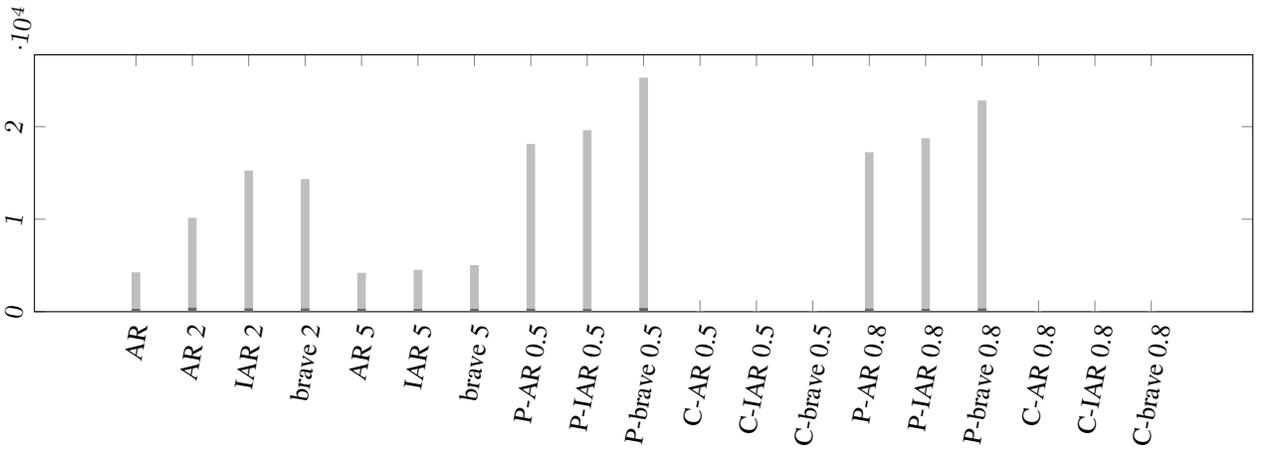
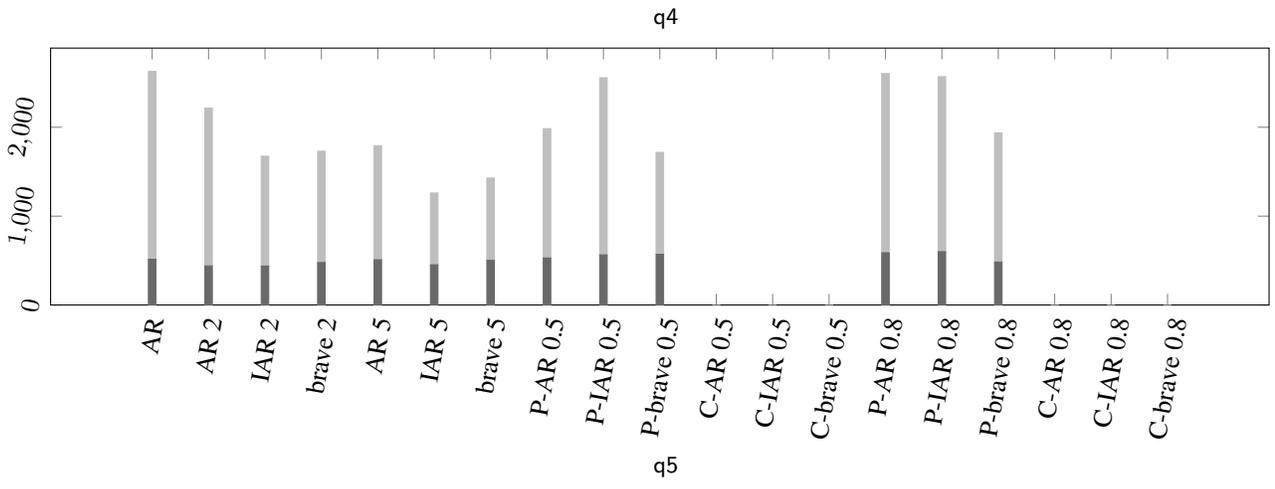
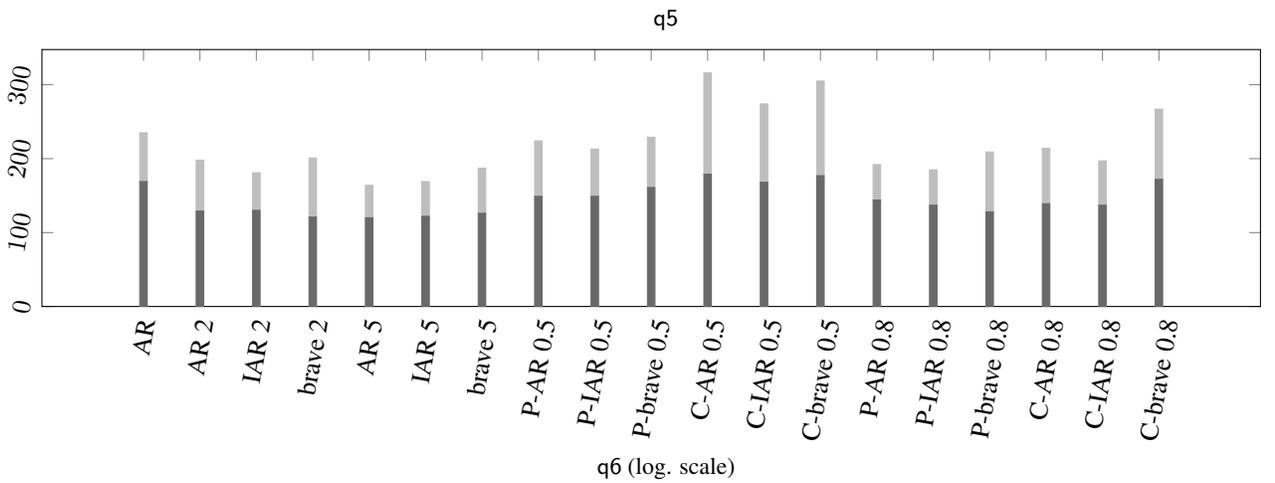
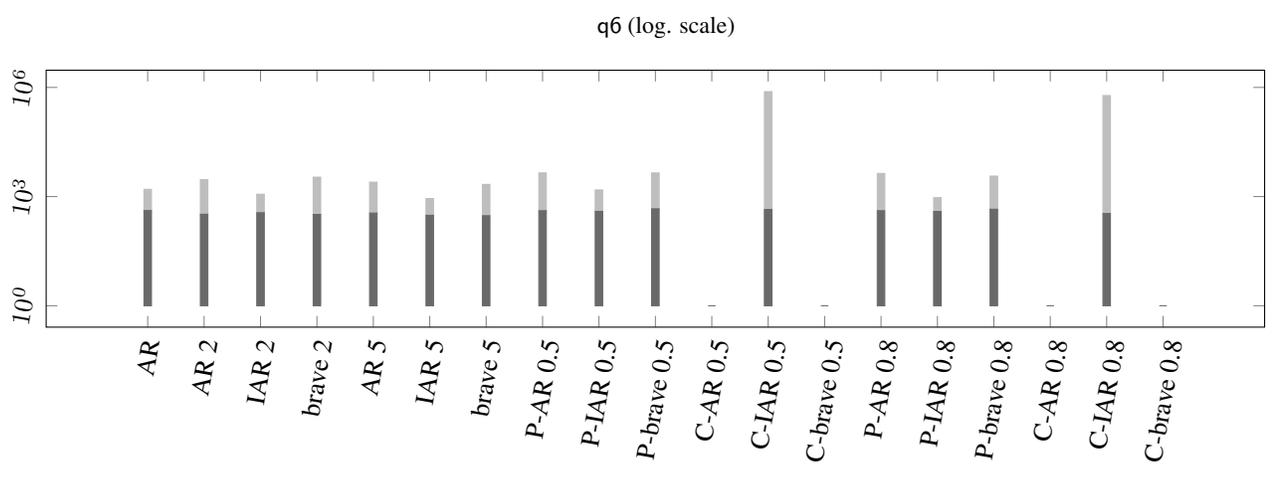
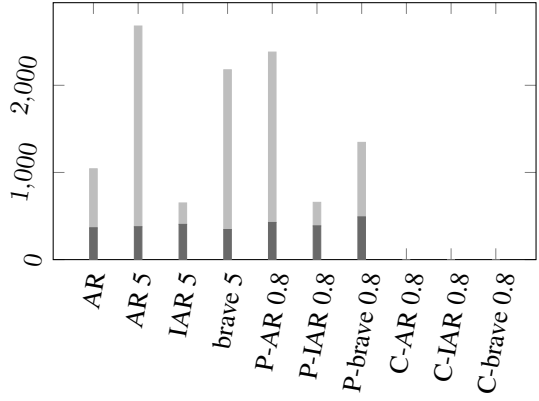
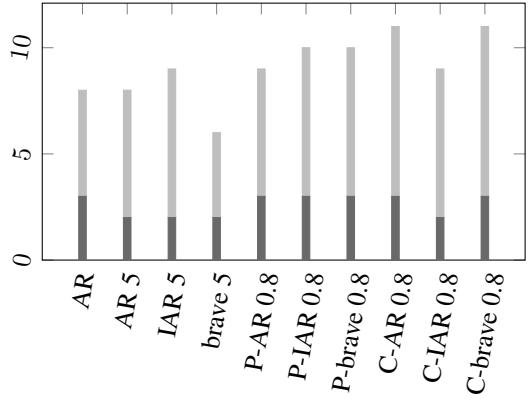
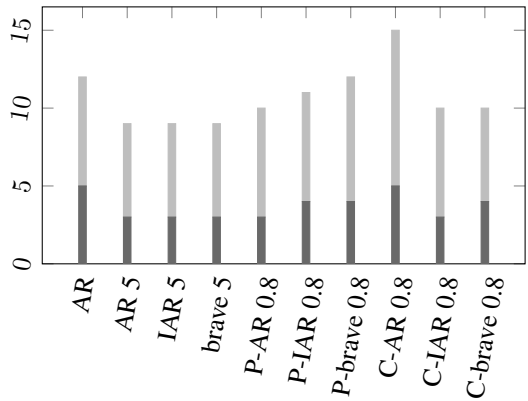
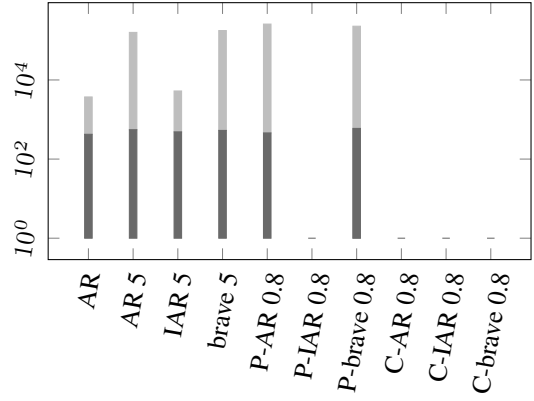
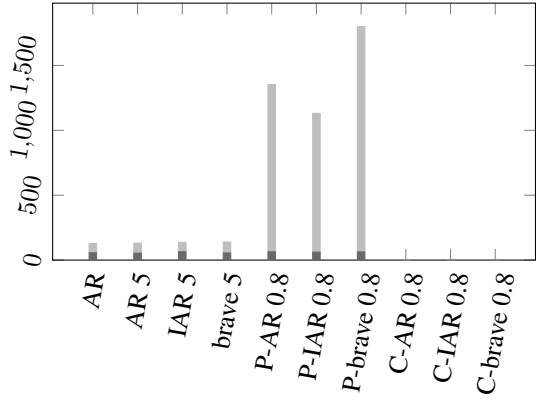
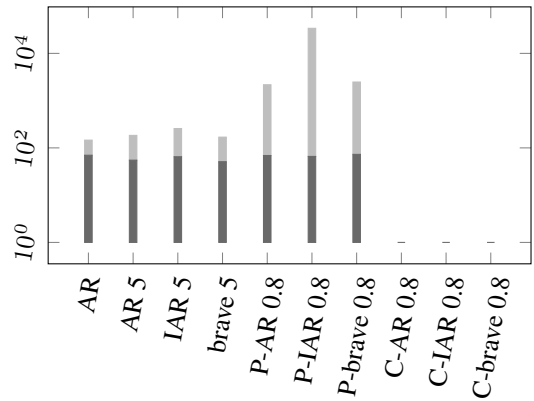
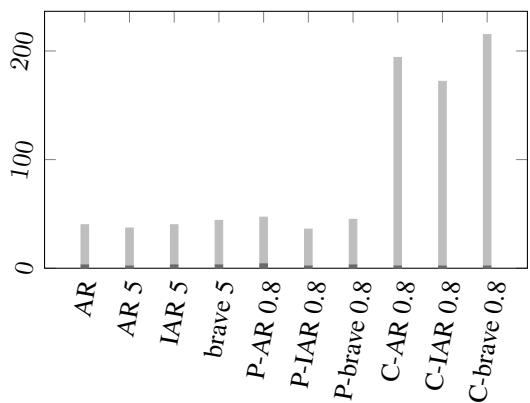
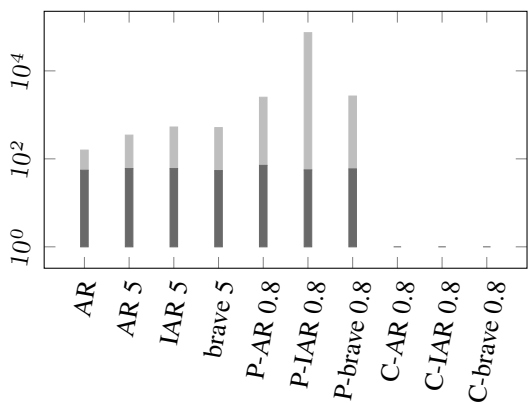
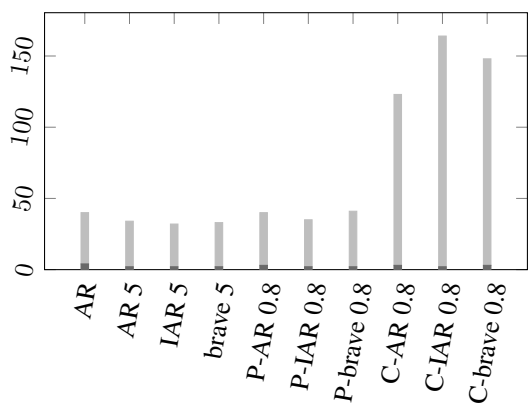
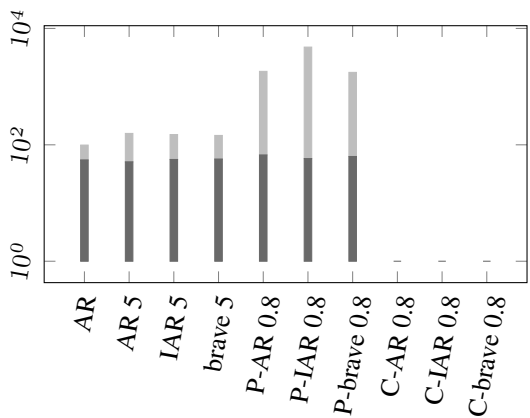
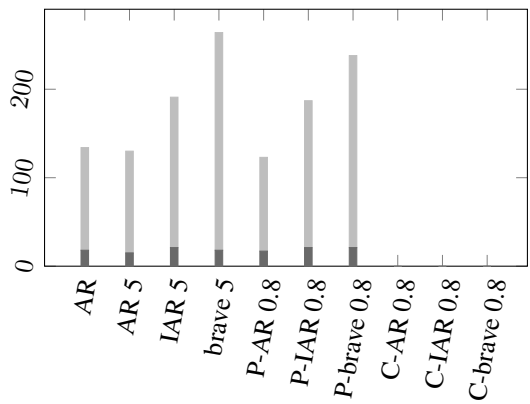
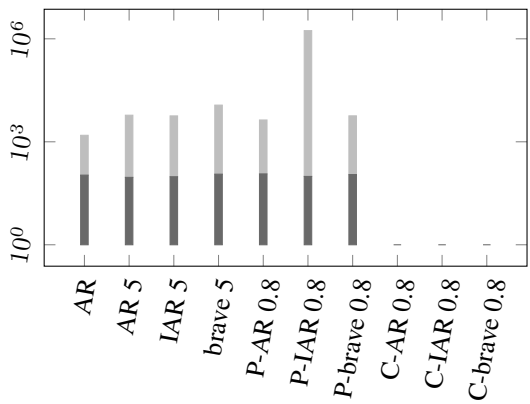
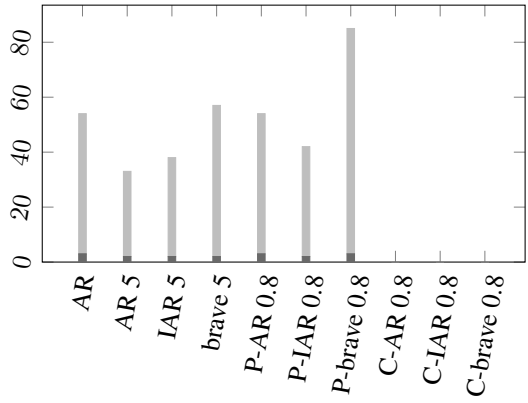
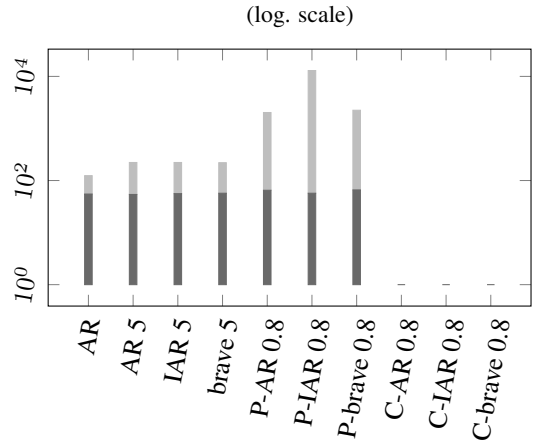
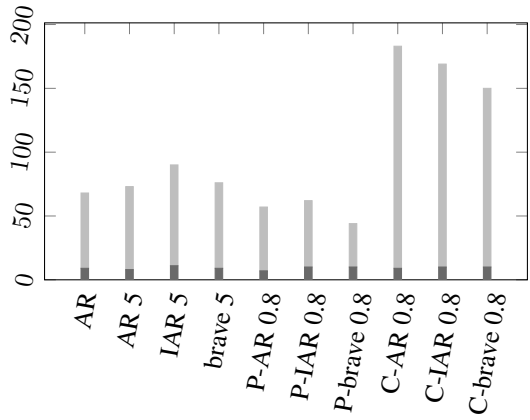
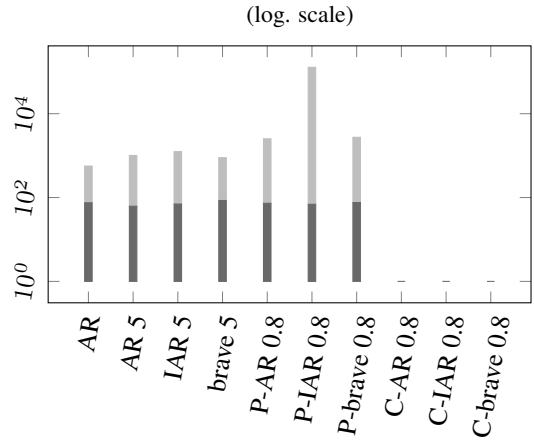
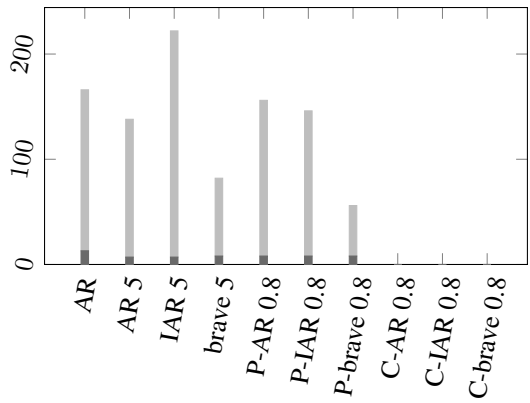
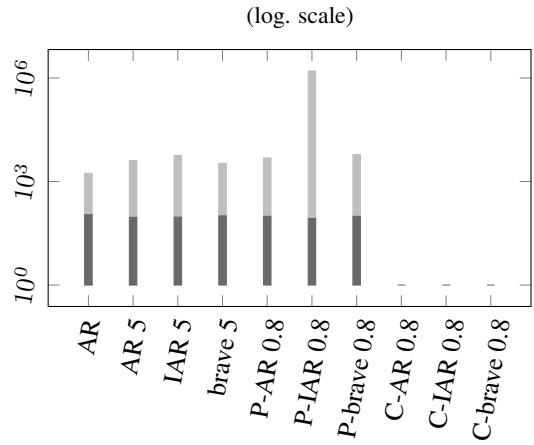
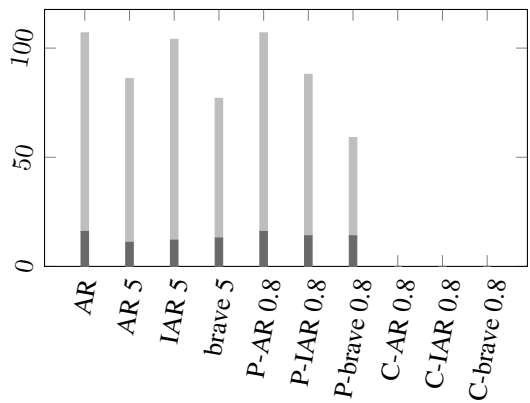
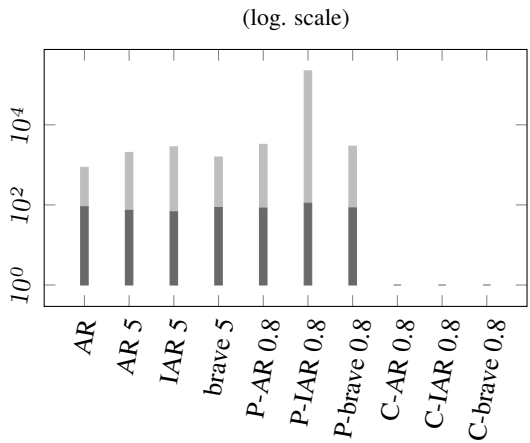
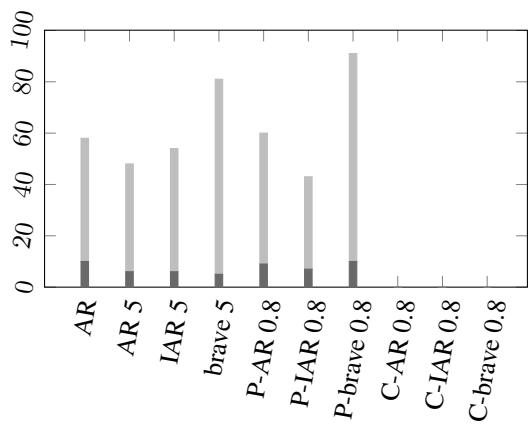
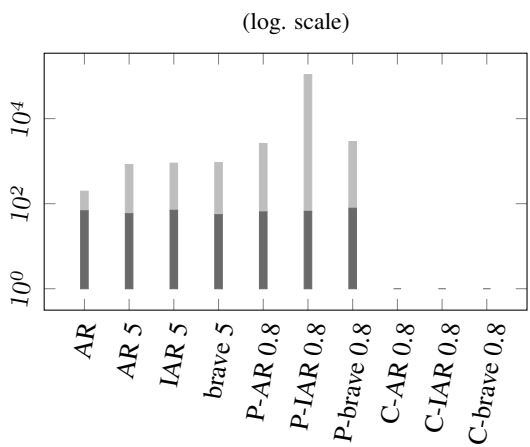
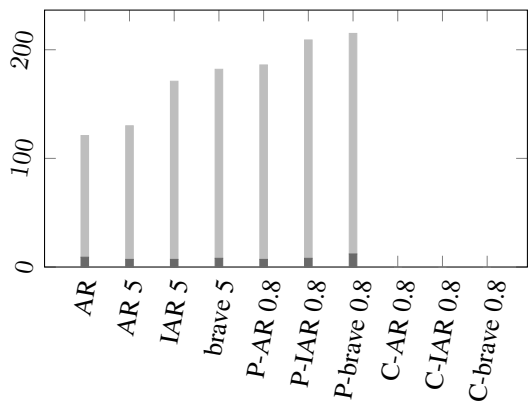
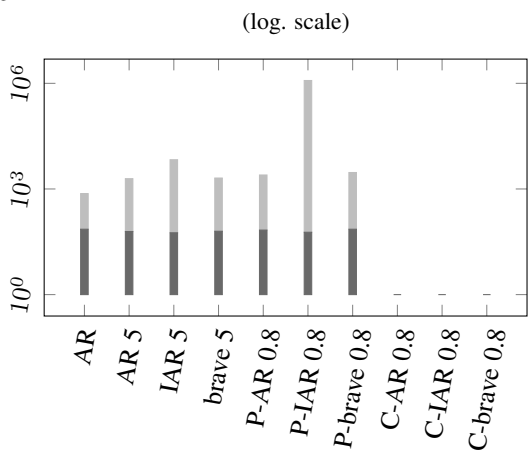
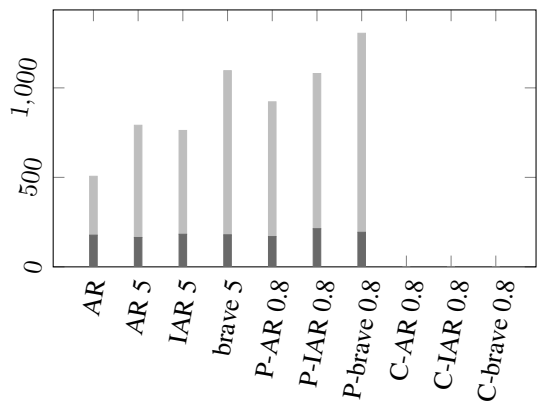
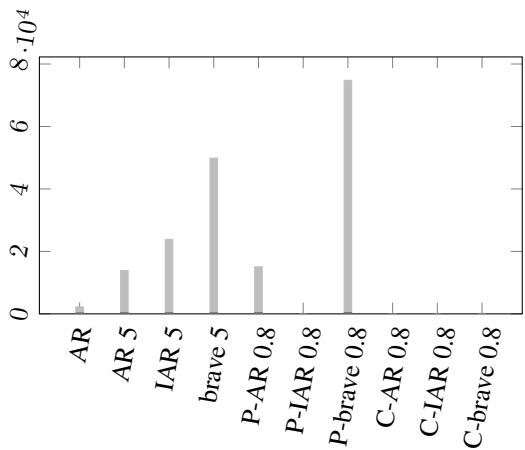
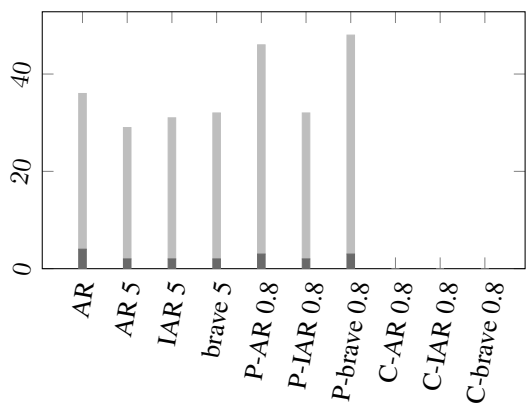
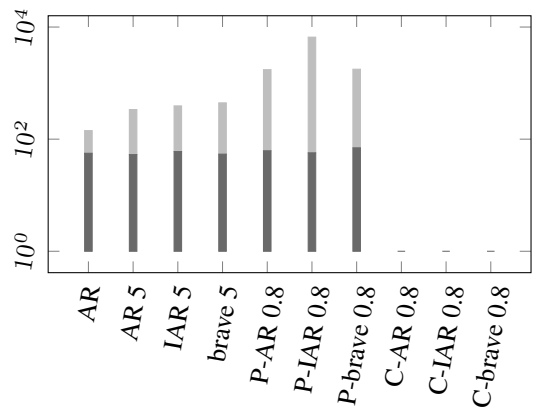
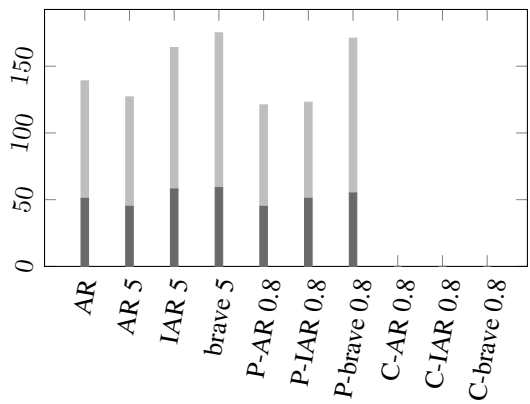
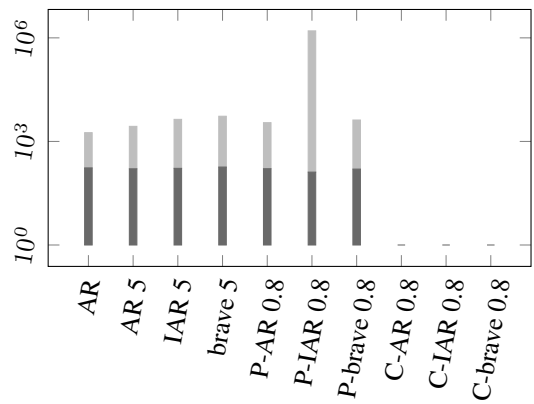
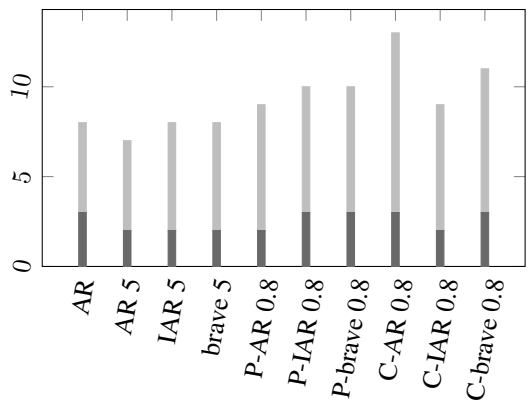
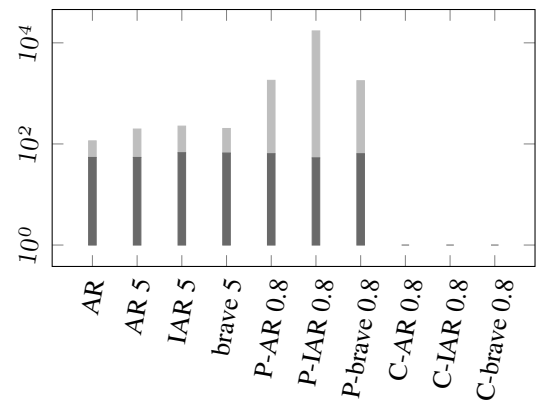
\input{appendix-expe-semantics-comp}

\clearpage
\subsection{Running Times \wrt the Proportion of Facts Involved in Some Conflicts or Size on CQAPri Benchmark}

In this section we consider the evolution of running time when the proportion of facts involved in some conflict or the data size grows for the CQAPri benchmark. We fix one parameter $X$ or $Y$ in \mn{uXcY} and look at the running time evolution when the other parameter increases. Remember that CQAPri benchmark ensures that $\mn{uXcY}\subseteq\mn{uXcY'}$ for $Y\leq Y'$ and $\mn{uXcY}\subseteq\mn{uX'cY}$ for $X\leq X'$. When fixing the data size we choose $X=20$, \ie about 2M facts, and when fixing the proportion of facts in conflicts we choose $Y=20$, \ie about 30\% of facts in conflict. 

We present the results for S-AR, and X-\sem in the case where the priority relation is score-structured, with $\varphi_\mn{P_{1}\text{-}max}$ and \cqaprienc encoding variants.  
For the X-IAR semantics, we choose to consider the three algorithms specific to X-IAR (or X-IAR and X-brave) and only the first generic algorithm \algosat for comparison (it was one of the two methods to compute X-IAR answers under optimal repairs in \cite{DBLP:phd/hal/Bourgaux16}), as these algorithms are not really well-suited for X-IAR.  

For readability, for each semantics and algorithm we first plot all queries, then repeat those with lower running time separately with fitting scales. 
When a query does not occur in the left-most graphic, it means that it runs out of time or memory for all considered datasets.

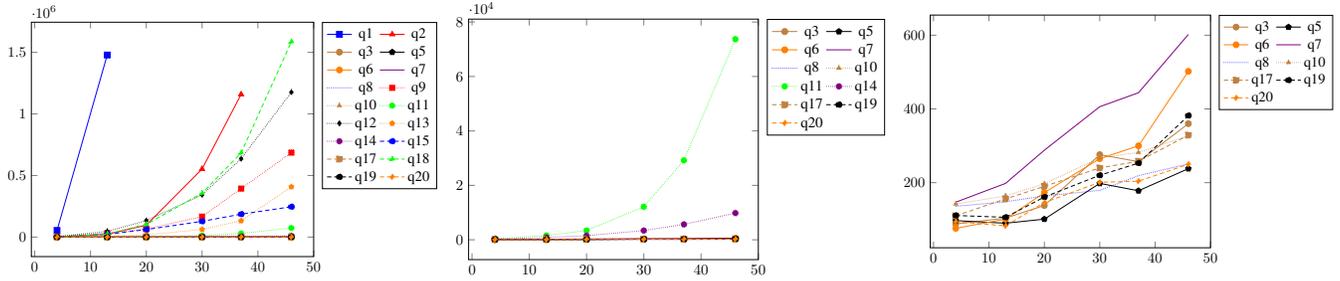
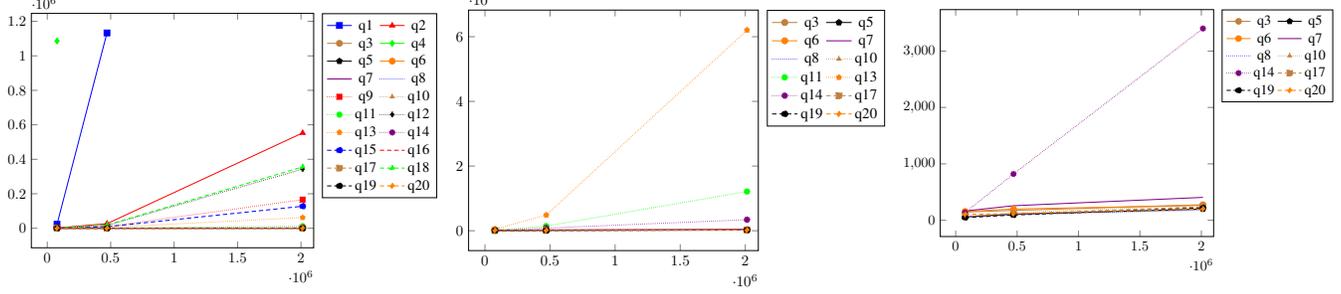
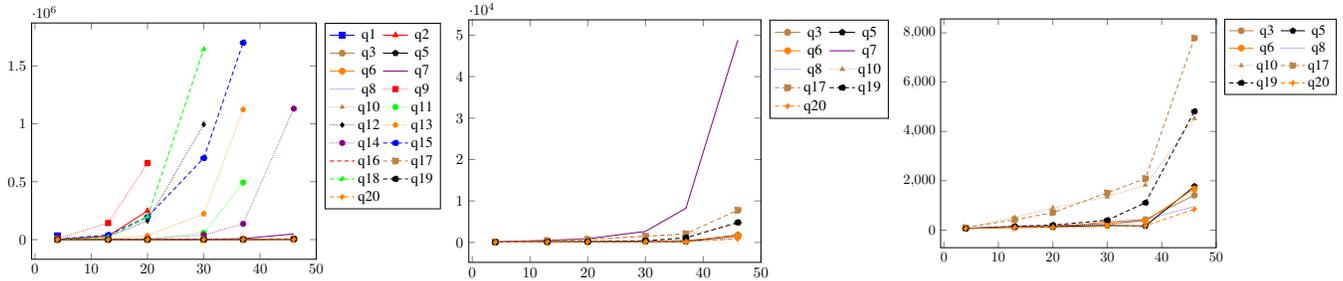
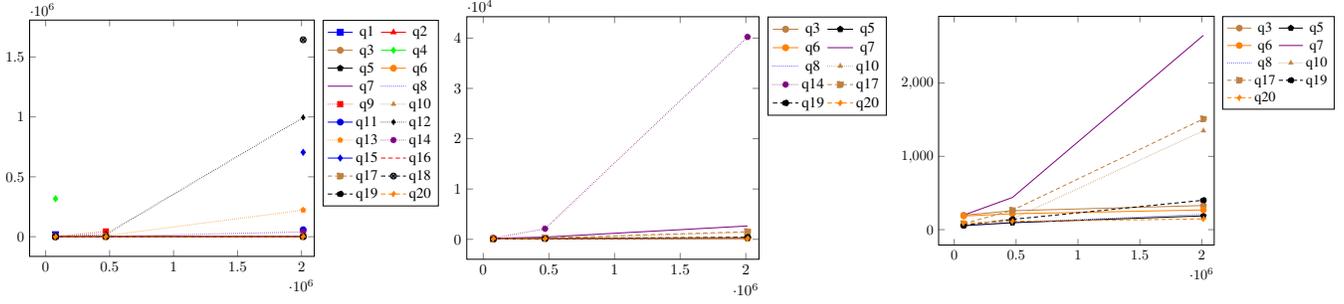
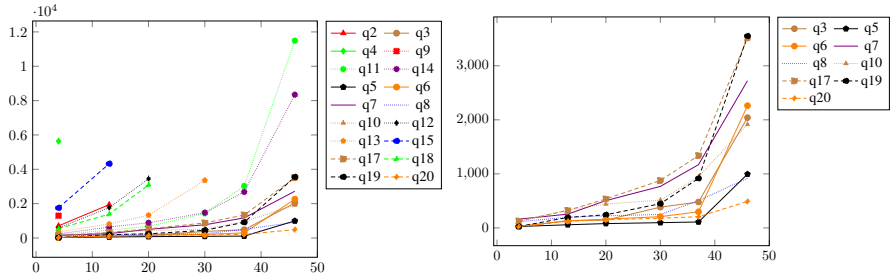
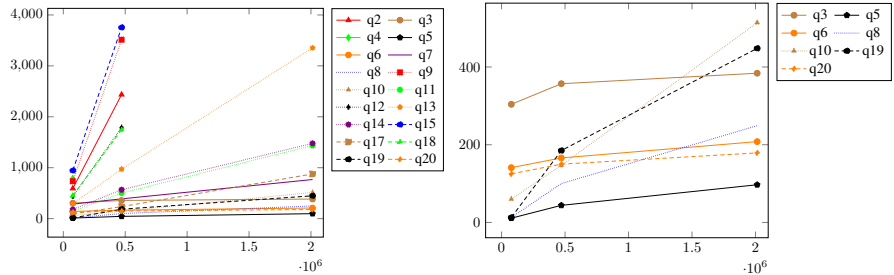
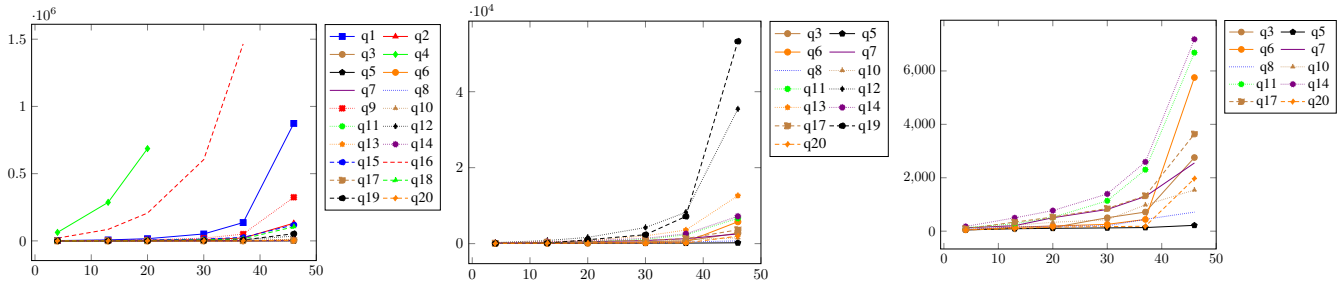
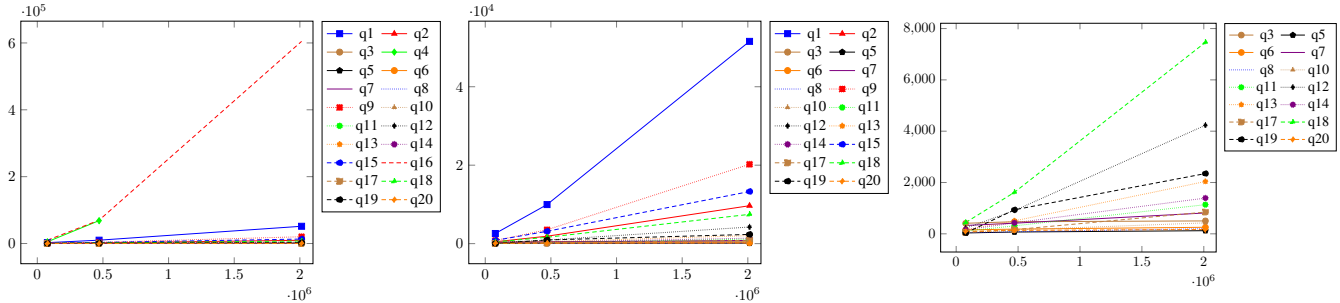
\input{appendix-expe-cqapri-times-AR}
\input{appendix-expe-cqapri-times-IAR}
\input{appendix-expe-cqapri-times-brave}

\clearpage
\subsection{Encodings Properties}

We pick a few queries and look at the size of the encodings build to answer them under the different semantics and encodings. 
For multi-answer encodings, $\Phi_{X\text{-}\sem}(q(\ans))=\Psi_{X\text{-}\sem}(q(\ans))\wedge\bigwedge_{\ans\in\mi{Potans}}x_{\ans}$. 
We did not find any blatant relationship between encoding size and running time, except as completion-optimal repairs are concerned with the blow-up of both encoding sizes and running times. We summarize some observations below.

\paragraph{S-AR v.s.\ P-AR} 
For S-AR, the CQAPri benchmark leads to rather small encodings for single answer, while the two other datasets, and especially the Food Inspection dataset, lead to bigger encodings. 
However, adding priorities increases a lot more the size of the encodings of the CQAPri benchmark (typically several orders of magnitude) than those of the Food Inspection and Physicians datasets, that mostly stay in the same order of magnitude, except for completion-optimal repairs.  
This is probably because of the kind of constraints in the different benchmarks. For the Food Inspection/Physicians datasets (keys and FD), all the facts reachable in the conflict graph from a given fact share the same predicate, and they are often also in conflict with the original fact (in all cases where the only constraint on the considered table is a key constraint). On the contrary, on CQAPri benchmark (concept and role disjointness axioms), the facts reachable from a given fact are often not in conflict with it.

\paragraph{ $\varphi_\mn{P_{1}\text{-}max}$  v.s.\ $\varphi_\mn{P_{2}\text{-}max}$ v.s.\ $\varphi_\mn{C\text{-}max}$ (AR, IAR and brave semantics)} 
\begin{itemize}
\item $\varphi_\mn{C\text{-}max}$ encoding leads to the biggest encodings, by far.
\item Number of variables of $\varphi_\mn{P_{1}\text{-}max}$ v.s.\ $\varphi_\mn{P_{2}\text{-}max}$:
\begin{itemize}
\item With \cavsatenc encoding (AR and IAR semantics), $\varphi_\mn{P_{1}\text{-}max}$  and $\varphi_\mn{P_{2}\text{-}max}$ encodings have exactly the same number of variables. 
\item With \cqaprienc encoding, $\varphi_\mn{P_{2}\text{-}max}$ encodings have less (or as many) variables than $\varphi_\mn{P_{1}\text{-}max}$ encodings. The difference is more important for AR and IAR semantics on CQAPri databases, especially \mn{u1c1}, \mn{u1c50} and \mn{u20c1}.
\end{itemize}
Intuitively $\varphi_\mn{P_{1}\text{-}max}$ encodings consider all facts reachable in the directed conflict graph while $\varphi_\mn{P_{2}\text{-}max}$ encodings consider `every other fact' along a directed conflict chain, which become all facts reachable in the directed conflict graph when we add the cause facts with \cavsatenc encoding. Food Inspection and Physicians datasets are less impacted by this because facts reachable in the conflict graph are often already in conflict with the cause. 
\item Number of clauses of $\varphi_\mn{P_{1}\text{-}max}$  v.s.\ $\varphi_\mn{P_{2}\text{-}max}$ encodings: depends on the dataset.
\begin{itemize}
\item On CQAPri databases with few conflicts (\mn{u1c1} and \mn{u20c1}, and especially on \mn{u1c1}), when \cqaprienc encoding is used (AR, IAR and brave) $\varphi_\mn{P_{1}\text{-}max}$ encodings are generally bigger than $\varphi_\mn{P_{2}\text{-}max}$ encodings, while when \cavsatenc encoding (AR and IAR) is used it is the opposite.

\item On Food Inspection and Physicians datasets and \mn{u20c50}, $\varphi_\mn{P_{2}\text{-}max}$ encodings are generally bigger than $\varphi_\mn{P_{1}\text{-}max}$ encodings.

\item On \mn{u1c50}, it depends on the query. 
\end{itemize}
\end{itemize}

\paragraph{\cqaprienc v.s.\ \cavsatenc encoding (AR and IAR semantics)} 
By construction \cavsatenc encodings are bigger than \cqaprienc encodings. 
The average and maximal numbers of variables and clauses increase more for AR and IAR with $\varphi_\mn{P_{2}\text{-}max}$ encoding than with $\varphi_\mn{P_{1}\text{-}max}$ encoding for which the difference is quite small. The reason is that $\varphi_\mn{P_{1}\text{-}max}$ encoding already considers all facts reachable in the directed conflict graph contrary to $\varphi_\mn{P_{2}\text{-}max}$ encoding, as explained above. 

We observe that the average and maximal numbers of variables and clauses increase in general more in proportion for \mn{u1c1}, \mn{u1c50} and \mn{u20c1} than \mn{u20c50}, Food Inspection and Physicians datasets.


\begin{table}[h]
\caption{Number of variables and clauses in the encoding for a single answer $\Phi_{S\text{-}AR}(q(\ans))$ ($[minimum \mid average \mid maximum]$) and in the encoding for a set of answers $\Phi_{S\text{-}AR}(PotAns)$, with \cqaprienc or \cavsatenc encoding. }
{\small
\setlength{\tabcolsep}{0pt}
\begin{tabular*}{\textwidth}{l @{\extracolsep{\fill}}r r r r r r r r}
\toprule
&  \multicolumn{2}{c}{Single, \cqaprienc}& \multicolumn{2}{c}{Single, \cavsatenc}&  \multicolumn{2}{c}{Set, \cqaprienc}& \multicolumn{2}{c}{Set \cavsatenc}\\
&\#var. & \#cl. &\#var. & \#cl. &\#var. & \#cl. &\#var. & \#cl. \\
\midrule
\mn{u1c1} \mn{q1}& $[1 \mid 1 \mid 3]$ & $[2 \mid 3 \mid 11]$ & $[2 \mid 2 \mid 10]$ & $[5 \mid 8 \mid 34]$ & 407&1,754 & 928 & 3,319\\
\mn{u1c1} \mn{q4}&$[1 \mid 1 \mid 2] $&$ [1 \mid 1 \mid 2]$ & $[2 \mid 2 \mid 4]$ & $[4 \mid 5 \mid 10]$ &5,655 & 11,332& 5,690& 11,437\\
\mn{u1c1} \mn{q7}&$[1 \mid 1 \mid 1]$ & $[1 \mid 1 \mid 1]$ & $[2 \mid 2 \mid 2]$ & $[4 \mid 4 \mid 4]$ & 2 & 2 &3 & 5\\
\mn{u1c1} \mn{q11}&$[1 \mid 1 \mid 2]$ & $[1 \mid 7 \mid 16]$ & $[3 \mid 8 \mid 17]$ & $[6 \mid 29 \mid 64]$ & 9& 33 &38 & 122\\
\mn{u1c1} \mn{q14}& $[1 \mid 1 \mid 1]$ & $[1 \mid 1 \mid 2]$ & $[2 \mid 2 \mid 3]$ & $[5 \mid 7 \mid 8]$ & 8 & 11 &15 &32 \\
\mn{u1c1} \mn{q17}&$[1 \mid 1 \mid 1]$ & $[20 \mid 20 \mid 20]$ & $[5 \mid 5 \mid 5]$ & $[32 \mid 32 \mid 32]$ & 2 & 21&6 & 33\\
\bottomrule
\end{tabular*}
}
\end{table}

\begin{table}
\caption{Number of variables and clauses in the encoding for a single answer $\Phi_{P\text{-}AR}(q(\ans))$ ($[minimum \mid average \mid maximum]$), with \cqaprienc or \cavsatenc encoding, for P$_1$ encoding (score-structured priority, $n=5$). }
{\small
\setlength{\tabcolsep}{0pt}
\begin{tabular*}{\textwidth}{l @{\extracolsep{\fill}}r r r r }
\toprule
&  \multicolumn{2}{c}{\cqaprienc}& \multicolumn{2}{c}{\cavsatenc}\\
&\#var. & \#cl. &\#var. & \#cl. \\
\midrule
\mn{u1c1} \mn{q1}&$[3 \mid 62 \mid 548]$ & $[15 \mid 148 \mid 1,203]$ & $[4 \mid 63 \mid 548]$ & $[16 \mid 151 \mid 1,207]$ \\
\mn{u1c1} \mn{q4}&$[1 \mid 7 \mid 17]$ & $[2 \mid 17 \mid 37]$ & $[2 \mid 7 \mid 17]$ & $[5 \mid 19 \mid 39]$\\
\mn{u1c1} \mn{q7}& \multicolumn{4}{c}{all answers are trivial}\\
\mn{u1c1} \mn{q11}& \multicolumn{4}{c}{all answers are trivial}\\
\mn{u1c1} \mn{q14}&$[8 \mid 12 \mid 17]$ & $[19 \mid 29 \mid 40]$ & $[9 \mid 13 \mid 17]$ & $[22 \mid 32 \mid 41]$ \\
\mn{u1c1} \mn{q17}& \multicolumn{4}{c}{all answers are trivial}\\
\bottomrule
\end{tabular*}
}
\end{table}

\begin{table}
\caption{Number of variables and clauses in the encoding for a set of answers $\Phi_{P\text{-}AR}(PotAns)$, with \cqaprienc or \cavsatenc encoding, for P$_1$ encoding (score-structured priority, $n=5$). }
{\small
\setlength{\tabcolsep}{0pt}
\begin{tabular*}{\textwidth}{l @{\extracolsep{\fill}}r r r r }
\toprule
& \multicolumn{2}{c}{ \cqaprienc}& \multicolumn{2}{c}{\cavsatenc}\\
&\#var. & \#cl. &\#var. & \#cl.  \\
\midrule
\mn{u1c1} \mn{q1}& 1,854 &4,726 & 2,054 & 5,407\\
\mn{u1c1} \mn{q4}& 2,112 & 4,240 &2,118 & 4,268\\
\mn{u1c1} \mn{q7}& \multicolumn{4}{c}{all answers are trivial}\\
\mn{u1c1} \mn{q11}& \multicolumn{4}{c}{all answers are trivial}\\
\mn{u1c1} \mn{q14}& 40 & 90&43 & 100\\
\mn{u1c1} \mn{q17}& \multicolumn{4}{c}{all answers are trivial}\\
\bottomrule
\end{tabular*}
}
\end{table}

\begin{table}
\caption{Number of variables and clauses in the encoding for a single answer $\Phi_{P\text{-}AR}(q(\ans))$ ($[minimum \mid average \mid maximum]$), with \cqaprienc or \cavsatenc encoding, for P$_2$ encoding (score-structured priority, $n=5$). }
{\small
\setlength{\tabcolsep}{0pt}
\begin{tabular*}{\textwidth}{l @{\extracolsep{\fill}}r r r r }
\toprule
&  \multicolumn{2}{c}{\cqaprienc}& \multicolumn{2}{c}{\cavsatenc}\\
&\#var. & \#cl. &\#var. & \#cl. \\
\midrule
\mn{u1c1} \mn{q1}&$[1 \mid 1 \mid 2]$ & $[6 \mid 64 \mid 550]$ & $[4 \mid 63 \mid 548]$ & $[19 \mid 172 \mid 1,363]$ \\
\mn{u1c1} \mn{q4}&$[1 \mid 1 \mid 1]$ & $[1 \mid 7 \mid 17]$ & $[2 \mid 7 \mid 17]$ & $[4 \mid 21 \mid 41]$ \\
\mn{u1c1} \mn{q7}& \multicolumn{4}{c}{all answers are trivial}\\
\mn{u1c1} \mn{q11}& \multicolumn{4}{c}{all answers are trivial}\\
\mn{u1c1} \mn{q14}&$[1 \mid 1 \mid 1]$ & $[9 \mid 13 \mid 18]$ & $[9 \mid 13 \mid 17]$ & $[23 \mid 34 \mid 45]$ \\
\mn{u1c1} \mn{q17}& \multicolumn{4}{c}{all answers are trivial}\\
\bottomrule
\end{tabular*}
}
\end{table}

\begin{table}
\caption{Number of variables and clauses in the encoding for a set of answers $\Phi_{P\text{-}AR}(PotAns)$, with \cqaprienc or \cavsatenc encoding, for P$_2$ encoding (score-structured priority, $n=5$). }
{\small
\setlength{\tabcolsep}{0pt}
\begin{tabular*}{\textwidth}{l @{\extracolsep{\fill}}r r r r }
\toprule
&  \multicolumn{2}{c}{ \cqaprienc}& \multicolumn{2}{c}{\cavsatenc}\\
&\#var. & \#cl. &\#var. & \#cl. \\
\midrule
\mn{u1c1} \mn{q1}& 272 & 2,663&2,054 & 5,844\\
\mn{u1c1} \mn{q4}& 2,057 & 4,147&2,118 & 4,284\\
\mn{u1c1} \mn{q7}& \multicolumn{4}{c}{all answers are trivial}\\
\mn{u1c1} \mn{q11}& \multicolumn{4}{c}{all answers are trivial}\\
\mn{u1c1} \mn{q14}& 6 & 43&43 & 107\\
\mn{u1c1} \mn{q17}& \multicolumn{4}{c}{all answers are trivial}\\
\bottomrule
\end{tabular*}
}
\end{table}

\begin{table}
\caption{Number of variables and clauses in the encoding for a single answer $\Phi_{C\text{-}AR}(q(\ans))$ ($[minimum \mid average \mid maximum]$), with \cqaprienc or \cavsatenc encoding (score-structured priority, $n=5$). }
{\small
\setlength{\tabcolsep}{0pt}
\begin{tabular*}{\textwidth}{l @{\extracolsep{\fill}}r r r r}
\toprule
&  \multicolumn{2}{c}{\cqaprienc}& \multicolumn{2}{c}{\cavsatenc}\\
&\#var. & \#cl. &\#var. & \#cl.\\
\midrule
\mn{u1c1} \mn{q1}&\multicolumn{4}{c}{time-out or out-of-memory}\\
\mn{u1c1} \mn{q4}&$[1 \mid 177 \mid 580]$ & $[2 \mid 1,251 \mid 4,750]$ & $[7 \mid 185 \mid 580]$ & $[14 \mid 1,281 \mid 4,752]$ \\
\mn{u1c1} \mn{q7}&\multicolumn{4}{c}{all answers are trivial}\\
\mn{u1c1} \mn{q11}&\multicolumn{4}{c}{all answers are trivial}\\
\mn{u1c1} \mn{q14}&$[129 \mid 333 \mid 582]$ & $[508 \mid 2,314 \mid 4,759]$ & $[163 \mid 379 \mid 582]$ & $[716 \mid 2,710 \mid 4,760]$ \\
\mn{u1c1} \mn{q17}&\multicolumn{4}{c}{all answers are trivial}\\
\bottomrule
\end{tabular*}
}
\end{table}
\begin{table}
\caption{Number of variables and clauses in the encoding for a set of answers $\Phi_{C\text{-}AR}(PotAns)$, with \cqaprienc or \cavsatenc encoding (score-structured priority, $n=5$). }
{\small
\setlength{\tabcolsep}{0pt}
\begin{tabular*}{\textwidth}{l @{\extracolsep{\fill}}r r r r }
\toprule
& \multicolumn{2}{c}{\cqaprienc}& \multicolumn{2}{c}{\cavsatenc}\\&\#var. & \#cl. &\#var. & \#cl. \\
\midrule
\mn{u1c1} \mn{q1}&\multicolumn{4}{c}{time-out or out-of-memory}\\
\mn{u1c1} \mn{q4} &10,774 &288,695& 12,430 & 373,677\\
\mn{u1c1} \mn{q7}&\multicolumn{4}{c}{all answers are trivial}\\
\mn{u1c1} \mn{q11}&\multicolumn{4}{c}{all answers are trivial}\\
\mn{u1c1} \mn{q14}& 2,748 & 49,574 &3,210 &62,715\\
\mn{u1c1} \mn{q17}&\multicolumn{4}{c}{all answers are trivial}\\
\bottomrule
\end{tabular*}
}
\end{table}

\begin{table}
\caption{Number of variables and clauses in the encoding for a single answer $\Phi_{P\text{-}brave}(q(\ans))$ ($[minimum \mid average \mid maximum]$) and in the encoding for a set of answers $\Phi_{P\text{-}brave}(PotAns)$, for P$_1$ encoding (score-structured priority, $n=5$). }
{\small
\setlength{\tabcolsep}{0pt}
\begin{tabular*}{\textwidth}{l @{\extracolsep{\fill}}r r r r}
\toprule
&  \multicolumn{2}{c}{Single}&   \multicolumn{2}{c}{Set}\\
&\#var. & \#cl. &\#var. & \#cl. \\
\midrule
\mn{u1c1} \mn{q1}&$[7 \mid 66 \mid 552]$ & $[16 \mid 151 \mid 1,207]$ & 2,885 & 5,407\\
\mn{u1c1} \mn{q4}&$[3 \mid 8 \mid 18]$ & $[5 \mid 19 \mid 39]$ & 4,164 & 6,742\\
\mn{u1c1} \mn{q7}&\multicolumn{4}{c}{all answers are trivial}\\
\mn{u1c1} \mn{q11}&\multicolumn{4}{c}{all answers are trivial}\\
\mn{u1c1} \mn{q14}&$[11 \mid 15 \mid 19]$ & $[22 \mid 32 \mid 41]$ &49 & 101 \\
\mn{u1c1} \mn{q17}&\multicolumn{4}{c}{all answers are trivial}\\
\bottomrule
\end{tabular*}
}
\end{table}

\begin{table}
\caption{Number of variables and clauses in the encoding for a single answer $\Phi_{P\text{-}brave}(q(\ans))$ ($[minimum \mid average \mid maximum]$) and in the encoding for a set of answers $\Phi_{P\text{-}brave}(PotAns)$, for P$_2$ encoding (score-structured priority, $n=5$). }
{\small
\setlength{\tabcolsep}{0pt}
\begin{tabular*}{\textwidth}{l @{\extracolsep{\fill}}r r r r}
\toprule
&  \multicolumn{2}{c}{Single}&   \multicolumn{2}{c}{Set}\\
&\#var. & \#cl. &\#var. & \#cl. \\
\midrule
\mn{u1c1} \mn{q1}&$[6 \mid 65 \mid 550]$ & $[6 \mid 26 \mid 166]$ &2,863 & 2,021 \\
\mn{u1c1} \mn{q4}&$[2 \mid 7 \mid 17]$ & $[3 \mid 5 \mid 10]$ & 4,153 & 6,615\\
\mn{u1c1} \mn{q7}&\multicolumn{4}{c}{all answers are trivial}\\
\mn{u1c1} \mn{q11}&\multicolumn{4}{c}{all answers are trivial}\\
\mn{u1c1} \mn{q14}&$[10 \mid 14 \mid 18]$ & $[6 \mid 8 \mid 10]$ &46 & 27 \\
\mn{u1c1} \mn{q17}&\multicolumn{4}{c}{all answers are trivial}\\
\bottomrule
\end{tabular*}
}
\end{table}

\begin{table}
\caption{Number of variables and clauses in the encoding for a single answer $\Phi_{C\text{-}brave}(q(\ans))$ ($[minimum \mid average \mid maximum]$) and in the encoding for a set of answers $\Phi_{C\text{-}brave}(PotAns)$ (score-structured priority, $n=5$). }
{\small
\setlength{\tabcolsep}{0pt}
\begin{tabular*}{\textwidth}{l @{\extracolsep{\fill}}r r r r}
\toprule
&  \multicolumn{2}{c}{Single}&   \multicolumn{2}{c}{Set}\\
&\#var. & \#cl. &\#var. & \#cl. \\
\midrule
\mn{u1c1} \mn{q1}&\multicolumn{4}{c}{time-out or out-of-memory} \\
\mn{u1c1} \mn{q4}&$[8 \mid 186 \mid 581]$ & $[14 \mid 1,281 \mid 4,752]$ &14,476 & 375,192 \\
\mn{u1c1} \mn{q7}&\multicolumn{4}{c}{all answers are trivial}\\
\mn{u1c1} \mn{q11}&\multicolumn{4}{c}{all answers are trivial}\\
\mn{u1c1} \mn{q14}&$[165 \mid 381 \mid 584]$ & $[716 \mid 2,711 \mid 4,760]$ & 3,216 & 62,716 \\
\mn{u1c1} \mn{q17}&\multicolumn{4}{c}{all answers are trivial}\\
\bottomrule
\end{tabular*}
}
\end{table}

\begin{table}
\caption{Number of variables and clauses in the encoding for a single cause $\Phi_{P\text{-}IAR}(\Cmc)$ ($[minimum \mid average \mid maximum]$), with \cqaprienc or \cavsatenc encoding, for P$_1$ encoding (score-structured priority, $n=5$). }
{\small
\setlength{\tabcolsep}{0pt}
\begin{tabular*}{\textwidth}{l @{\extracolsep{\fill}}r r r r}
\toprule
&  \multicolumn{2}{c}{\cqaprienc}& \multicolumn{2}{c}{\cavsatenc}\\
&\#var. & \#cl. &\#var. & \#cl. \\
\midrule
\mn{u1c1} \mn{q1}&$[3 \mid 26 \mid 521]$ & $[8 \mid 63 \mid 1,199]$ & $[4 \mid 27 \mid 523]$ & $[11 \mid 65 \mid 1,205]$  \\
\mn{u1c1} \mn{q4}&$[1 \mid 7 \mid 17]$ & $[2 \mid 17 \mid 37]$ & $[2 \mid 7 \mid 17]$ & $[5 \mid 19 \mid 39]$ \\
\mn{u1c1} \mn{q7}& \multicolumn{4}{c}{all answers are trivial}\\
\mn{u1c1} \mn{q11}&\multicolumn{4}{c}{all answers are trivial}\\
\mn{u1c1} \mn{q14}&$[8 \mid 12 \mid 17]$ & $[18 \mid 28 \mid 39]$ & $[9 \mid 13 \mid 17]$ & $[21 \mid 31 \mid 40]$ \\
\mn{u1c1} \mn{q17}&\multicolumn{4}{c}{all answers are trivial}\\
\bottomrule
\end{tabular*}
}
\end{table}

\begin{table}
\caption{Number of variables and clauses in the encoding for a single cause $\Phi_{P\text{-}IAR}(\Cmc)$ ($[minimum \mid average \mid maximum]$), with \cqaprienc or \cavsatenc encoding, for P$_2$ encoding (score-structured priority, $n=5$). }
{\small
\setlength{\tabcolsep}{0pt}
\begin{tabular*}{\textwidth}{l @{\extracolsep{\fill}}r r r r }
\toprule
&  \multicolumn{2}{c}{\cqaprienc}& \multicolumn{2}{c}{\cavsatenc}\\
&\#var. & \#cl. &\#var. & \#cl. \\
\midrule
\mn{u1c1} \mn{q1}&$[1 \mid 1 \mid 2]$ & $[3 \mid 26 \mid 520]$ & $[4 \mid 27 \mid 523]$ & $[12 \mid 75 \mid 1,361]$  \\
\mn{u1c1} \mn{q4}&$[1 \mid 1 \mid 1]$ & $[1 \mid 7 \mid 17]$ & $[2 \mid 7 \mid 17]$ & $[4 \mid 21 \mid 41]$  \\
\mn{u1c1} \mn{q7}& \multicolumn{4}{c}{all answers are trivial}\\
\mn{u1c1} \mn{q11}&\multicolumn{4}{c}{all answers are trivial}\\
\mn{u1c1} \mn{q14}&$[1 \mid 1 \mid 1]$ & $[8 \mid 12 \mid 17]$ & $[9 \mid 13 \mid 17]$ & $[22 \mid 33 \mid 44]$  \\
\mn{u1c1} \mn{q17}&\multicolumn{4}{c}{all answers are trivial}\\
\bottomrule
\end{tabular*}
}
\end{table}

\begin{table}
\caption{Number of variables and clauses in the encoding for a single cause $\Phi_{C\text{-}IAR}(\Cmc)$ ($[minimum \mid average \mid maximum]$), with \cqaprienc or \cavsatenc encoding (score-structured priority, $n=5$). }
{\small
\setlength{\tabcolsep}{0pt}
\begin{tabular*}{\textwidth}{l @{\extracolsep{\fill}}r r r r}
\toprule
&  \multicolumn{2}{c}{\cqaprienc}& \multicolumn{2}{c}{\cavsatenc}\\
&\#var. & \#cl. &\#var. & \#cl. \\
\midrule
\mn{u1c1} \mn{q1}&\multicolumn{4}{c}{time-out or out-of-memory}\\
\mn{u1c1} \mn{q4}&$[1 \mid 177 \mid 580]$ & $[2 \mid 1,251 \mid 4,750]$ & $[7 \mid 185 \mid 580]$ & $[14 \mid 1,281 \mid 4,752]$\\
\mn{u1c1} \mn{q7}& \multicolumn{4}{c}{all answers are trivial}\\
\mn{u1c1} \mn{q11}&\multicolumn{4}{c}{all answers are trivial}\\
\mn{u1c1} \mn{q14}&$[129 \mid 333 \mid 582]$ & $[507 \mid 2,313 \mid 4,758]$ & $[163 \mid 379 \mid 582]$ & $[715 \mid 2,709 \mid 4,759]$\\
\mn{u1c1} \mn{q17}&\multicolumn{4}{c}{all answers are trivial}\\
\bottomrule
\end{tabular*}
}
\end{table}

\end{document}